\documentclass[longauth]{aa}  

\usepackage{graphicx}
\usepackage{txfonts}
\usepackage{siunitx}
\usepackage{caption}
\usepackage{subcaption}
\usepackage[utf8]{inputenc}
\usepackage{amsmath}
\usepackage[english]{babel}
\usepackage{float}
\usepackage{url}
\usepackage{fancyhdr}
\usepackage[version=4]{mhchem}
\usepackage{textcomp, gensymb}
\usepackage{lscape}
\usepackage{adjustbox}
\usepackage{longtable}
\usepackage{pdflscape}
\usepackage{natbib}
\usepackage[colorlinks=true, citecolor=blue, linkcolor=blue, urlcolor=blue]{hyperref}
\bibpunct{(}{)}{;}{a}{}{,}             
\newcommand{\matrixtk}{{\fontfamily{pcr}\selectfont MATRIX}\,}

\begin{document}

   \title{TOI-4438\,b: a transiting mini-Neptune amenable to atmospheric characterization}
   \titlerunning{TOI-4438\,b}
   \subtitle{}
     
\author{
E.\,Goffo\inst{\ref{i:unito},\ref{i:tls}},
P.\,Chaturvedi\inst{\ref{i:tls}},
F.\,Murgas\inst{\ref{i:iac}, \ref{i:ull}},
G.\,Morello\inst{\ref{i:swe},\ref{i:iac},\ref{i:iaa}},
J.\,Orell-Miquel\inst{\ref{i:iac}, \ref{i:ull}},
L.\,Acu\~na\inst{\ref{i:max}},
L.\,Pe\~na-Mo\~nino\inst{\ref{i:iaa}},
E.\,Pall\'e\inst{\ref{i:iac},\ref{i:ull}},
A.\,P.\,Hatzes\inst{\ref{i:tls}},
S.\,Gerald\'ia-Gonz\'alez\inst{\ref{i:iac}, \ref{i:ull}},
F.\,J.\,Pozuelos\inst{\ref{i:iaa}},
A.\,F.\,Lanza\inst{\ref{i:inafcat}},
D.\,Gandolfi\inst{\ref{i:unito}},
J.\,A.\,Caballero\inst{\ref{i:esac}},
M.\,Schlecker\inst{\ref{i:ari}},
M.\,P\'erez-Torres\inst{\ref{i:iaa},\ref{i:capa},\ref{i:cyp}},
N.\,Lodieu\inst{\ref{i:iac},\ref{i:ull}},
A.\,Schweitzer\inst{\ref{i:ham}},
C.\,Hellier\inst{\ref{keele}},
S.\,V.\,Jeffers\inst{\ref{i:maxgot}},
C.\,Duque-Arribas\inst{\ref{i:ucm}},
C.\,Cifuentes\inst{\ref{i:esac}}, 
V.\,J.\,S.\,B\'ejar\inst{\ref{i:iac},\ref{i:ull}},
M.\,Daspute\inst{\ref{i:ariel}},
F.\,Dubois\inst{\ref{i:ver},\ref{i:astrolab}},
S.\,Dufoer\inst{\ref{i:ver}},
E.\,Esparza-Borges\inst{\ref{i:iac},\ref{i:ull}},
A.\,Fukui\inst{\ref{i:meg}, \ref{i:iac}},
Y.\,Hayashi\inst{\ref{i:tokio}},
E.\, Herrero\inst{\ref{i:ieec}},
M.\,Mori\inst{\ref{i:tokio}},
N.\,Narita\inst{\ref{i:meg}, \ref{i:mitaka}, \ref{i:iac}},
H.\,Parviainen\inst{\ref{i:iac},\ref{i:ull}},
L.\,Tal-Or\inst{\ref{i:ariel}},
S.\,Vanaverbeke\inst{\ref{i:ver},\ref{i:astrolab}, \ref{i:be3}},
I.\,Hermelo\inst{\ref{i:caha}},
P.\,J.\,Amado\inst{\ref{i:iaa}},  
S.\,Dreizler\inst{\ref{i:got}},
Th.\,Henning\inst{\ref{i:max}},
J.\,Lillo-Box\inst{\ref{i:esac}},
R.\,Luque\inst{\ref{i:chi},\ref{i:iaa}},
M.\,Mallorqu\'in\inst{\ref{i:iac},\ref{i:ull}},
E.\,Nagel\inst{\ref{i:got}},
A.\,Quirrenbach\inst{\ref{i:hei}}, 
S.\,Reffert\inst{\ref{i:hei}},
A.\,Reiners\inst{\ref{i:got}}, 
I.\,Ribas\inst{\ref{i:ieec},\ref{i:ice}},
P.\,Sch\"ofer\inst{\ref{i:iaa}},
H.\,M.\,Tabernero\inst{\ref{i:ucm},\ref{i:cab1}},
M.\,Zechmeister\inst{\ref{i:got}}
}

\authorrunning{E.\,Goffo et al.}

\institute{
\label{i:unito}Dipartimento di Fisica, Universit\`a degli Studi di Torino, via Pietro Giuria 1, I-10125, Torino, Italy
\and
\label{i:tls}Th\"uringer Landessternwarte Tautenburg, 07778 Tautenburg, Germany  
\and
\label{i:iac}Instituto de Astrof\'isica de Canarias (IAC), Calle V\'ia L\'actea s/n, 38205 La Laguna, Tenerife, Spain
\and
\label{i:ull}Departamento de Astrof\'isica, Universidad de La Laguna (ULL), 38206 La Laguna, Tenerife, Spain
\and        
\label{i:swe}Department of Space, Earth and Environment, Chalmers University of Technology, SE-412 96 Gothenburg, Sweden
\and  
\label{i:max}Max-Planck-Institut f\"ur Astronomie, K\"onigstuhl 17, 69117 Heidelberg, Germany
\and
\label{i:iaa}Instituto de Astrof\'isica de Andaluc\'ia (IAA-CSIC), Glorieta de la Astronom\'ia s/n, 18008 Granada, Spain
\and
\label{i:inafcat}INAF – Osservatorio Astrofisico di Catania, Via S. Sofia 78, 95123 Catania, Italy
\and
\label{i:esac}Centro de Astrobiología, (CSIC-INTA), ESAC Campus, Camino bajo del castillo s/n, 28692 Villanueva de la Cañada, Madrid, Spain
\and
\label{i:ari}Steward Observatory and Department of Astronomy, The University of Arizona, Tucson, AZ 85721, USA
\and        
\label{i:ham}Hamburger Sternwarte, Gojenbergsweg 112, 21029 Hamburg, Germany
\and
\label{keele}Astrophysics Group, Keele University, Staffordshire ST5 5BG, United Kingdom
\and
\label{i:maxgot}Max Planck Institut f\"ur Sonnensystemforschung, Justus-von-Liebig-Weg 3, 37077 G\"ottingen, Germany
\and 
\label{i:ucm}Departamento de F\'isica de la Tierra y Astrof\'isica and IPARCOS-UCM (Instituto de F\'isica de Part\'iculas y del Cosmos de la UCM), Facultad de Ciencias F\'isicas, Universidad Complutense de Madrid, 28040 Madrid, Spain
\and
\label{i:ariel}Department of Physics, Ariel University, Ariel 40700, Israel
\and
\label{i:ver}Vereniging Voor Sterrenkunde, Oude Bleken 12, 2400 Mol, Belgium
\and
\label{i:astrolab}AstroLAB IRIS, Provinciaal Domein “De Palingbeek”, Verbrande-molenstraat 5, 8902 Zillebeke, Ieper, Belgium
\and
\label{i:meg}Komaba Institute for Science, The University of Tokyo, 3-8-1 Komaba, Meguro, Tokyo 153-8902, Japan
\and
\label{i:tokio}Department of Multi-Disciplinary Sciences, Graduate School of Arts and Sciences, The University of Tokyo, 3-8-1 Komaba, Meguro, Tokyo, 153-8902, Japan
\and 
\label{i:ieec}Institut d’Estudis Espacials de Catalunya (IEEC), Calle Gran Capita 2-4, 08034, Barcelona, Spain
\and
\label{i:mitaka}Astrobiology Center, 2-21-1 Osawa, Mitaka, Tokyo 181-8588, Japan
\and
\label{i:be3}Centre for Mathematical Plasma-Astrophysics, Department of Mathematics, KU Leuven, Celestijnenlaan 200B, 3001 Heverlee, Belgium
\and
\label{i:caha}Centro Astron\'omico Hispano en Andaluc\'ia (CAHA), Observatorio de Calar Alto, Sierra de los Filabres, 04550 G\'ergal, Spain
\and
\label{i:got}Institut f\"ur Astrophysik und Geophysik, Georg-August-Universit\"at G\"ottingen, Friedrich-Hund-Platz 1, 37077 G\"ottingen, Germany
\and
\label{i:chi}Department of Astronomy \& Astrophysics, University of Chicago, Chicago, IL 60637, USA
\and
\label{i:hei}Landessternwarte, Zentrum f\"ur Astronomie der Universit\"at Heidelberg, K\"onigstuhl 12, 69117 Heidelberg, Germany  
\and
\label{i:ice}Institut de Ci\'encies de l’Espai (CSIC-IEEC), Campus UAB, de Can Magrans s/n, 08193 Bellaterra, Barcelona, Spain
\and
\label{i:cab1}Centro de Astrobiolog\'ia (CSIC-INTA), Carretera de Ajalvir km 4, 28850 Torrej\'on de Ardoz, Madrid, Spain  
\and
\label{i:capa}Center for Astroparticles and High Energy Physics (CAPA), Universidad de Zaragoza, E-50009 Zaragoza, Spain
\and
\label{i:cyp}School of Sciences, European University Cyprus, Diogenes street, Engomi, 1516 Nicosia, Cyprus
}
        
\date{Received 30 December 2023 /  Accepted dd Month 2024}

\abstract{
We report the confirmation and mass determination of a mini-Neptune transiting the M3.5\,V star TOI-4438 (G 182-34) every 7.44 days. A transit signal was detected with NASA's \textit{TESS} space mission in the sectors 40, 52, and 53. In order to validate the planet TOI-4438\,b and to determine the system properties, we combined \textit{TESS} data with high-precision radial velocity measurements from the CARMENES spectrograph, spanning almost one year, and ground-based transit photometry.
We found that TOI-4438\,b has a radius of $R_\mathrm{b}$\,=\,2.52\,$\pm$\,0.13 R$_\oplus$ (5\% precision), which together with a mass of $M_\mathrm{b}$\,=\,5.4\,$\pm$\,1.1~M$_\oplus$ (20\% precision), results in a bulk density of $\rho_\mathrm{b}$\,=\,1.85$^{+0.51}_{-0.44}$\,g\,cm$^{-3}$ ($\sim$28\% precision), aligning the discovery with a volatile-rich planet.
Our interior structure retrieval with a pure water envelope yields a minimum water mass fraction of 46\% (1$\sigma$).
TOI-4438\,b is a volatile-rich mini-Neptune with likely H/He mixed with molecules, such as water, CO$_2$, and CH$_4$.
The primary star has a $J$-band magnitude of 9.7, and the planet has a high transmission spectroscopy metric (TSM) of 136\,$\pm$\,13. Taking into account the relatively warm equilibrium temperature of $T_\mathrm{eq}$\,=\,435\,$\pm$\,15~K, and the low activity level of its host star, TOI-4438\,b is one of the most promising mini-Neptunes around an M dwarf for transmission spectroscopy studies. 
}
   
\keywords{planetary systems -- planets and satellites: individual: TOI-4438\,b  --  planets and satellites: atmospheres -- methods: radial velocity -- techniques:  spectroscopic -- stars: low-mass}
\maketitle 

\section{Introduction}
\label{sec:introduction}
 
Precise and accurate planetary mass and radius measurements are necessary to enable the atmospheric characterization of exoplanets. Current and forthcoming telescopes such as the James Webb Space Telescope (\textit{JWST}), the Atmospheric Remote-sensing Infrared Exoplanet Large-survey (\textit{Ariel}), and the Extremely Large Telescope ELT will expand our frontiers of planetary science, allowing us to explore the composition of the upper gaseous layer of exoplanets. Selecting the best targets amenable to future atmospheric characterization is therefore crucial. 

Since the beginning of this century, the \textit{CoRoT} \citep{Baglin2006}, \textit{Kepler} \citep{Borucki2010}, \textit{K2} \citep{Howell2014}, and \textit{TESS} \citep{Ricker2015} space-based telescopes have opened our eyes to several thousands of transiting exoplanets in our galaxy.
Surprisingly, most of these planets have no counterparts within our solar system, since their sizes are between the ones of Earth and Neptune \citep[1.0\,$\lesssim\,R_\mathrm{p}$\,$\lesssim$\,4.0\,R$_{\oplus}$;][]{2013ApJS..204...24B}. Small planets with radii 1.8$\lesssim\,R_\mathrm{p}$\,$\lesssim$\,4.0\,R$_{\oplus}$, the so-called mini-Neptunes, seem to host hydrogen-dominated atmospheres.  
The composition of the outer atmosphere as well as the  internal planetary structure are still
not well understood, as they could be explained by several degenerate models \citep{Zeng2019PNAS..116.9723Z}. Atmospheric follow-up observations are needed to solve the discrepancy between various composition models. 
With the methodology proposed by \cite{Kempton2018PASP..130k4401K}, one can calculate the expected signal-to-noise ratio (S/N) of transmission spectra for a transiting planet, based on the strength of spectral features and the brightness of the host star, assuming cloud-free atmospheres, namely the transmission spectroscopy metric (TSM).
The TSM is used for determining which transiting exoplanets are the best targets for atmospheric characterization via transmission spectroscopy, especially with the Near Infrared Imager and Slitless Spectrograph (NIRISS) on \textit{JWST}.

Small planets including mini-Neptunes happen to be common around M-type dwarf stars \citep{2013ApJ...767...95D, 2021A&A...653A.114S,2023A&A...670A.139R}, as found also in data from the \textit{TESS} mission \citep{chaturvedi2022, espinoza2022, 2022A&A...666A...4K, 2023A&A...675A.177G, 2023A&A...677A..38B}.
Follow-up observations focused on M dwarfs have their own advantages as these are abundant compared to F, G, and K stars. Their smaller stellar radii and masses produce deeper transit signals (the transit depth $\delta$ is inversely proportional to $R_\star^2$) and larger Doppler reflex motions (the radial velocity semi-amplitude $K_\star$ is inversely proportional to $M_\star^{2/3}$), increasing our capability to detect small low-mass planets \citep[see, e.g.,][]{2023A&A...670A.139R}.

Here, we present the characterisation of TESS object of interest (TOI)-4438\,b, a transiting mini-Neptune orbiting the M3.5\,V star TOI-4438 (G 182-34). We used CARMENES precise radial velocities (RVs) and ground-based transit observations to confirm the planet and to derive its fundamental parameters, such as mass, radius, orbital period, semi-major axis, and equilibrium temperature.
TOI-4438\,b with a score of $\sim$136 scales very high on TSM, thereby, making it a promising target for future atmospheric observations with \textit{JWST}. 
This work is part of the \textit{TESS} follow-up program within the CARMENES observations survey \citep[see, e.g.,][]{2023A&A...675A.177G, 2023A&A...670A..84K, 2023arXiv231010244M, 2023A&A...677A.182M, 2023A&A...678A..80P}.

\section{Observations}
\label{sec:Observations}

\subsection{TESS photometry}\label{tess}

\textit{TESS} observed TOI-4438 in sector 40, from 25 June to 23 July 2021, with CCD\#2 of camera 1 with an integration time of 2 min. TOI-4438 had more observations taken in sector 52, from 19 May to 12 June 2022, and in sector 53, from 13 June to 8 July 2022. The observations were done again at the same cadence of 2 min with CCD\#1 of camera 1 in sector 52, and with CCD\#2 of camera 1 in sector 53.  We retrieved the \textit{TESS} data from the Mikulski Archive for Space Telescopes (MAST)\footnote{ \url{https://mast.stsci.edu/portal/Mashup/Clients/Mast/Portal.html}.} The data were processed by Science Processing Operations Center (SPOC, \citealp{SPOC}) pipeline using the Presearch Data Conditioning Simple Aperture Photometry (PDCSAP) algorithm. For details, refer \cite{2012Stumpe, 2014PASP..126..100S, 2012Smith}.  

\begin{figure*}
     \centering
     \begin{subfigure}[b]{0.33\textwidth}
         \centering
         \includegraphics[width=\textwidth]{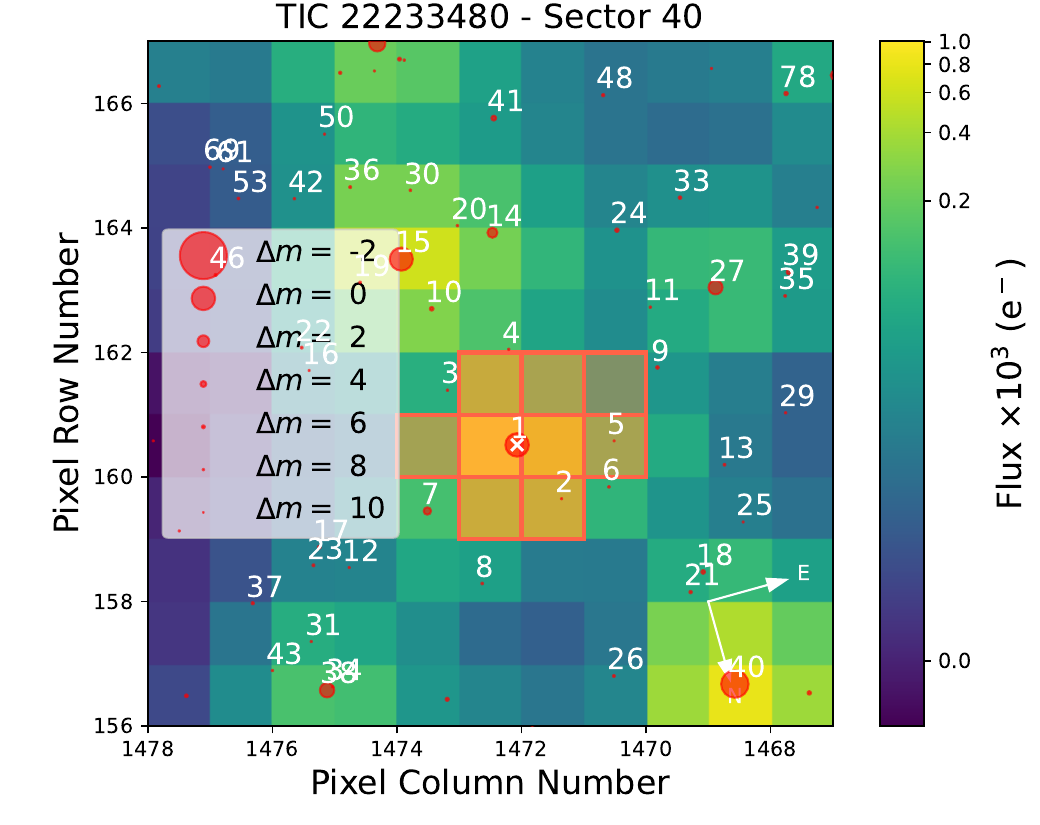}
     \end{subfigure}
     \hfill
     \begin{subfigure}[b]{0.33\textwidth}
         \centering
         \includegraphics[width=\textwidth]{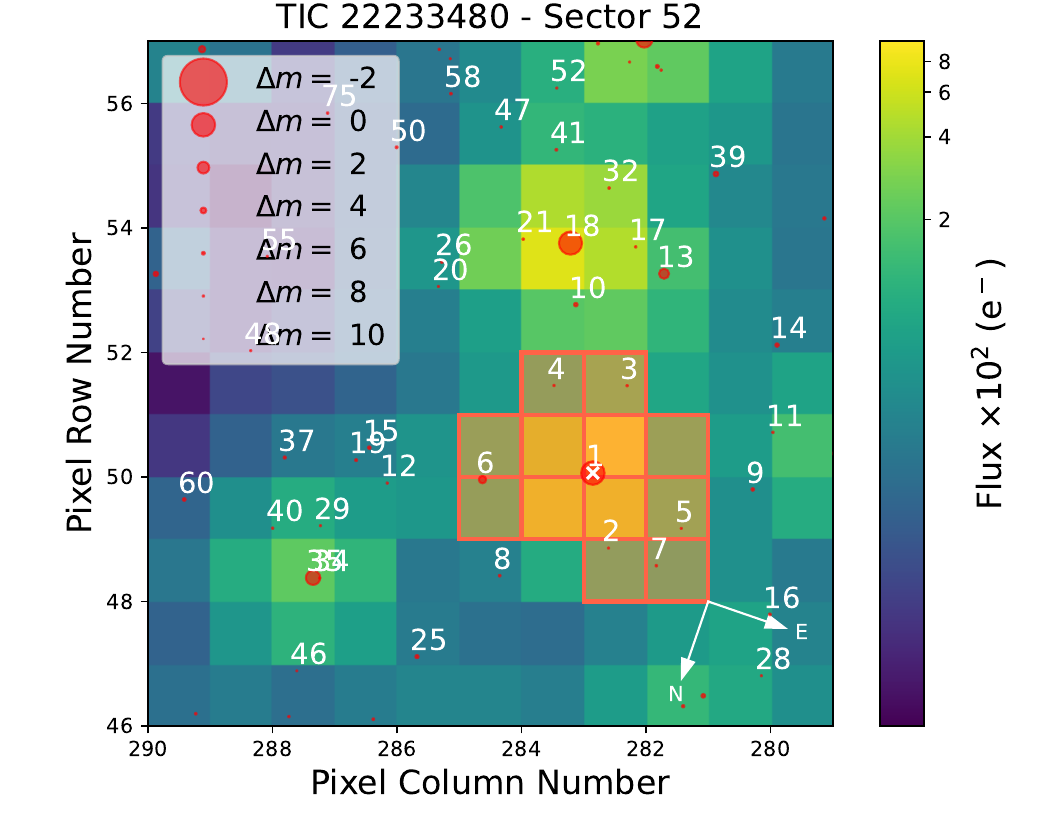}
     \end{subfigure}
     \hfill
     \begin{subfigure}[b]{0.33\textwidth}
         \centering
         \includegraphics[width=\textwidth]{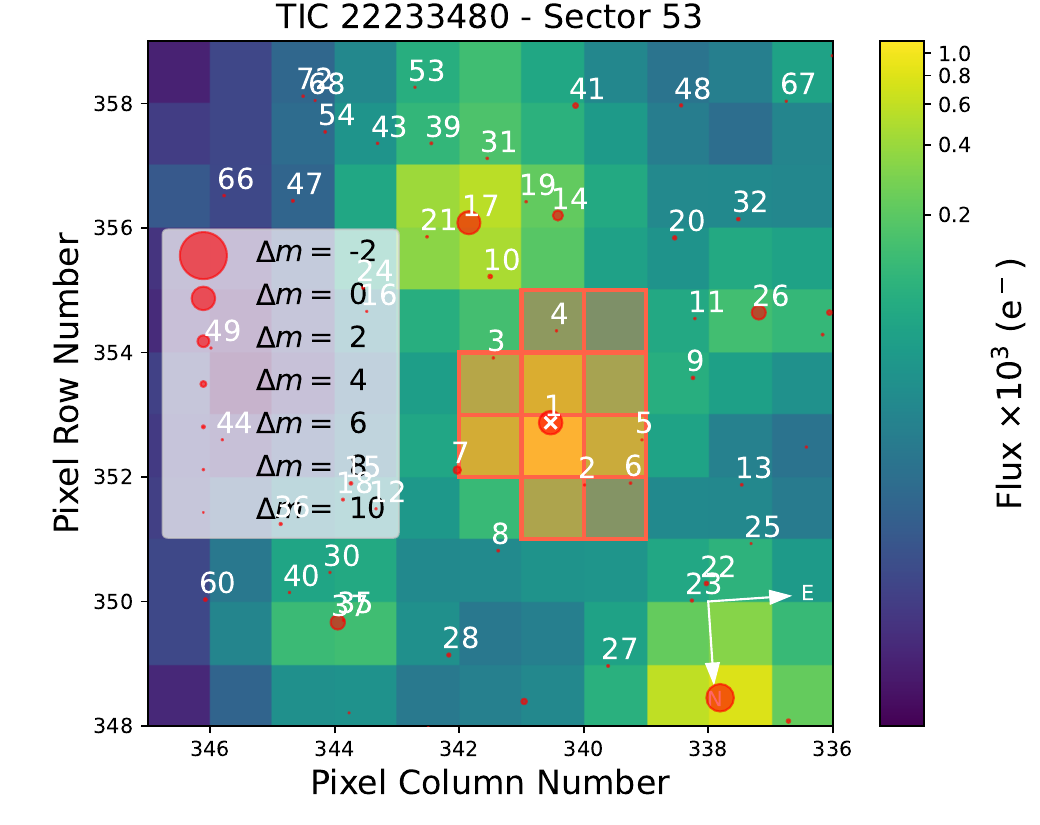}
     \end{subfigure}
        \caption{TPFs  of TOI-4438 in TESS sectors 40, 52, and 53. The red bordered squares mark the optimal photometric aperture used to extract the SAP fluxes. The different sizes of the red circles show the $G$-band magnitudes from $Gaia$ DR3 for all nearby stars with respect to TOI-4438 (marked with an "$\times$'') up to $\Delta G$\,=\,10 mag fainter.}
        \label{fig:TPF_Gaia}
\end{figure*}

\begin{figure*}
     \centering
     \begin{subfigure}[b]{0.9\textwidth}
         \centering
         \includegraphics[width=\textwidth]{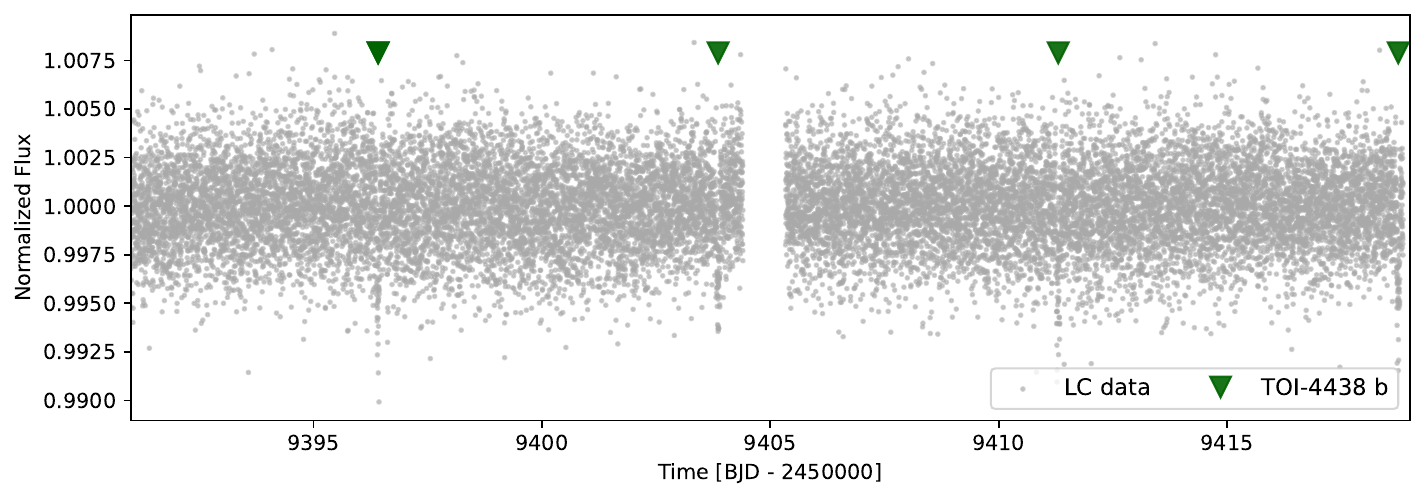}
     \end{subfigure}
     \begin{subfigure}[b]{0.9\textwidth}
         \centering
         \includegraphics[width=\textwidth]{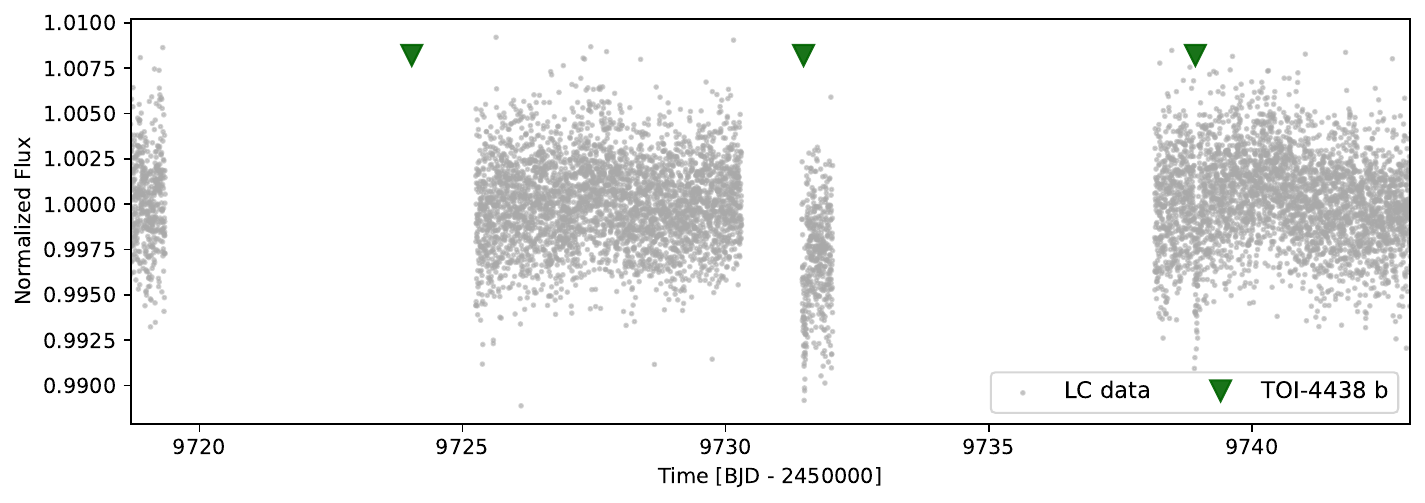}
     \end{subfigure}
     \begin{subfigure}[b]{0.9\textwidth}
         \centering
         \includegraphics[width=\textwidth]{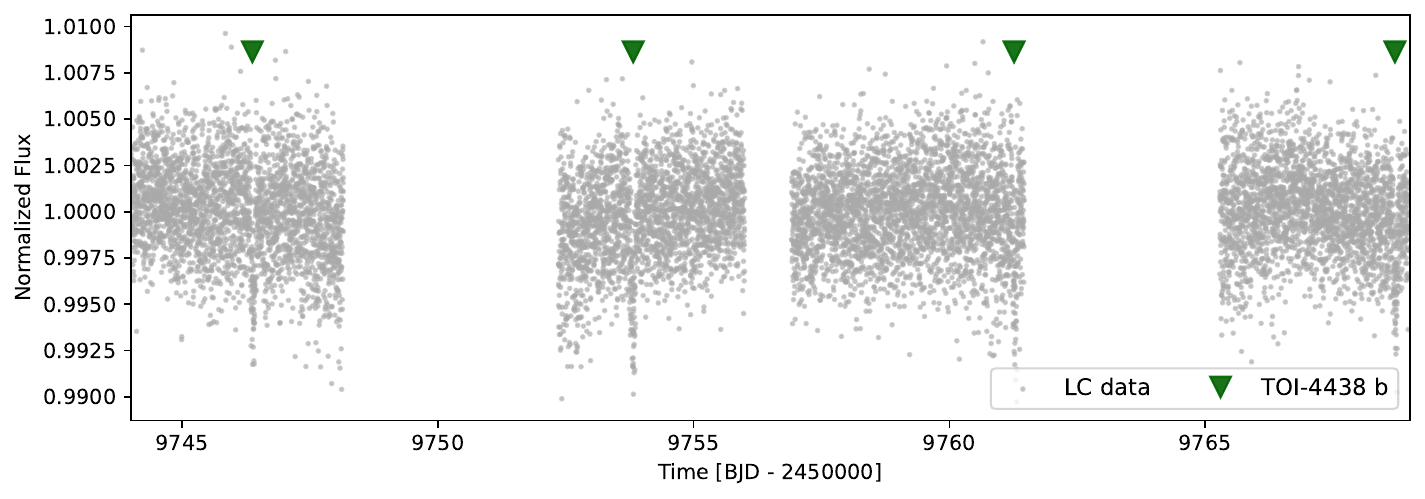}
     \end{subfigure}
        \caption{\textit{TESS} PDCSAP light curve of TOI-4438 from sectors 40, 52, and 53. The green triangles mark the timings of the transits of TOI-4438\,b.}
        \label{fig:TESS_lc_sectors}
\end{figure*}

In order to correct for the photometric contamination caused by \textit{TESS}'s large pixel size of 21\arcsec, a photometric mask is used to extract the \textit{TESS} Simple Aperture Photometry (SAP) light curve.
We plotted the target pixel files (TPFs; Fig.\,\ref{fig:TPF_Gaia}) of TOI-4438 using the code \texttt{tpfplotter}\footnote{\url{https://github.com/jlillo/tpfplotter}.} \citep{aller2020}. The TPFs include all the \textit{Gaia} Data Release 3 (DR3, \citealt{gaiadr3}) sources, down to $\Delta G$\,=\,10\,mag fainter than the target star and spatially located within the field of view of the \textit{TESS} aperture. No bright sources fall inside the photometric aperture of TOI-4438.
For the analysis presented in this paper, we used the \textit{TESS} contamination corrected PDCSAP light curves (Fig.\,\ref{fig:TESS_lc_sectors}).

\subsection{CARMENES radial velocity follow-up}\label{sec:rv}

We spectroscopically monitored TOI-4438 with the Calar Alto high-Resolution search for M dwarfs with Exoearths with Near-infrared and optical Echelle Spectrographs (CARMENES) installed at the 3.5\,m telescope at Calar Alto Observatory, Almer\'ia, Spain. The instrument comprises two highly stable, fiber-fed spectrographs \citep{CARMENES,CARMENES18}, one covering the wavelength range 520--960\,nm (VIS) with a resolving power of R\,$\approx$\,94\,600, the other covering the near-infrared range 960--1710\,nm (NIR), with a resolving power of R\,$\approx$\,80\,400.

We collected 34 high-resolution spectra between 21 May 2022 and 2 May 2023.  
Depending on the sky condition and observing constraints, each of the spectra was exposed between 700 and 1800~s. The resultant average S/N was $\sim$47 at 740\,nm. We followed the standard \texttt{caracal} pipeline \citep{2014A&A...561A..59Z} to reduce the CARMENES data.

The precise relative RVs were computed using the {\texttt{serval}}\footnote{\url{https://github.com/mzechmeister/serval}.} pipeline \citep{2018A&A...609A..12Z} with the standard corrections for barycentric and secular motions. The ancillary corrections applied for the nightly drift by incorporating the nightly zero-points and the atmospheric effects using the telluric lines are described by \cite{Nagel23}. In our analysis we also used the {\texttt{serval}} produced spectral activity indicators, namely, the chromatic index (CRX), the differential line width (dLW), the H$\alpha$ index, the Ca~{\sc{ii}} infrared triplet (IRT) index, and the Na~{\sc{i}}\,D doublet index.

We discarded two data points in the CARMENES VIS RV time series due to their low signal-to-noise ratio (S/N\,$<$\,21). The final CARMENES dataset includes 32 measurements with a root mean square (RMS) of\,$\approx$\,2\,m\,s$^{-1}$. 
We did not include the NIR RVs in our further analysis, given their low signal-to-noise ratio.
Table\,\ref{tab:CARMV_RV_VIS} lists the CARMENES VIS RV measurements with the time information in the Barycentric Julian Date- Barycentric Dynamical Time (BJD$_{\mathrm{TDB}}$) units.

\subsection{Ground-based photometry}

\subsubsection{MuSCAT2}\label{muscat2}

TOI-4438 was observed on three nights, namely, on 17 May 2022, 2 May 2023, and 8 July 2023, with the multi-band imager MuSCAT2 \citep{Narita2019} mounted on the 1.5\,m Telescopio Carlos S\'{a}nchez (TCS) at Teide Observatory, Tenerife, Spain. MUSCAT2 is equipped with four 1k$\times$1k pixel CCDS, with each having a field of view of 7.4\arcmin$\times$7.4\arcmin. The cameras can simultaneously observe in the $g'$, $r'$, $i'$, and $z_s$ bands. We observed two complete transits and one partial transit of TOI-4438\,b in each band. The times for exposure varied between 10\,s to 30\,s. The data reduction were carried out using the standard pipeline as described by \cite{Parviainen2019, Parviainen2015}.

\subsubsection{SuperWASP}

The SuperWASP transit search \citep{2006PASP..118.1407P} observed 
the field of TOI-4438 in 2004 and then every year from 2007 to 2010. The eight cameras have each 2048$\times$2048 CCDs with f/1.8 lenses. A total of 71\,000 photometric data points were obtained over a period of of $\sim$\,120 nights in the visible passband of 400--700\,nm.  At $V$\,=\,13.7\,mag, TOI-4438 is relatively faint for WASP, though it is the brightest star in the 48\arcsec\ extraction aperture.  

\subsubsection{TJO}

We carried out a dedicated photometry monitoring campaign of TOI-4438 on 49 different nights from April to September 2023 with the 0.8\,m Telescopi Joan Oró (TJO; \citealt{2010SPIE.7740E..3KC}) at the Montsec Observatory, Lleida, Spain. A total of 285 data points were observed using the Johnson $R$ filter of the LAIA imager on a 4k$\times$4k CCD with a 30\arcmin\ field of view. 
The standard reduction process was followed using the {\tt{AstroImageJ}}\citep{2017AJ....153...77C} package and the {\tt{icat}} pipeline \citep{2006IAUSS...6E..11C} of the TJO. We used indigenous pipelines for removal of outliers for observations made during poor sky conditions. We obtained an effective rms of $\sim$9 ppt in the $R$ filter.

\subsubsection{LCOGT}

We performed photometric monitoring of TOI-4438 in the $V$ band between 2 August and 6 November 2023, using the SBIG camera of one of the 40\,cm telescope of Las Cumbres Observatory Global Telescope \citep[LCOGT;][]{Brown2013} providing a field of view of 29.2\arcmin$\times$19.5\arcmin. We took three exposures of 600\,s in each of the 71 different observing epochs during 96 d. 
The observing conditions were mostly good and the average seeing varied from 2.0 to 6.0\arcsec. 
Raw data were processed using the {\tt{BANZAI}} pipeline \citep{McCully2018} and differential aperture photometry of TOI-4438 was performed using the {\tt{AstroImageJ}} software. 
An optimal aperture of 8 pixels ($\sim$ 4.5\arcsec) was adopted to minimize the dispersion of the differential light curves. 

We used the 1\,m LCOGT telescopes from 9 June to 4 November 2023, to observe TOI-4438 in the $B$ filter with an exposure time of 120 s until JD\,=\,2460156, and of 77 s afterwards. This produced 88 images, out of which nine were found to be bad.  
The final light curve has an rms of 0.57\% $\sim$ with 79 data points, spanning 148 d. 

\subsubsection{ASAS-SN}

TOI-4438 was photometrically monitored by the All-Sky Automated Survey for Supernovae (ASAS-SN; \citealp{Shappee2014, Kochanek2017}). ASAS-SN consists of a network of 24 robotic telescopes with a diameter of 14\,cm, distributed around the world. We retrieved the ASAS-SN light curve from the webpage of the project\footnote{\url{https://asas-sn.osu.edu/}.}. There are 246 $V$-band photometry measurements taken for the star between 3 May 2013 and 25 September 2018 with a rms scatter of $\sim$0.021 mag.

\subsubsection{e-EYE}

TOI-4438 was also observed by e-EYE (Entre Encinas y Estrellas)\footnote{\url{www.e-eye.es/}.} 16\arcsec\ ODK corrected-Dall-Kirkham reflector, located at Fregenal de la Sierra in Badajoz, Spain. $V$ band observations were taken between 12 September and 13 November 2023 with a Kodak KAF-16803 CCD chip mounted on ASA DDM85. The images were reduced using the {\texttt{LesvePhotometry} package\footnote{\url{www.dppobservatory.net}.}. The rms of the data was $\sim$0.006 mag.}

\subsubsection{ADONIS}

TOI-4438 was observed by the privately owned  ADONIS observatory in the $V$ filter from 30 September to 6 December 2023. The telescope used is 0.25 m skywatcher Quattro Newtonian telescope with a Moravian CCD camera.
{\texttt{LesvePhotometry} is used to perform the aperture photometry.  

\section{Stellar properties}

The star TOI-4438 was originally discovered as a high-proper motion star (34th in the plate 182 in Hercules) in the northern hemisphere at the Lowell Observatory by \cite{GICLAS196623}. The star has been identified as G~182--34 \citep{1971lpms.book.....G} and has been tabulated by a number of catalogs on proper motion \citep{1979lccs.book.....L, 2005AJ....129.1483L, 2016ApJ...817..112S}, parallax \citep{1988AJ.....95..237D, 2002AJ....124.1170D, 1995gcts.book.....V}, or the solar neighborhood \citep{2002AJ....123.2806R, 2011AJ....142..138L, 2013MNRAS.435.2161F, 2021A&A...649A...6G}.

\subsection{Stellar parameters}
\label{Star_param}

We derived the stellar parameters as discussed in \citet{schweitzer2019}. The total luminosity was computed by summing over the spectral energy distribution with  {\tt{VOSA}} \citep{VOSA}. Broadband photometry from passbands Johnson ($B$) to AllWISE ($W4$) and astrometry from \textit{Gaia} DR3 catalog were utilized. 
The stellar atmospheric parameters ($T_{\rm eff}$, $\log{g}$, and [Fe/H]) were derived with the {\tt SteParSyn}\footnote{\url{https://github.com/hmtabernero/SteParSyn/}.} code \citep{tab22} using the line list and model grid described by \citet{marfil21}.
The mass of the star was derived using the the linear mass-radius relation \citep{schweitzer2019} whereas the radius of the star was derived from Stefan-Boltzmann's law assuming the bolometric luminosity and the spectroscopic temperature.
The stellar parameters of TOI-4438 are listed in Table\,\ref{tab:stellar_parameters}.
The effective temperature of $T_\mathrm{eff}$\,=\,3422\,$\pm$\,81~K and surface gravity of $\log{g_\star}$\,=\,4.68\,$\pm$\,0.04 (cgs) match the $\sim$ M3\,V spectral-type from the spectro-photometric relations of \cite{2013ApJS..208....9P} and \cite{cifuentes2020} and, the M3.5$\pm$0.5\,V spectral type from low-resolution optical spectroscopy by \cite{2003AJ....126.3007R}.

\begin{table}
\caption{Stellar parameters of TOI-4438.}\label{tab:stellar_parameters}
\centering
\begin{tabular}{lcr}
\hline\hline
\noalign{\smallskip}
Parameter & Value & Reference\\
\noalign{\smallskip}
\hline
\noalign{\smallskip}
Identifiers & G 182-34 & Gic71 \\
& Karmn J18012+355 & Cab16\\
 & TIC~22233480  & TIC\\
\noalign{\smallskip}
\hline
\noalign{\smallskip}
Sp. type & M3.5\,V & Reid03 \\
\noalign{\smallskip}
\hline
\noalign{\smallskip}
$\alpha$ (ICRS, epoch 2016.0) & 18:01:16.14 & \textit{Gaia} DR3\\
$\delta$ (ICRS, epoch 2016.0) & +35:35:41.6  & \textit{Gaia} DR3\\
$\mu_a\,\cos\delta$ (mas\,yr$^{-1}$)  & +29.212 $\pm$ 0.017 & \textit{Gaia} DR3\\
$\mu_{\delta}$ (mas\,yr$^{-1}$)  & --517.512 $\pm$ 0.018 & \textit{Gaia} DR3\\
$\varpi$ (mas) & 33.2642\,$\pm$\,0.0154 & \textit{Gaia} DR3\\
$d$ (pc) & 30.0623\,$\pm$\, 0.0139 & \textit{Gaia} DR3\\
$\gamma$ (km\,s$^{-1}$) & --35.81 $\pm$ 0.47& \textit{Gaia} DR3\\
\noalign{\smallskip}
\hline
\noalign{\smallskip}
$T_{\text{eff}}$ [K] & 3422\,$\pm$\,81 & This work \\
$\log{g_\star}$ [cgs] & 4.68\,$\pm$\,0.04 & This work \\
{[Fe/H]} [dex] & --0.19\,$\pm$\,0.05 & This work \\
\noalign{\smallskip}
\hline
\noalign{\smallskip}
$L_{\star}$ [$L_{\odot}$] &  0.01706\,$\pm$\,0.00007 & This work\\
$R_{\star}$ [$R_{\odot}$] & 0.372\,$\pm$\,0.018 & This work \\
$M_{\star}$ [$M_{\odot}$] & 0.368\,$\pm$\,0.021 & This work \\
$\rho_{\star}$ [g\,cm$^{-3}$] & 10.06\,$\pm$\,1.62 & This work\\ 
$v\sin\,i_{\star}$ [km\,s$^{-1}$] & $<$\,2 & This work\\
$P_{\text{rot}}$ [d] & 68\,$\pm$\,6 & This work\\
Age [Gyr] & 5.1\,$\pm$\,2.8 & This work\\
\noalign{\smallskip}
\hline
\noalign{\smallskip}
$G$ [mag] & 12.50\,$\pm$\,0.68 & \textit{Gaia} DR3\\
$T$ [mag] & 11.2695\,$\pm$\,0.0073  & TIC\\
$J$ [mag] & 9.695\,$\pm$\,0.021 & 2MASS\\
$K_s$ [mag] & 8.869 $\pm$ 0.018 & 2MASS\\
\noalign{\smallskip}
\hline
\end{tabular}
\tablebib{Gic71: \cite{1971lpms.book.....G}; Cab16: \cite{2016SPIE.9910E..0EC};
TIC: \citet{Stassun2019};
\textit{Gaia} DR3: \cite{gaiadr3}; 
2MASS: \citet{skrutskie2006};  Reid03: \cite{2003AJ....126.3007R}.
}
\end{table}

\subsection{Stellar activity}
\label{sec:activity}

The broad wavelength range coverage offered by the CARMENES VIS and NIR channels allows us to study a total of sixteen spectral activity indicators. 
These are pEW(H$\alpha$) (from which we derived $\log{{\rm H}\alpha/L_{\rm bol}}$ as \citealt{2018A&A...614A..76J}), HeD3, NaD, Ca IRT-a,-b,-c, He~{\sc i}$\lambda$10830, Pa $\beta$, CaH2, CaH3, TiO 7050, TiO 8430, TiO 8860, VO 7436, VO 7942, and FeH Wing-Ford, and they were computed following the methods described by \cite{2019A&A...623A..44S}. In addition, the output parameters from \texttt{serval} also include the CRX and dLW parameters for both channels \citep{2018A&A...609A..12Z}, resulting in twenty activity diagnostics.

In all CARMENES spectra we found that H$\alpha$ is in absorption, indicating that TOI-4438 is a weakly active star with undetectable levels of activity-induced emission in the H$\alpha$ line.
We also did not detect H$\alpha$ emission due to flaring in the CARMENES spectra. 
\cite{2018A&A...614A..76J} showed that only about one third of M3 dwarfs are H$\alpha$ active. Given the M3.5\,V spectral type, the low activity level of TOI-4438 is therefore not surprising. We also computed Pearson’s $r$ correlation coefficient between the measured RVs and the spectral activity indicators as described in \cite{2020Sci...368.1477J} and confirmed the star to be least active.

\subsection{Rotation period of the star}
\label{rot_star}

We inspected the SuperWASP light curves for rotational modulations as discussed in \citet{2011PASP..123..547M}.
There appears to be significant power at long periods (see Fig.~\ref{fig:wasp}), although the periodograms are hard to interpret. In each observation season, the amplitude of the modulations is variable, and the data covered only about two cycles. There are likely to be profile and phase changes between years.  We found a rotational modulation with an amplitude of $\sim$\,6 mmag, and a period of 68\,$\pm$\,6 d. The first harmonic is also present at $\sim$\,34 d in the years 2004 and 2010.

In addition, we computed the generalized Lomb-Scargle (GLS; \citealp{2009Zechmeister}) periodograms of the TJO, LCOGT (in $V$- and $B$- filters), ASAS-SN, e-EYE, and ADONIS light curves of TOI-4438, as shown in Fig.\,\ref{fig:GLS_photo}. 
The periodogram of the TJO light curve shows its highest peak at $\sim$34 d, and a second high peak at $\sim$64 d, consistent with the SuperWASP results. The same applies to the periodogram of the LCOGT data ($V$-filter), where we found the highest signal at 68\,$\pm$\,10 d, and the ASAS-SN photometry, which shows a modulation at around 57\, d. LCOGT $B$-filter data show a peak at 22\,$\pm$\,1 d with a false alarm probability of $<$0.1\%, which could be related to the second harmonic of the stellar rotation period. Since the e-EYE and ADONIS data covered a shorter baseline, the periodograms displayed no significant signals around 68 or 34 d.
We concluded that TOI-4438 has a rotation period of 68\,$\pm$\,6 d, as derived from SuperWASP photometry.

\begin{figure}[ht!]
\includegraphics[clip, trim=18.3cm 6.5cm 3.0cm 6.5cm, width=1.\textwidth]{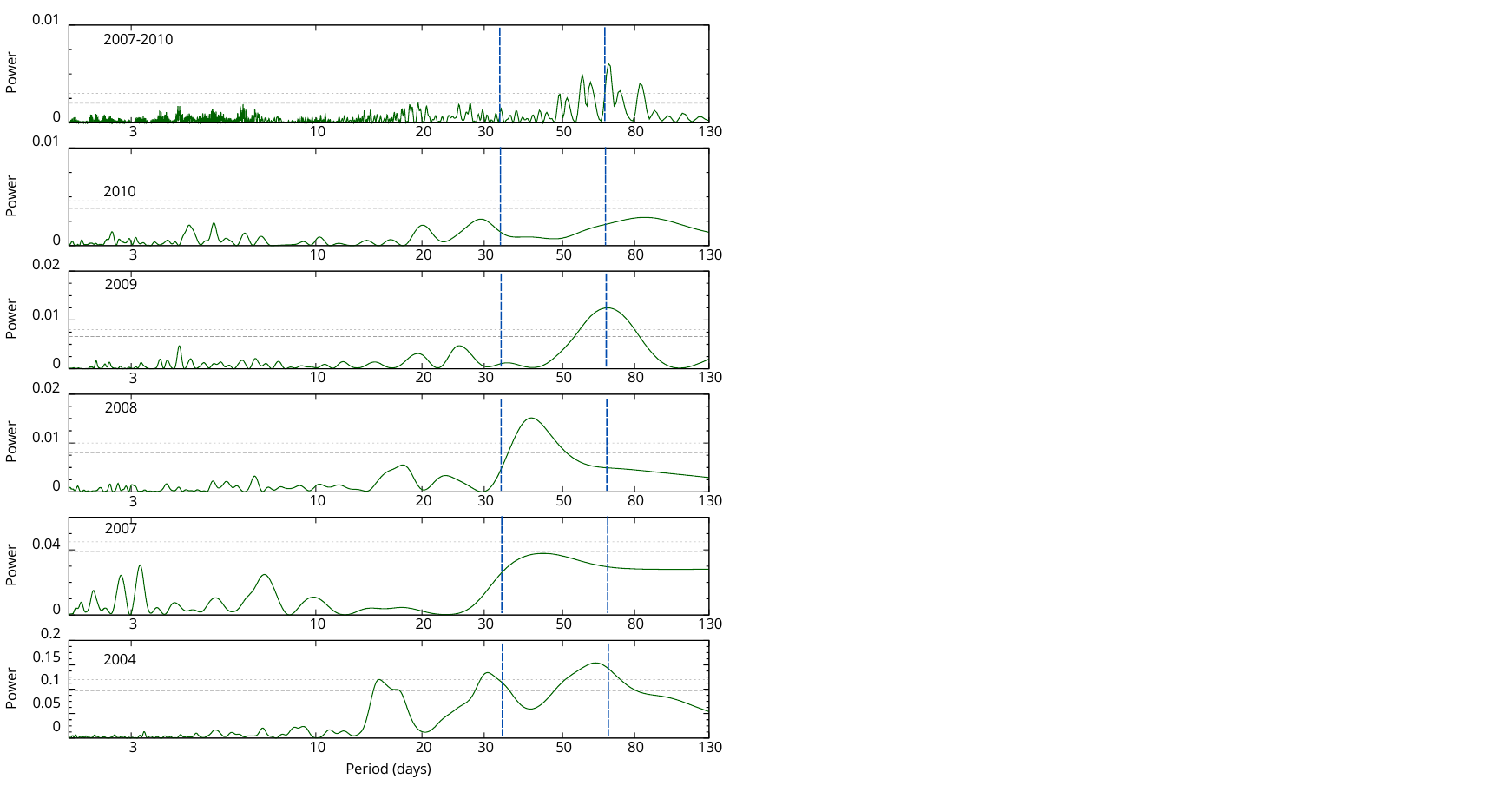}
  \caption{Periodogram of the SuperWASP data for TOI-4438 from each year and (top) the periodogram of 2007--2010 combined. The horizontal dashed lines indicate the estimated 10\%- and 1\%-likelihood false-alarm levels. The blue vertical dashed lines mark the stellar rotation period and its first harmonic.}
\label{fig:wasp}
\end{figure}

\begin{figure}
\includegraphics[clip, trim=15.cm 1.5cm 3.6cm 1.8cm, width=0.75\textwidth]{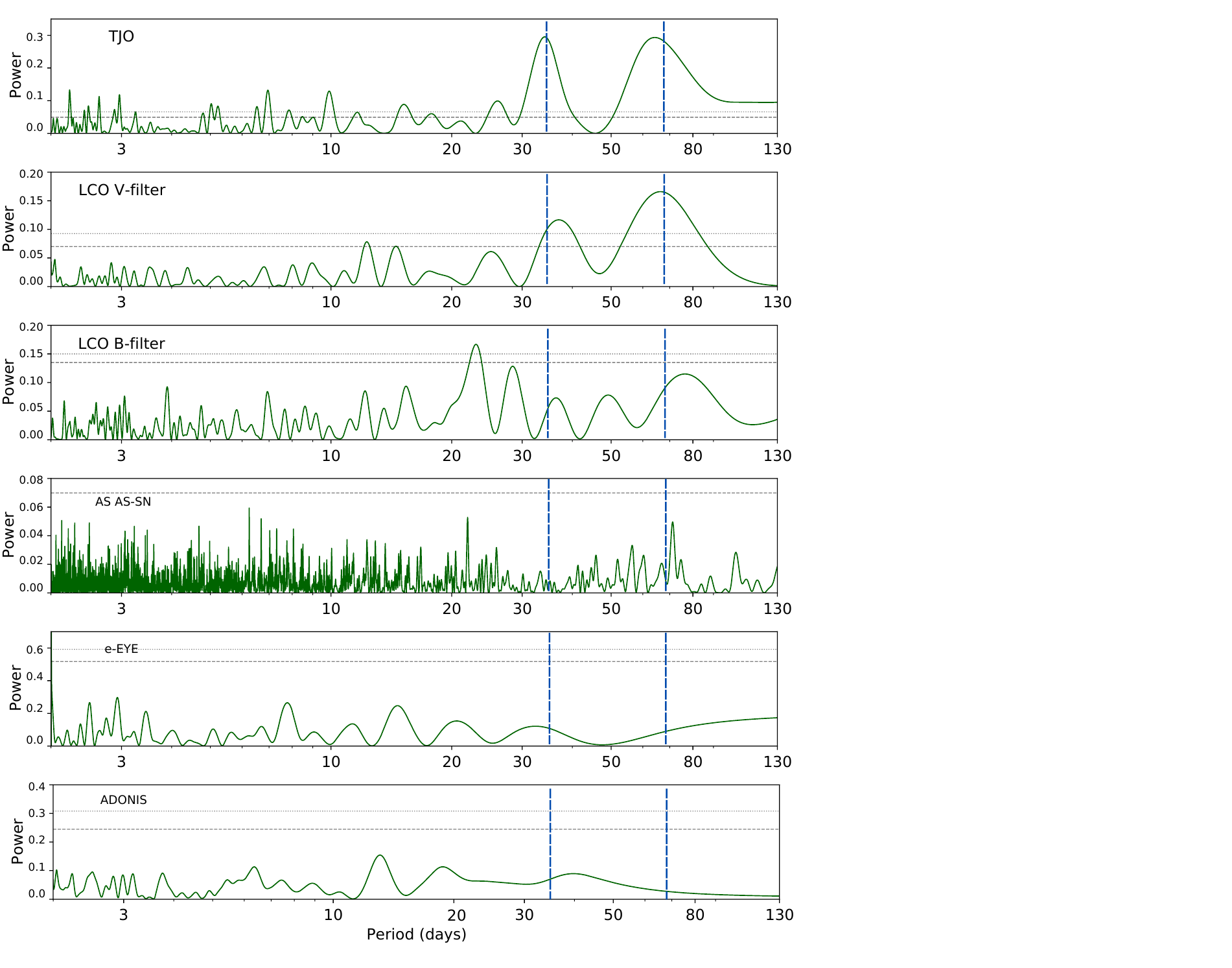}    
        \caption{Periodograms of the TJO, LCOGT ($V$ and $B$), ASAS-SN, e-EYE, and ADONIS data for TOI-4438. The horizontal dashed lines indicate the estimated 10\%- and 1\%-likelihood false-alarm levels. The blue vertical dashed lines mark the stellar rotation period and its first harmonic.}
        \label{fig:GLS_photo}
\end{figure}

\subsection{Age}
\label{sec:age}
With a rotation period of 68\,$\pm$\,6 d, TOI-4438 is very likely an old star. From the relations of \cite{2023ApJ...947L...3B}, a maximum rotation period of 25 d is detected for members of the Praesepe cluster (500--700 Myr; \citealt{2019A&A...628A..66L}). The rotation period of TOI-4438 is at least twice longer, so the age is expected to be much older, although the relation is not linear. Similarly, \cite{2020ApJ...904..140C} showed that field stars with spectral types between K0 and M1--M2 and ages of 6\,$\pm$\,2 Gyr have rotation periods around 35--45\,d, again much shorter than TOI-4438.
Eventually, we used the rotation--age relations applicable to M2.5--6.5 dwarfs from the recent study by \cite{2023ApJ...954L..50E} and derived an age of 5.1\,$\pm$\,2.8\,Gyr.

To constrain the age of an M-type dwarf we can also look for wide, higher-mass, gravitationally bound companions.
We searched for possible co-moving common proper motion companions to TOI-4438 in the $Gaia$ Data Release 3 (DR3) in a radius of five degrees. The search was based on the proximity in parallax, using a very conservative interval of 3.25 pc, 200 times more than the error bar on the parallax reported in DR3. Ten sources were returned but none of them shows a proper motion in right ascension consistent with the large motion of TOI-4438 ($\mu_{\alpha} \cos\delta$\,$\approx$\,$-$517.5 mas\,yr$^{-1}$). 
Moreover, we checked several parameters that may indicate departure from a single star astrometric model, such as the renormalized unit weight error (RUWE), the excess noise of the source and its significance, and the percent of successful-IPD windows with more than one peak.
None of them point toward the presence of an unseen companion around TOI-4438.
This is consistent with the null results on TOI-4438 of the laser-guided, adaptive-optics, Robo-AO M-dwarf multiplicity survey of \cite{2020AJ....159..139L}.

\section{Analysis}
\label{sec:Analysis}

\subsection{Transit light curve modeling}
\label{tronly}

Prior to performing a joint analysis of the photometric and Doppler time series (Sect.~\ref{jfit}), we performed a preliminary analysis of the available transit light curves to check whether the different passbands provide consistent transit depths. With this aim, we used the code \texttt{pyaneti} \citep{2018Barragan,2022Barragan} to model the \textit{TESS} light curve from sectors 40, 52, and 53 (Sect.\,\ref{tess}), and MuSCAT2 ground-based photometry (Sect.\,\ref{muscat2}).
The \textit{TESS} data included in the analysis are subsets of 8~h of the PDCSAP light curves, centered around each transit epoch, and detrended by fitting a second-order polynomial to the out-of-transit data.

Based on the period and ephemeris reported by SPOC (P$_\mathrm{orb,b}$\,=\,7.446303\,$\pm$\,0.000023\,d and T$_\mathrm{0,b}$\,=\,2459396.40958\,$\pm$\,0.00075 BJD$_\mathrm{TDB}$), we allowed the period to vary between 7.2 and 7.7 d, and the epoch of reference transit between BJD$_\mathrm{TDB}$\,=\,2459396.16 and 2459396.66 d.
We adopted the parameterization proposed by \cite{Anderson_2010} for the eccentricity $e$ and the argument of periastron of the stellar orbit $\omega_\star$ (i.e., $\sqrt{e}\sin{\omega_\star}$ and $\sqrt{e}\cos{\omega_\star}$).
\texttt{pyaneti} uses \cite{2002Mandel&Agol}’s limb-darkened quadratic model for the transit light curve and the limb-darkening parameterization proposed by \cite{kipping02013}.
We adopted Gaussian priors for the linear and quadratic limb-darkening coefficients ($q_1$, $q_2$) using the values calculated with the \texttt{Python Limb Darkening Toolkit}\footnote{\url{https://github.com/hpparvi/ldtk}.} (PyLDTk, \citealp{Parviainen2015b}) for \textit{TESS} and MuSCAT2 bands. 
We set uniform priors for all of the remaining parameters.
A photometric jitter term was added to account for any instrumental noise not included in the nominal uncertainties.
We sampled the parameter space with 500 walkers using the Markov chain Monte Carlo (MCMC) ensemble sampler algorithm implemented in \texttt{pyaneti}.
We created the posterior distributions using the last 5000 iterations of the converged chains with a thin factor of 10, giving a distribution of 250\,000 data points for each sampled parameter.
We followed the procedure and the convergence test as described by \cite{2018Barragan}.

In order to check for uniform transit depths in the \textit{TESS} and MuSCAT2 bands, and determine the chromaticity of the transits in each available filter, we performed two analyses: (1) we allowed an independent scaled planetary radius $R_\mathrm{b}/R_{\star}$ for each passband; (2) we modeled a single scaled planetary radius, namely assuming the same transit depth in all passbands.
Table~\ref{table:LC_fit_only} lists the adopted priors and posterior values of the sampled parameters.
For the parameter estimates and their 1$\sigma$ uncertainties, we took the median and  the 68.3\% credibility interval of the posterior distributions. 
Fig.\,\ref{fig:TOI-4438b_LC_folded} shows the phase-folded \textit{TESS} and MuSCAT2 transit light curves of TOI-4438\,b. 
We measured an orbital period of $P_\mathrm{orb, b}$\,=\,7.4462800\,$\pm$\,0.0000087\,d, an epoch of reference transit of $T_\mathrm{0, b}$\,=\,2459396.4116\,$\pm$\,0.0006~d (BJD$_\mathrm{TDB}$), and a radius of $R_\mathrm{b}$\,=\,2.51\,$\pm$\,0.12\,$R_{\oplus}$ from the unique scaled planet radius~fit~(case~2).

Fig.~\ref{fig:post_ditrib_rp} displays the posterior distributions for the sampled scaled planetary radii $R_\mathrm{p}/R_{\star}$ for each passband. We found that the posterior distributions for the \textit{TESS} and MuSCAT2 bands overlap, except for the $g'$ and $r'$ filters.

We also found a color dependence, given the deeper transits in the $g'$ and $r'$ filters. Assuming an astrophysical origin, this color trend could be caused by contamination from a cooler dwarf \citep{Parviainen2019} or unocculted star spots \citep{Ballerini2012}. Potential inaccuracies due to the stellar limb-darkening parameterization \citep{Morello2017} or planetary atmospheric signatures \citep{Chen2021} should be negligible compared to the error bars. 
In order to check the hypothesis of astrophysical contamination, we fitted the measured planetary radii in multiple bands by assuming a unique planet-to-star radius ratio $R_\mathrm{b}/R_{\star}$ and different star spots configurations modeled with \texttt{ExoTETHyS} \citep{Morello2021}.
The stellar spectrum was modeled as the sum of two components, namely the quiet photosphere and the spots, characterized by different values of $T_{\text{eff}}$ and $\log{g_\star}$, and neglecting stellar limb-darkening effects \citep{Cracchiolo2021,Thompson2023}. We adopted the \texttt{PHOENIX-COND} and \texttt{PHOENIX-DRIFT} libraries of stellar spectra \citep{Claret2012,Claret2013,Husser2013}.  
Technical details are given in Appendix \ref{app:starspot_transpec}.
It turned out that a spot filling factor of $\sim$40$\%$ and temperature contrast of $\sim$500 K are needed to explain the different measurements between $g'$ and \textit{TESS} filters, but they struggle to maintain the uniform transit depth observed in $i'$, $z_s$, and \textit{TESS}. Furthermore, these values would denote a very high level of stellar activity, in contrast with all other indicators (see Sect.~\ref{sec:activity}). This scenario is mathematically equivalent to that of a cooler M-type contaminant with similar brightness to that of the host star, following the model by \cite{Morello2023}. Such a bright contaminant is also unlikely to be unseen (see Sect.~\ref{sec:age}).
Another possibility is that the transit measurements in $g'$ and $r'$ bands are affected by systematic bias, due to the low S/N of the MuSCAT2 light curves obtained with these filters. Therefore, we ignored the bluer $g'$ and $r'$, and used $i'$ and $z_s$ only for the following analysis. We discuss an alternative solution including all bands and potential color contamination in Appendix \ref{app:starspot_transpec}.

\begin{table*}[ht!] 
\caption{TOI-4438\,b parameters from the transit light curve modeling with \texttt{pyaneti}.$^{(a)}$}
\label{table:LC_fit_only} 
\begin{adjustbox}{width=0.9\linewidth,center}
\begin{tabular}{lccc}        
\hline\hline                 
\noalign{\smallskip}
Parameter & Prior & Derived value (multi-radii) & Derived value (single radius) \\
\noalign{\smallskip}
\hline
\textbf{\textit{Model parameters}} & & \\
\noalign{\smallskip}
$P_\mathrm{orb, b}$ [d] & $\mathcal{U}$[7.2, 7.7]  & 7.4462801 $\pm$ 0.0000091  & 7.4462800 $\pm$ 0.0000087 \\ 
$T_\mathrm{0,b}$ [BJD$_\mathrm{TDB}-$2,450,000] &  $\mathcal{U}$[9396.16, 9396.66]  & 9396.4116 $\pm$ 0.0007 & 9396.4116 $\pm$ 0.0006 \\ 
$R_\mathrm{b}/R_\star$ &  $\mathcal{U}$[0.001, 0.09] & ...  & 0.0619$^{+0.0006}_{-0.0005}$ \\
$R_\mathrm{b}/R_\star \mathrm{TESS}$ &  $\mathcal{U}$[0.001, 0.09] &  0.0606 $\pm$ 0.0012 & ... \\
$R_\mathrm{b}/R_\star \mathrm{MuSCAT2\,g'}$ &  $\mathcal{U}$[0.001, 0.09] & 0.0678 $\pm$ 0.0026 & ... \\
$R_\mathrm{b}/R_\star \mathrm{MuSCAT2\,r'}$ &  $\mathcal{U}$[0.001, 0.09] & 0.0661 $\pm$ 0.0019 & ... \\
$R_\mathrm{b}/R_\star \mathrm{MuSCAT2\,i'}$ &  $\mathcal{U}$[0.001, 0.09] & 0.0622 $\pm$ 0.0013 & ... \\
$R_\mathrm{b}/R_\star \mathrm{MuSCAT2\,z_s}$ &  $\mathcal{U}$[0.001, 0.09] & 0.0624 $\pm$ 0.0010 & ... \\
$b_\mathrm{b}$ & $\mathcal{U}$[0, 1] & 0.38$^{+0.16}_{-0.24}$ & 0.15$^{+0.14}_{-0.10}$ \\
\noalign{\smallskip}
$a/R_{\star}$ & $\mathcal{U}$[1, 50] & 28.6$^{+4.0}_{-4.9}$ & 30 $\pm$ 5\\
\noalign{\smallskip}
$\sqrt{e_\mathrm{b}}\sin{\omega_\mathrm{\star,b}}$ & $\mathcal{U}$[--1, 1]  & --0.13$^{+0.38}_{-0.25}$ & --0.14$^{+0.40}_{-0.27}$ \\
\noalign{\smallskip}
$\sqrt{e_\mathrm{b}}\cos{\omega_\mathrm{\star,b}}$ & $\mathcal{U}$[--1, 1]  & --0.004 $\pm$ 0.447 & --0.02 $\pm$ 0.47 \\
\noalign{\smallskip}
\hline
\noalign{\smallskip}
\textbf{\textit{Derived parameters}} & & &\\
$R_\mathrm{b}$\,[$R_\oplus$] & ...  & ...  &  2.51 $\pm$ 0.12 \\
$R_\mathrm{b,TESS}$\,[$R_\oplus$] & ...  & 2.46 $\pm$ 0.13 & ... \\
$R_\mathrm{b,MuSCAT2\,g'}$\,[$R_\oplus$] & ...  & 2.75 $\pm$ 0.17 & ... \\
$R_\mathrm{b,MuSCAT2\,r'}$\,[$R_\oplus$] & ...  &  2.68 $\pm$ 0.15  & ... \\
$R_\mathrm{b,MuSCAT2\,i'}$\,[$R_\oplus$] & ...  & 2.52 $\pm$ 0.13 & ... \\
$R_\mathrm{b,MuSCAT2\,z_s}$\,[$R_\oplus$] & ...  &  2.53 $\pm$ 0.13  & ... \\
$a_\mathrm{b}$ [au] & ... & 0.0492$^{+0.0074}_{-0.0086}$  & 0.053 $\pm$ 0.009 \\
\noalign{\smallskip}
$e_\mathrm{b}$ & ...  & 0.21$^{+0.24}_{-0.14}$ &  0.23$^{+0.25}_{-0.15}$  \\
\noalign{\smallskip}
$\omega_{\star,\mathrm{b}}$ [deg] & ...  & --34$^{+134}_{-111}$ & --37$^{+139}_{-108}$ \\
\noalign{\smallskip}
$i_\mathrm{b}$ [deg] & ...  & 89.22$^{+0.50}_{-0.53}$ & 89.71$^{+0.35}_{-0.21}$\\
\noalign{\smallskip}
$\tau_{14,\mathrm{b}}$ [h] & ... & 2.02 $\pm$ 0.03 & 2.01 $\pm$ 0.02 \\
$T_\mathrm{eq,b}$ [K] $^{(b)}$ & ...  & 453$^{+46}_{-30}$  & 438$^{+45}_{-32}$ \\
\noalign{\smallskip}
$S_\mathrm{b}$ [S$_{\oplus}$] & ... & 7.04$^{+3.3}_{-1.7}$ & 6.1$^{+2.9}_{-1.6}$ \\
\noalign{\smallskip}
\hline
\noalign{\smallskip}
\textbf{\textit{Additional model parameters}} & &  \\
\textit{Parameterized limb-darkening coefficients}& &  \\
$q_\mathrm{1,TESS}$ & $\mathcal{N}$[0.276, 0.1]  & 0.30 $\pm$ 0.08 & 0.30 $\pm$ 0.08 \\
$q_\mathrm{2,TESS}$ & $\mathcal{N}$[0.261, 0.1]  &  0.26 $\pm$ 0.10 & 0.26 $\pm$ 0.10 \\
$q_\mathrm{1,MuSCAT2\,g'}$ & $\mathcal{N}$[0.629, 0.1]  &  0.64 $\pm$ 0.10 & 0.64 $\pm$ 0.10 \\
$q_\mathrm{2,MuSCAT2\,g'}$ & $\mathcal{N}$[0.312, 0.1]  & 0.35 $\pm$ 0.10  & 0.35 $\pm$ 0.10 \\
$q_\mathrm{1,MuSCAT2\,r'}$ & $\mathcal{N}$[0.572, 0.1]  &  0.55 $\pm$ 0.16 & 0.55 $\pm$ 0.16 \\
$q_\mathrm{2,MuSCAT2\,r'}$ & $\mathcal{N}$[0.317, 0.1]  & 0.33 $\pm$ 0.10  &  0.33 $\pm$ 0.10  \\
$q_\mathrm{1,MuSCAT2\,i'}$ & $\mathcal{N}$[0.334, 0.1]  &  0.37 $\pm$ 0.08 & 0.37 $\pm$ 0.08 \\
$q_\mathrm{2,MuSCAT2\,i'}$ & $\mathcal{N}$[0.260, 0.1]  & 0.31 $\pm$ 0.10  & 0.31 $\pm$ 0.10 \\
$q_\mathrm{1,MuSCAT2\,z_s}$ & $\mathcal{N}$[0.241, 0.1]  &  0.16 $\pm$ 0.08  &  0.16 $\pm$ 0.08\\
$q_\mathrm{2,MuSCAT2\,z_s}$ & $\mathcal{N}$[0.240, 0.1]  &  0.19 $\pm$ 0.10 &  0.19 $\pm$ 0.10 \\
\noalign{\smallskip}
\hline
\noalign{\smallskip}
\textit{Jitter terms} & & \\ $\sigma_\mathrm{TESS}$ & $\mathcal{J}$[0,100]  & 0.000058 $\pm$ 0.000054  & 0.000058 $\pm$ 0.000054\\
$\sigma_{MuSCAT2\,g'}$ & $\mathcal{J}$[0,100]  & 0.00021 $\pm$ 0.00020  & 0.00021 $\pm$ 0.00020\\
$\sigma_\mathrm{MuSCAT2\,r'}$ & $\mathcal{J}$[0,100]  & 0.00030 $\pm$ 0.00026  & 0.00030 $\pm$ 0.00026\\
$\sigma_\mathrm{MuSCAT2\,i'}$ & $\mathcal{J}$[0,100]  & 0.00014 $\pm$ 0.00012 & 0.00014 $\pm$ 0.00012\\
$\sigma_\mathrm{MuSCAT2\,z_s}$ & $\mathcal{J}$[0,100]  & 0.00018 $\pm$ 0.00026  & 0.00018 $\pm$ 0.00026\\
\noalign{\smallskip}
\hline
\end{tabular}
\end{adjustbox}
\tablefoot{ $^a$ $\mathcal{U}[a,b]$ refers to uniform priors between $a$ and $b$; $\mathcal{N}[a,b]$ refers to Gaussian priors with mean $a$ and standard deviation $b$; $\mathcal{J}[a,b]$ refers to Jeffrey's priors between $a$ and $b$. Inferred parameters and uncertainties are defined as the median and the 68.3\% credible interval of their posterior distributions.
$^b$ Assuming zero albedo and uniform redistribution of heat.}
\end{table*}

\begin{figure*}
     \centering
     \begin{subfigure}[b]{0.33\textwidth}
         \centering
         \includegraphics[width=\textwidth]{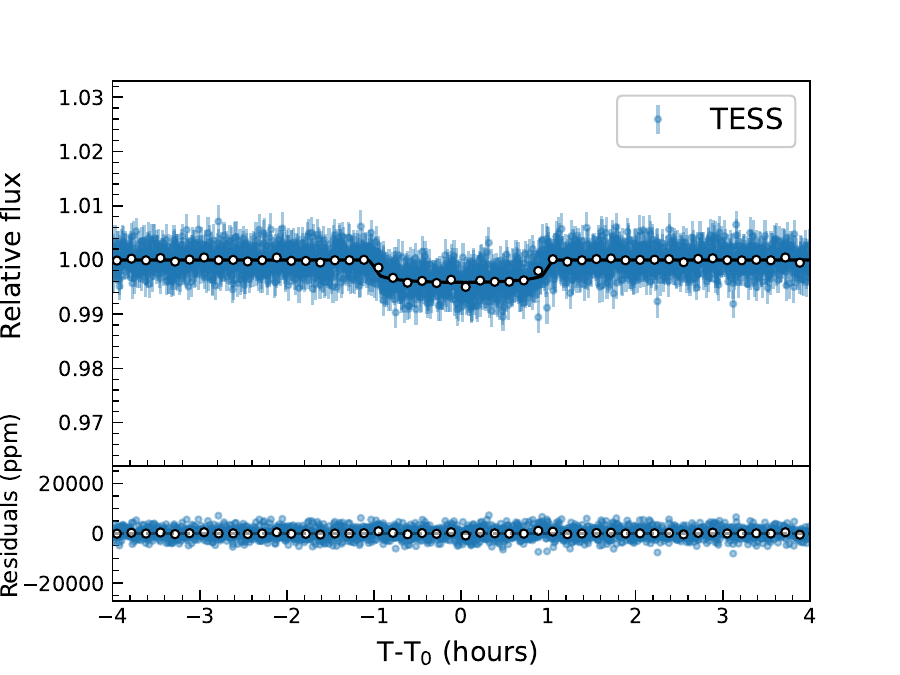}
         \caption{}
     \end{subfigure}
     \begin{subfigure}[b]{0.33\textwidth}
         \centering
         \includegraphics[width=\textwidth]{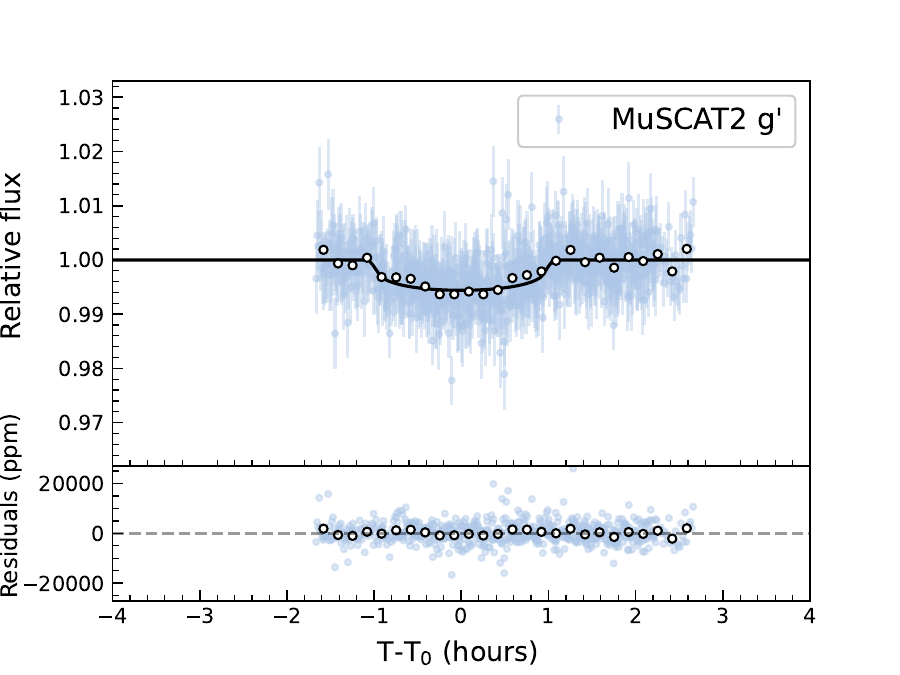}
         \caption{}
     \end{subfigure}
     \begin{subfigure}[b]{0.33\textwidth}
         \centering
         \includegraphics[width=\textwidth]{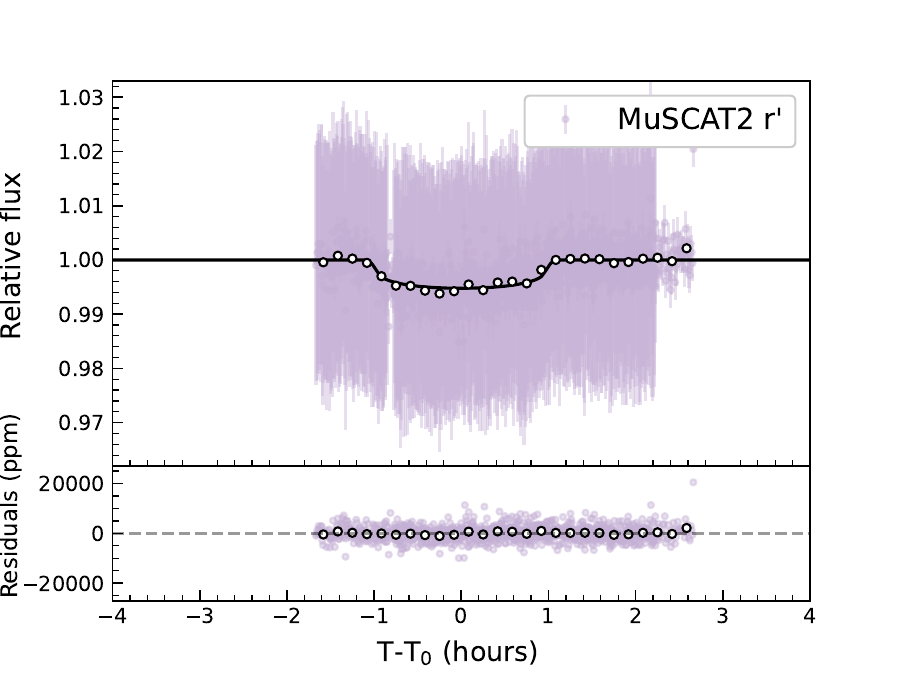}
         \caption{}
     \end{subfigure}
     \begin{subfigure}[b]{0.33\textwidth}
         \centering
         \includegraphics[width=\textwidth]{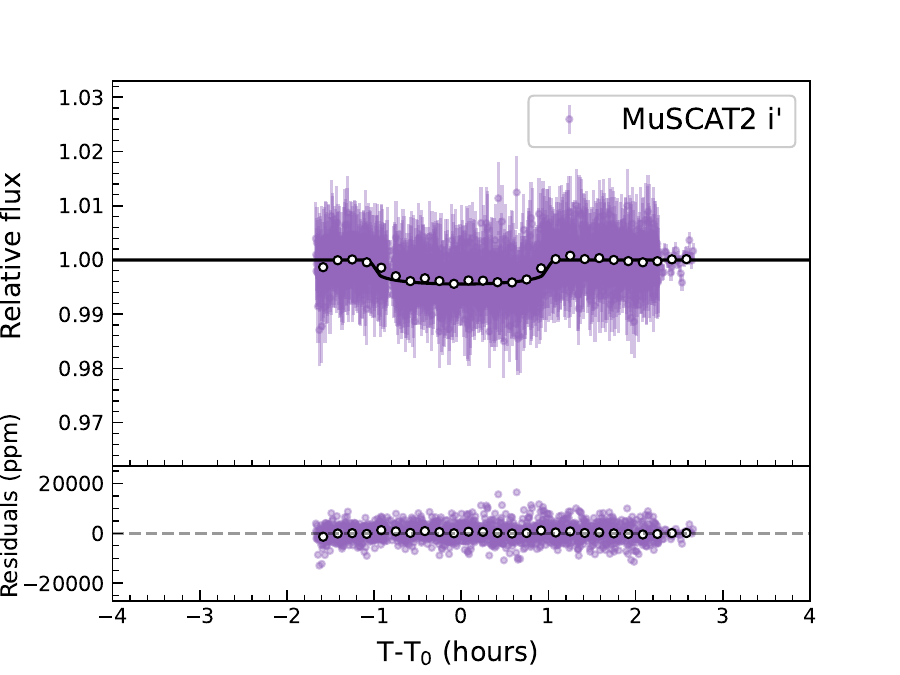}
         \caption{}
     \end{subfigure}
     \begin{subfigure}[b]{0.33\textwidth}
         \centering
         \includegraphics[width=\textwidth]{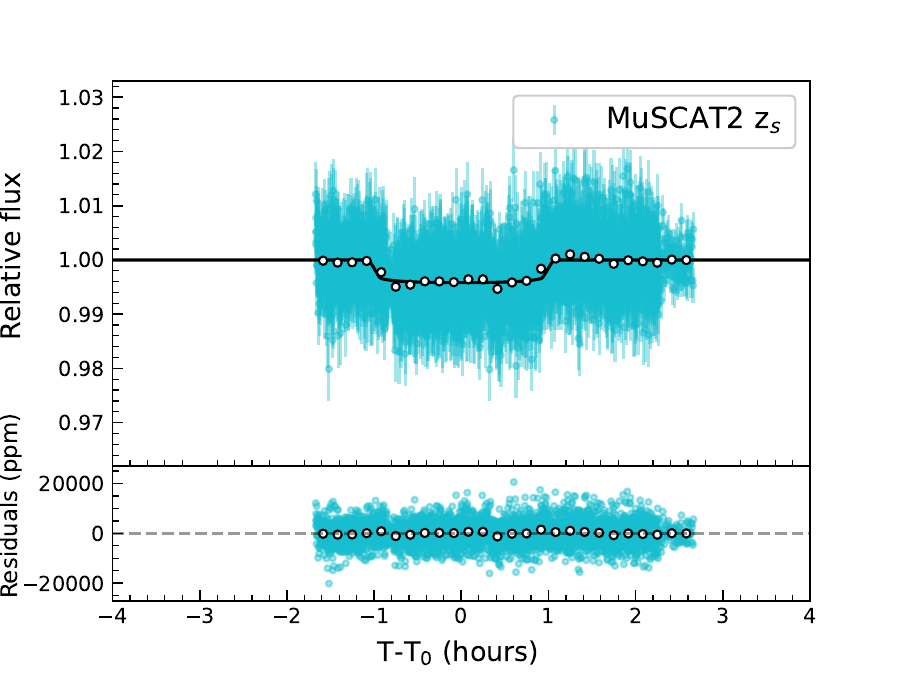}
         \caption{}
     \end{subfigure}
        \caption{Phase-folded transit light curves of TOI-4438\,b from the (a) \textit{TESS}, and (b) $g'$, (c) $r'$, (d) $i'$, and (e) z$_s$ MuSCAT2 bands. Data are shown as filled circles with their nominal uncertainties. The 10-min binned data points are marked with white circles.}
        \label{fig:TOI-4438b_LC_folded}
\end{figure*}

\begin{figure}[ht!]
     \centering
     \begin{subfigure}[b]{0.48\textwidth}
         \centering
         \includegraphics[width=\textwidth]{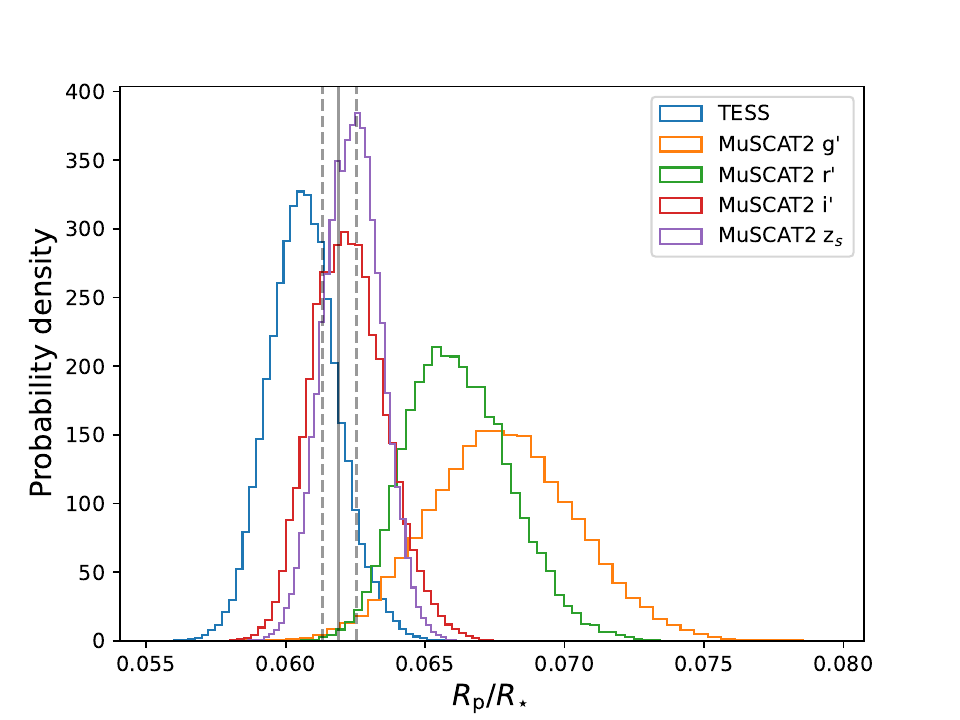}
         \label{fig:postdist}
     \end{subfigure}
     \hfill
        \caption{Posterior distributions of the sampled scaled planetary radii $R_\mathrm{b}/R_{\star}$ for the \textit{TESS} and MuSCAT2 $g'$, $r'$, $i'$, $z_s$ bands. The solid and dotted, vertical, gray lines mark the median and the 68.3\% credibility interval of the $R_p/R_{\star}$'s posterior distribution, as derived in case 2 (see Sect.~\ref{tronly}). 
        }
        \label{fig:post_ditrib_rp}
\end{figure}

\subsection{Additional planet search and detection limits}

We adopted the box least-square ({\tt BLS}; \citep{2002A&A...391..369K,2016A&C....17....1H} method to look for additional transits in the PDCSAP \textit{TESS} light curves.
Before applying the BLS we detrended the data using the package \texttt{citlalique}\footnote{\url{https://github.com/oscaribv/citlalicue}}, which allows to detrend light curves using Gaussian processes (GPs) to remove low frequency signals in the light curves \citep{2022Barragan}.

We used the {\tt BoxLeastSquares} class of the {\tt astropy.timeseries} python package. Whenever the program identified a transit signal, we removed the in-transit data points and ran the algorithm again to search for additional transiting planets. The BLS periodograms of the \textit{TESS} light curves were evaluated over periods from 0.5 d to half the total length of the observations, namely $\sim$190\,d. The first transit signal was identified at 7.44 d, corresponding to the orbital period of TOI-4438\,b. We found no other signals in the data corresponding to additional transiting planets.

To identify possible additional companions that might perturb the orbit of TOI-4438\,b, we analysed the available transit light curves for transit timing variation (TTVs). We used the ten \textit{TESS} transits from sectors 40, 52, and 53, and the two full transits observed with MuSCAT2 on 2 May and 8 July 2023. We co-added the data from MuSCAT2 passbands $z_s$ and $i'$ prior to the analysis. We individually modeled the mid-time of each transit with the code \texttt{pyaneti} \citep{2022Barragan}. We found no significant TTV in the $\sim$770 d baseline of our observations, as shown in Fig.\,\ref{fig:TTV}.

\begin{figure}[h]
\centering
\includegraphics[width=0.5\textwidth]{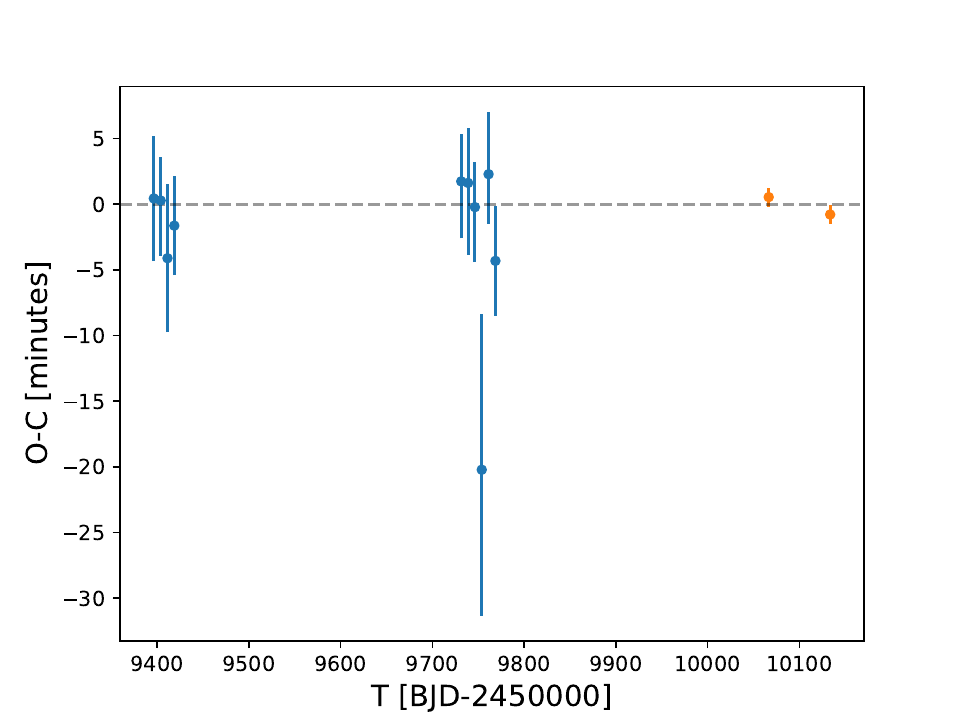}
\caption{Transit-timing variations of TOI-4438\,b. Blue points correspond to TESS transits while the orange points correspond to the full MuSCAT2 transits.  \label{fig:TTV}
        }
\end{figure}

The lack of extra signals attributable to transiting planets in the \textit{TESS} data might be due to one of the following scenarios \cite[see, e.g.,][]{wells2021,schanche2022,delrez2022,pozuelos2023}: 
(1) no other planets exist in the system; (2) they do exist, but they do not transit; (3) they do exist and transit the host star, but have orbital periods longer than the ones explored in this study; (4) they do exist and transit, but the photometric precision of the data is not high enough to detect them. 
Scenarios (1) and (2) might be further explored by employing RV follow-up data as discussed in Sect.~\ref{sec:rv}. Scenario (3) can be tested when \textit{TESS} will reobserve TOI-4438 in sectors 79 and 80 (June-July 2024), extending the available time baseline\footnote{See the \textit{TESS}-point Web Tool available at \url{https://heasarc.gsfc.nasa.gov/wsgi-scripts/TESS/TESS-point_Web_Tool/TESS-point_Web_Tool/wtv_v2.0.py/}.}.
To evaluate scenario (4), we studied the detection limits of the current \textit{TESS} photometry conducting injection-and-recovery experiments with the \matrixtk code\footnote{The \matrixtk (\textbf{M}ulti-ph\textbf{A}se \textbf{T}ransits \textbf{R}ecovery from \textbf{I}njected e\textbf{X}oplanets) code is open access on GitHub at \url{https://github.com/PlanetHunters/tkmatrix}.} \citep{matrix2023}. We used the \matrixtk code to inject synthetic planets over the PDCSAP light curves corresponding to the three \textit{TESS} sectors used in this study, combining a unique set of radius, orbital period, and phase for each synthetic planet. 
We explored the $R_{\mathrm{planet}}$--$P_{\mathrm{planet}}$ parameter space in the ranges of 0.5--3.0\,R$_{\oplus}$ with steps of 0.25\,R$_{\oplus}$, and 0.5--15.0 d with steps of 0.25 d. Moreover, for each combination of $R_{\mathrm{planet}}$--$P_{\mathrm{planet}}$ \matrixtk produces five different phases, that is, different values of $T_{0}$. In total, we explored 3000 scenarios. For simplicity, the injected planets had impact parameters and eccentricities equal to zero. 

Once the synthetic planets were injected in \matrixtk, we detrended the light curves using a bi-weight filter with a window size of 0.5 d, which was found to be the optimal value to recover the known planet TOI-4438\,b. Moreover, existing transits in the data were masked out to evaluate the ease of finding TOI-4438\,b-like planets.
In \matrixtk, a synthetic planet is recovered when its epoch matches the injected epoch with 1~hour accuracy, and its period is within 5\,\% of the injected period. 
Since we injected the synthetic signals in the PDCSAP light curve, these signals were not affected by the PDCSAP systematic corrections; hence, the detection limits that we found represent the most optimistic scenario \citep[see, e.g.,][]{pozuelos2020}.

The resulting detectability map from this injection-and-retrieval experiment is shown in Fig.~\ref{fig:recovery}. On the one hand, we found that transiting Earth- and sub-Earth size planets would remain unnoticed for the complete set of periods explored with recovery rates ranging from 50 to 0$\%$. On the other hand, mini-Neptunes such as TOI-4438\,b, seem to be easily detectable, with recovery rates higher than 80$\%$ for all the periods, that is, the presence of additional transiting planets in the system with sizes larger than 1.5~R$_\oplus$ and with orbital periods between 0.5 and 15.0 d seems unlikely.

\begin{figure}
\includegraphics[width=\columnwidth]{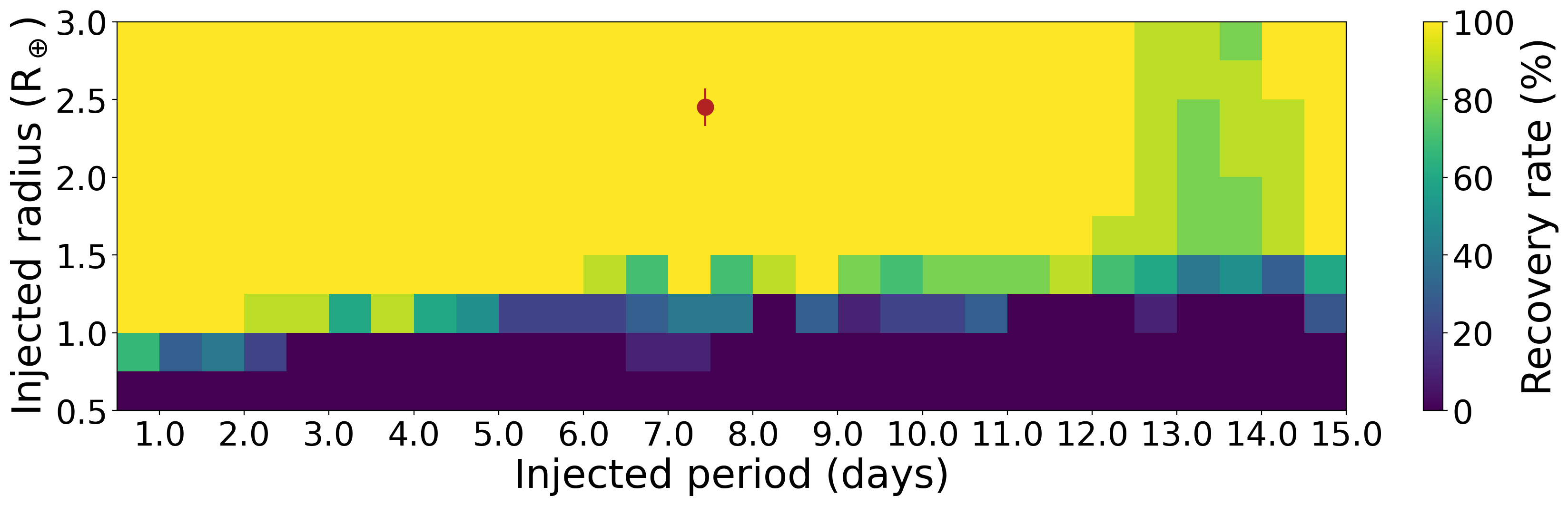}
\caption{The resulting detectability map from the injection-and-retrieval experiment. Each pixel corresponds to ten of the 3000 different scenarios, that is, ten light curves with injected planets having different $P_{\mathrm{planet}}$, $R_{\mathrm{planet}}$, and $T_{0}$. Larger recovery rates are presented in yellow and green colors, while lower recovery rates are shown in blue and darker hues. The red dot refers to TOI-4438\,b, which has a recovery rate of 100$\%$, highlighting the ease of detecting mini-Neptunes in this \textit{TESS} data set with orbital periods shorter than 15 d.
} \label{fig:recovery}
\end{figure}

\subsection{Frequency analysis of the RV data}
\label{freq_analysis}

We performed a frequency analysis to search for periodic signals due to orbiting planets or stellar activity.  We computed the GLS periodograms of the CARMENES RVs and activity indicators, as shown in Fig.~\ref{fig:GLS_TOI-4438}.
We estimated the false alarm probability (FAP) using the bootstrap method as described by \cite{murdoch_1993}. We calculated the periodogram of 10$^6$ time series obtained by randomly shuffling the data and their error bars, keeping the time values fixed. We determined that a peak is significant when it has a FAP\,$<$\,0.1\% \citep{1996CochranHatzes}.

The GLS periodogram of the CARMENES RVs in the VIS band (Fig.~\ref{fig:GLS_TOI-4438}, upper panel) shows the highest peak at $f_1$\,=\,0.13\,d$^{-1}$, corresponding to a period of about 7.44\,d. This signal is not significant (FAP\,$<$\,10\%) within the frequency range 0.0--0.3 d$^{-1}$. However, the peak was found at a known frequency, namely the orbital frequency of the planet, providing strong evidence that the signal is related to TOI-4438\,b. We estimated the FAP of the peak at orbital frequency of TOI-4438\,b using the "windowing bootstrap'' method described by \citet{2019dmde.book.....H} and following the procedure done in \cite{2024MNRAS.527.11138}. 
We found a FAP\,=\,0.002\% for $f_1$, which confirms the planetary nature of the signal discovered by \textit{TESS}.

We also used a pre-whitening technique \citep{2010Hatzes} to find additional signals. We subtracted the 7.44 d signal from the VIS RVs. 
The periodogram of the RV residuals following the subtraction of the signal at $f_1$ (second panel) shows no other significant signal. 

\begin{figure*}[ht!]
\begin{center}
\includegraphics[clip, trim=0.9cm 1.7cm 0.0cm 2.1cm, width=0.8\textwidth]{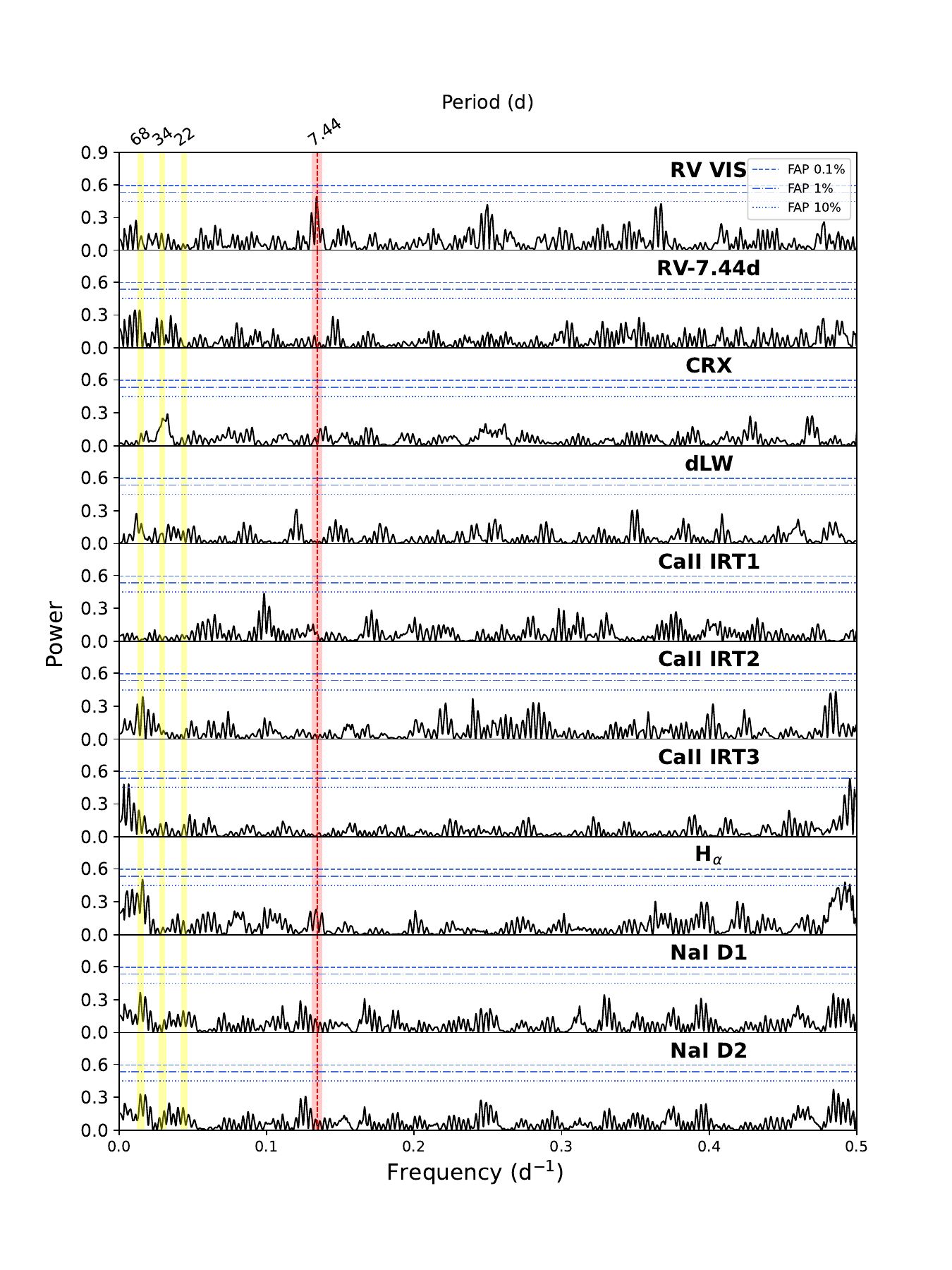}
\caption{GLS periodograms of the CARMENES VIS RV measurements (\textit{upper panel}), and RV residuals after subtracting $f_{1}$ (\textit{second  panel}). The periodograms of the activity indicators are shown in the remaining panels. The 10\%, 1\%, and 0.1\% FAPs are shown with horizontal blue lines. The red vertical line marks the orbital frequency of the transiting planet TOI-4438\,b. The yellow vertical lines mark the stellar rotation period at $\sim$\,68 d, and its first and second harmonics. \label{fig:GLS_TOI-4438}}
\end{center}
\end{figure*}

Finally, we performed a frequency analysis of the activity indicators obtained using the \texttt{serval} pipeline. 
The GLS periodograms show peaks at periods $>$\,30 d, with a FAP\,$>$\,10\% (lower panels of Fig.\,\ref{fig:GLS_TOI-4438}). Although not significant, peaks at $\sim$68 d and $\sim$34 d are present in the periodograms of the CRX, the \ion{Ca}{ii} IRT2, H$\alpha$, and \ion{Na}{i} D indices. The peaks appear at the periods observed in the ground-based photometry analysis of Sect.\,\ref{rot_star}, corroborating the stellar rotational period, and its first and second harmonic at 68 d, 34 d, and 22 d, respectively.

\subsection{Joint fit of the transit light curves and RV measurements}
\label{jfit}

We performed a joint modeling combining \textit{TESS}, MuSCAT2 $i'$ and $z_s$ transit light curves, and CARMENES VIS RV measurements, following the results presented in Sect.\,\ref{tronly} and \ref{freq_analysis}. 
We used the code \texttt{pyaneti} to perform an MCMC analysis. We followed the procedures described previously in Sect.\,\ref{tronly}, adopting uniform priors for the orbital period and reference transit epoch of TOI-4438\,b. We used the same parameterizations and prior distributions as in Sect.\,\ref{tronly}.
We performed two analyses: (i) we assumed a circular orbit fixing $\sqrt{e}\sin{\omega_\star}$ and $\sqrt{e}\cos{\omega_\star}$ to zero; (ii) we fitted for an eccentric orbit.
In a preliminary analysis we used a uniform prior for the stellar density $\rho_{\star}$.  The modeling of the transit light curves gave a mean stellar density consistent with the density of Table.~\ref{tab:stellar_parameters}, which corroborates the planetary nature of the transit signals.
We then sampled for the stellar density $\rho_{\star}$ using a Gaussian prior on the stellar mass and radius, derived in Sect.~\ref{Star_param}, and used wide uniform priors for the remaining model parameters.
We fitted for photometric and RV jitter terms to account for possible signals not captured by our models, or instrumental noise not included in the nominal error bars.

We explored the parameter space with 500 Markov chains initialized randomly inside the prior ranges. Once all chains converged, we used the last 5000 iterations of the converged chains with a thin factor of 10 to create the posterior distributions, leading to a distribution of 250\,000 data points for each sampled parameter. This procedure ensured a homogeneous sampling of the parameter space.
We followed the same procedure and convergence test as described by \cite{2018Barragan}.

Results and prior ranges of the fitted and derived parameters are reported in Table\,\ref{table:joint_fit}.
Since the planet is not close enough to the star to expect tidal circularization and there are no other reasons to assume so (Sect.\,\ref{timescale_circularization}), we chose the eccentric model as the best-fit, although the resulting eccentricity is consistent with zero.

Figs.~\ref{fig:rv_plot} and \ref{fig:rv_plot_pholded} display the CARMENES VIS time series and the RV curve phase-folded at the orbital period of TOI-4438\,b, respectively, along with the best-fitting Doppler model.
We found an RV semi-amplitude variation of $K_\mathrm{b}$\,=\,3.50\,$\pm$\,0.72~m\,s$^{-1}$, and a planetary mass of $M_\mathrm{b}$\,=\,5.4\,$\pm$\,1.1~M$_\oplus$ (20\% precision). We measured a radius of $R_\mathrm{b}$\,=\,2.52\,$\pm$\,0.13 R$_\oplus$ (5\% precision). Combining the planetary mass and radius, we calculated a mean density of $\rho_\mathrm{b}$\,=\,1.85$^{+0.51}_{-0.44}$~g\,cm$^{-3}$ ($\sim$28\% precision).

Although the results obtained with \texttt{pyaneti} are robust, we performed a second joint analysis of the light curves and RVs with \texttt{juliet} \citep{2019MNRAS.490.2262E} as a sanity check. The derived planetary parameters are consistent within errors with those presented in Table\,\ref{jfit}. The \texttt{juliet} planetary radius and mass are $R_\mathrm{b} = 2.54\,\pm$ 0.12\,R$_{\oplus}$ ($\sim$5\,\%), and ${M_\mathrm{b}} = 5.7\,\pm$ 1.1\,M$_{\oplus}$ ($\sim$20\,\%), respectively.

\begin{table*}[ht!] 
\caption{TOI-4438\,b parameters from the joint RVs and transit modeling with \texttt{pyaneti}.$^{(a)}$}
\label{table:joint_fit} 
\begin{adjustbox}{width=0.8\linewidth,center}
\begin{tabular}{lccc}        
\hline\hline                 
\noalign{\smallskip}
Parameter & Prior & Derived value ($e=0$) & Derived value ($e\neq0$) \\    
\hline                 
\noalign{\smallskip}
\textbf{\textit{Model parameters}} & & & \\
$P_\mathrm{orb, b}$ [d] & $\mathcal{U}$[7.2,7.7]  & 7.44628 $\pm$ 0.000009 & 7.44628 $\pm$ 0.000009 \\ 
$T_\mathrm{0,b}$ [BJD$_\mathrm{TDB}-$2,450,000] &  $\mathcal{U}$[9396.16, 9396.66]  & 9396.41178 $\pm$ 0.00066 & 9396.41168 $\pm$ 0.00065\\ 
$R_\mathrm{b}/R_\star$ &  $\mathcal{U}$[0.001,0.09] & 0.0613 $\pm$ 0.0007 & 0.062 $\pm$ 0.001 \\
$b_\mathrm{b}$ & $\mathcal{U}$[0,1] & 0.18$^{+0.15}_{-0.13}$ & 0.39$^{+0.16}_{-0.24}$ \\
\noalign{\smallskip}
$\sqrt{e_\mathrm{b}}\sin{\omega_\mathrm{\star,b}}$ & $\mathcal{U}$[--1,1]  & 0.0 & --0.31$^{+0.20}_{-0.14}$ \\
\noalign{\smallskip}
$\sqrt{e_\mathrm{b}}\cos{\omega_\mathrm{\star,b}}$ & $\mathcal{U}$[--1,1]  & 0.0 & 0.11$^{+0.14}_{-0.19}$ \\
\noalign{\smallskip}
$K_\mathrm{b}$ [m\,s$^{-1}$] & $\mathcal{U}$[0,50]   & 3.23 $\pm$ 0.69 & 3.50 $\pm$ 0.72 \\
\noalign{\smallskip}
\hline
\noalign{\smallskip}
\textbf{\textit{Derived parameters}} & &  \\
$M_\mathrm{b}$\,[$M_\oplus$] & ...  & 5.1 $\pm$ 1.1 & 5.4 $\pm$ 1.1 \\
$R_\mathrm{b}$\,[$R_\oplus$] & ...  & 2.48 $\pm$ 0.12 & 2.52$\pm$ 0.13 \\
$\rho_\mathrm{b}$ [g\,cm$^{-3}$] & ...  & 1.80$^{+0.53}_{-0.44}$ & 1.85$^{+0.51}_{-0.44}$ \\
\noalign{\smallskip}
$a_\mathrm{b}$ [au] & ...  & 0.0509 $\pm$ 0.0028 & 0.0534 $\pm$ 0.0037  \\
$e_\mathrm{b}$ & ...  & 0.0 & 0.14 $\pm$ 0.09  \\
$\omega_{\star,\mathrm{b}}$ [deg] & ...  & 0.0 & --70$^{+36}_{-32}$ \\
\noalign{\smallskip}
$i_\mathrm{b}$ [deg] & ...  & 89.64$^{+0.25}_{-0.31}$ & 89.34$^{+0.39}_{-0.21}$  \\
\noalign{\smallskip}
$\tau_{14,\mathrm{b}}$ [h] & ... & 2.002 $\pm$ 0.018 & 2.021$^{+0.033}_{-0.025}$  \\
\noalign{\smallskip}
$T_\mathrm{eq,b}$ [K] ($a$=0) & ...  & 446 $\pm$ 13 & 435 $\pm$ 15 \\
$T_\mathrm{eq,b}$ [K] ($a$=0.6) & ...  & 354 $\pm$ 7 & 347 $\pm$ 5 \\
$S_\mathrm{b}$ [S$_{\oplus}$] & ... & 6.57$^{+0.81}_{-0.68}$ & 6.00$^{+0.90}_{-0.77}$ \\
\noalign{\smallskip}
\hline
\noalign{\smallskip}
\textbf{\textit{Additional model parameters}} & &  \\
$\rho_{\star}$ [g\,cm$^{-3}$]& $\mathcal{N}$[10.06,1.62] & 8.92$^{+0.49}_{-0.99}$ & 10.1$^{+1.7}_{-1.4}$\\
$\gamma_\mathrm{CARMENES}$ [m\,s$^{-1}$] & $\mathcal{U}$[--2,2] & --0.2 $\pm$ 0.5 & --0.3 $\pm$ 0.5 \\
RV jitter term $\sigma_\mathrm{RV,CARMENES}$ [m\,s$^{-1}$] & $\mathcal{J}$[0,100]  & 1.48$^{+0.72}_{-0.79}$  & 1.33$^{+0.71}_{-0.80}$ \\
Phot. jitter term $\sigma_\mathrm{TESS}$ & $\mathcal{J}$[0,100]  & 0.000060 $\pm$ 0.000055 & 0.000060 $\pm$ 0.000055 \\
Phot. jitter term $\sigma_\mathrm{MuSCAT2\,i'}$ & $\mathcal{J}$[0,100]  & 0.00014 $\pm$ 0.00011 & 0.00013 $\pm$ 0.00011 \\
Phot. jitter term $\sigma_\mathrm{MuSCAT2\,z_s}$ & $\mathcal{J}$[0,100]  & 0.00018 $\pm$ 0.00016 & 0.00017 $\pm$ 0.00016\\
\noalign{\smallskip}
\hline
\end{tabular}
\end{adjustbox}
\end{table*}

\begin{figure*}[!t]
\centering
\includegraphics[width=1.0\textwidth]{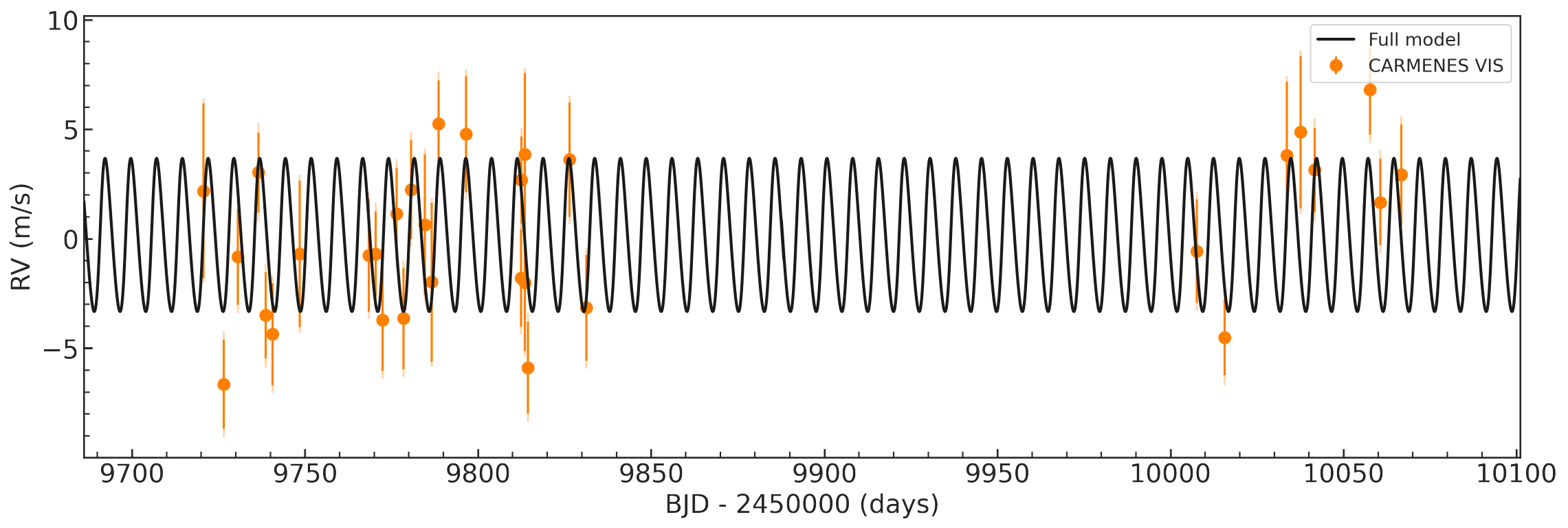}
\caption{CARMENES RV time series of TOI-4438 along with the best-fitting model. Data are shown as orange filled circles with their nominal uncertainties and semitransparent error bar extensions accounting for the jitter term. \label{fig:rv_plot}
        }
\end{figure*}

\begin{figure}[!t]
\centering
\includegraphics[width=0.5\textwidth]{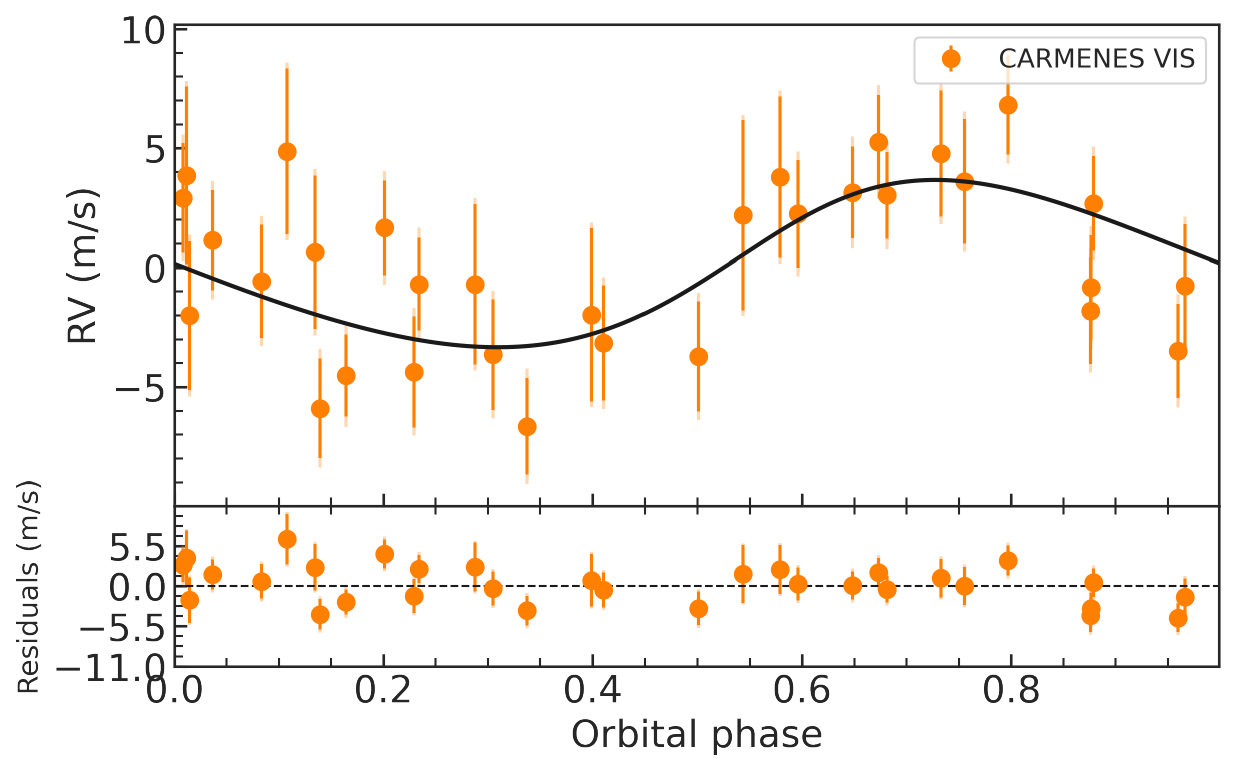}
\caption{Same as Fig.\,\ref{fig:rv_plot} but phase-folded at the orbital period of TOI-4438\,b. \label{fig:rv_plot_pholded}}
\end{figure}

\subsection{Tidal damping of the eccentricity and its consequences }\label{timescale_circularization} 

We estimated the current eccentricity decay timescale of TOI-4438b adopting the model by \citet{Leconteetal10}. We assumed that its current value is $e=0.14$, although its actually measured value is compatible with zero at the 2$\sigma$ level. To apply the tidal model, we expressed the product of the constant tidal time lag $\Delta t_{\rm p}$, adopted by \cite{Leconteetal10}, by the Love number of the planet $k_{2,\rm p}$: 
\begin{equation}
    k_{2, \rm p} \Delta t_{\rm p} =  \frac{3}{2 Q^{\prime}_{\rm p} n},
\end{equation} 
where $Q^{\prime}_{\rm p}$ is the modified tidal quality factor of the planet itself, $n=2\pi/P$ is the orbit mean motion, and $P$ the orbital period.

The rheology of mini-Neptune planets is unknown, thus we investigated two extreme cases. The former assumed $Q^{\prime}_{\rm p} = 10^{5}$, which is comparable to the modified tidal quality factor of Uranus or Neptune as estimated by for example \citet{TittemoreWisdom90} and \citet{Ogilvie14}, and is appropriate for a body where the dissipation of the tidal energy occurs in a fluid interior; the latter assumed that tides are mainly dissipated inside a rocky core encompassing the whole mass of the planet $5.4$~M$_{\oplus}$, but having a radius of 1.3 R$_{\odot}$. This hypotetical core radius corresponds to the  peak of the smallest radius component of the distribution of the transiting planets as observed by Kepler \citep{fulton2017rvalley}. According to the model by \cite{2017ApJ...847...29O} such a radius corresponds to the mean size of the bare core that remains after the gaseous envelope of a mini-Neptune has been removed by photoevaporation after the first few 100~Myr after its host star settled on the main sequence. 
In the latter case, we adopted $Q^{\prime}_{\rm p} =300$ as in the case of our Earth \citep{2009ApJ...707.1000H} and neglected the small mass and tidal dissipation in the fluid outer envelope of the planet in comparison with that in its rocky core. 
The present eccentricity tidal decay timescale was found to be $\tau_{\rm e} \equiv |e/(de/dt)| \sim 24$~Gyr in the former case, while $\tau_{\rm e} \sim 2$~Gyr in the latter case. We noted that tidal dissipation inside the star has a negligible effect on the evolution of the eccentricity because of the relatively large separation of the planet and its small mass. Our results on the eccentricity decay timescale are critically dependent on the $Q^{\prime}_{\rm p}$ of the planet, and therefore, it is difficult to reach a conclusion on the possibility that the current eccentricity is primordial or is still excited, for example, by the gravitational perturbation of a distant companion or by other effects \citep[see, e.g.,][]{Correiaetal20}. 

An interesting consequence of the eccentric orbit of TOI-4438b is its expected pseudosynchronization with the orbital motion, that is, its rotation is predicted to be faster than the orbital mean motion $n$ because tides tend to synchronize its rotation with the orbital velocity at periastron where they are stronger. Using the formalism of \citet{Leconteetal10}, we predicted a rotation period of the planet of 6.7 d, that is, 1.12 times shorter than its orbital period. Such a state of pseudosynchronization is reached after a timescale of $\sim 0.26$~Myr for $Q^{\prime}_{\rm p} =10^{5}$ or 5.6~kyr for $Q^{\prime}_{\rm p} = 300$, that is, on timescales much shorter than the system lifetime.  
We expect tides to dissipate energy inside the planet due to its eccentric orbit and pseudosynchronous rotation. The maximum dissipated power is predicted for $Q^{\prime}_{\rm p} =300$ and is $\sim 6.3 \times 10^{16}$~W giving a heat flux of $\sim 20$~W~m$^{-2}$ at the top of its atmosphere with a radius of $2.5$~R$_{\oplus}$. Such a flux is significantly larger than in the case of Jupiter, where the heat flux from the interior of the planet is $\sim 5.4$~W~m$^{-2}$ \citep{Guillotetal04}. Therefore, it can play a relevant role in the atmospheric dynamics of the planet. On the other hand, adopting $Q^{\prime}_{\rm p} = 10^{5}$, we found a dissipated power of $5.2 \times 10^{15}$~W and a surface flux of only $\sim 1.6$~W~m$^{-2}$.

\section{Discussion}
\label{sec:Discussion}

TOI-4438~b has a mass of $M_b$\,=\,5.4\,$\pm$\,1.1\,M$_{\oplus}$, and a radius of $R_b$\,=\,2.52\,$\pm$\,0.13\,R$_{\oplus}$, with precision of 20\% and 5\%, respectively. Its bulk density of $\rho_b$\,=\,1.85$^{+0.51}_{-0.44}$\,g\,cm$^{-3}$, is close to Neptune's of 1.638\,g\,cm$^{-3}$ and 32\% that of Earth's density.
We calculated an equilibrium temperature of 435 K, assuming zero albedo.
TOI-4438~b belongs to the small group ($\lesssim$\,50) of planets around M-dwarf stars whose masses and radii are known with a precision better than 25\%. Fig.\,\ref{fig:mr_plot} shows the mass-radius diagram with different theoretical composition models \citep{2016Zeng}.
TOI-4438~b lies in the mini-Neptunes domain. 

The position of TOI-4438~b in the mass-radius diagram is consistent with a high volatile content; it joins a group of low-density planets abundantly predicted in planet population syntheses based on the core accretion paradigm \citep[e.g.,][]{Venturini2020,Burn2021,Schlecker2021,Schlecker2021b}, where they correspond to planets with high ($\sim$50\,\%) water ice mass fractions that accreted the bulk of their mass outside the water ice line \citep[e.g.,][]{Mordasini2018a,Bitsch2021}.
However, the planet's bulk density is also compatible with an extended atmosphere.

With an instellation of $S_\mathrm{b}$ = 6.00$^{+0.90}_{-0.77}\,S_\oplus$, TOI-4438~b is orbiting significantly closer to its host star than the predicted inner edge of the habitable zone~\citep{Kopparapu2014}.
As such, it may either have undergone an instellation-induced runaway greenhouse transition~\citep[e.g.,][]{Kasting1988,Nakajima1992} and lost all its water to space, or currently be in a post-runaway state and host an extended steam atmosphere.
The latter scenario is consistent with the planet's low bulk density~\citep{Turbet2020,Dorn2021} and gains credibility through the expected long durations of runaway greenhouse phases in planets orbiting M~dwarfs~\citep{Luger2015}.
With a derived instellation of a few times that of Earth, TOI-4438\,b is well within the runaway greenhouse regime but close to the habitable zone inner edge. As such, and through the precise density estimate presented here, TOI-4438\,b contributes to the currently small sample of planets suited to probe the habitable zone inner edge discontinuity, a predicted demographic imprint of the habitable zone inner edge in the radius--density distribution of small exoplanets~\citep{Turbet2019,Schlecker2023}.

\begin{figure*}[!t]
\centering
\includegraphics[width=1.0\textwidth]{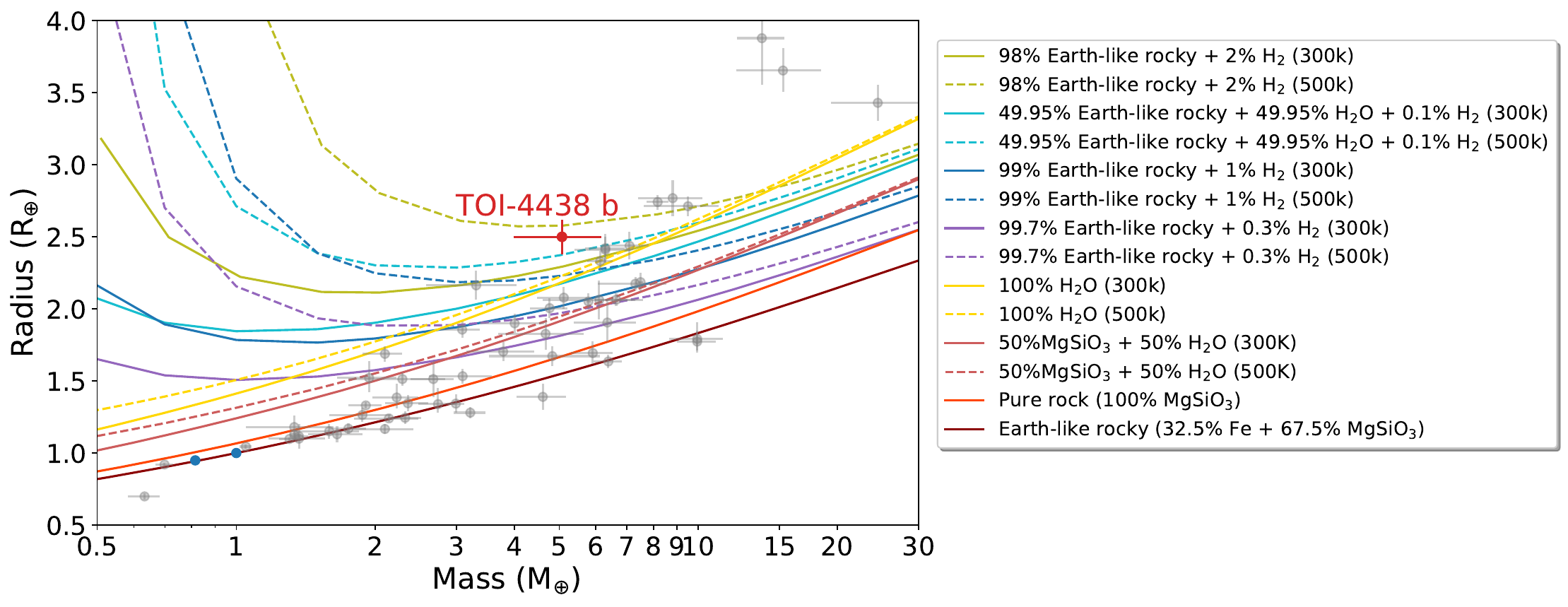}
\caption{Mass-radius diagram for well-characterized planets (R$<4$\,R$_{\oplus}$, M$<13$\,M$_{\oplus}$) around M-dwarf stars (T$_\mathrm{eff}<$4000\,K), as retrieved from the Transiting Extrasolar Planet Catalogue \citep{2011Southworth}.  The masses and radii are known with a precision better than 30\% and 10\%, respectively. The theoretical composition models from \cite{Zeng2019PNAS..116.9723Z} are displayed from bottom to top, with solid and dashed curves. The location of TOI-4438\,b is marked with a red dot. \label{fig:mr_plot}
        }
\end{figure*}

\subsection{Interior modeling}
\label{interior}

Given the planet mass and radius obtained in our transit and radial velocity analyses, we derived the interior composition of TOI-4438 b. The forward model favoured an interior structure model for water-dominated sub-Neptunes \citep{Acuna21,Aguichine21}. The interior structure of the planet was assumed to consist of three layers: a Fe-rich core, a silicate mantle, and a water layer. The phases of water present in the top layer were determined self-consistently with the irradiation received by the planet from its host star and an atmospheric model. To compute the temperature in the deep water layer, we used the $k$-correlated atmospheric model initially introduced in \cite{Marcq17} and \cite{Pluriel19}, with up-to-date opacity and equation of state data from \cite{Acuna23}. The atmospheric model computed the temperature profile and the radius contribution of the upper water envelope (from 300\,bar to 20\,mbar). The interior and the atmospheric models were coupled self-consistently assuming radiative-convective equilibrium in the atmosphere, and with an iterative algorithm \citep{Acuna21}. A water-dominated upper envelope was inferred with 99\% H$_2$O and 1\% CO$_{2}$. 

TOI-4438\,b's instellation is high enough to have molecules such as water in gas and supercritical phases. Fig.\, \ref{fig:interior_corner} shows the posterior distribution functions of the free parameters in our interior composition retrieval. Due to the low density of the planet and its distance from the star, the refractory elements probably consist of silicates instead of a Si and Fe mixture \citep{Aguichine20}. Therefore, in our analysis we set the Fe mass fraction to zero. We obtained a water mass fraction of $0.62^{+0.34}_{-0.16}$. With a composition between 46\% and 96\% H$_2$O (1$\sigma$), TOI-4438 b is volatile-rich. Moreover, it is very likely to have H/He atmosphere mixed with molecules, such as H$_2$O, CO$_{2}$, and CH$_{4}$. In the Solar System, comets and other small bodies with up to 80\% of pure water have been found \citep{McKay19}. However, low-mass planets with more than 50\% water in mass are not a natural outcome of planetesimal accretion and other planet formation mechanisms \citep{Kimura_Ikoma20,Miguel20}. Thus, atmospheric characterization is required to break the degeneracy between the mass and the composition of the envelope, and to determine whether TOI-4438 b has a massive, high metallicity envelope (i.e H$_2$O envelope scenario), or a less massive, low metallicity atmosphere (i.e H/He-dominated envelope scenario).

\begin{figure}[!t]
\centering
\includegraphics[width=0.4\textwidth]{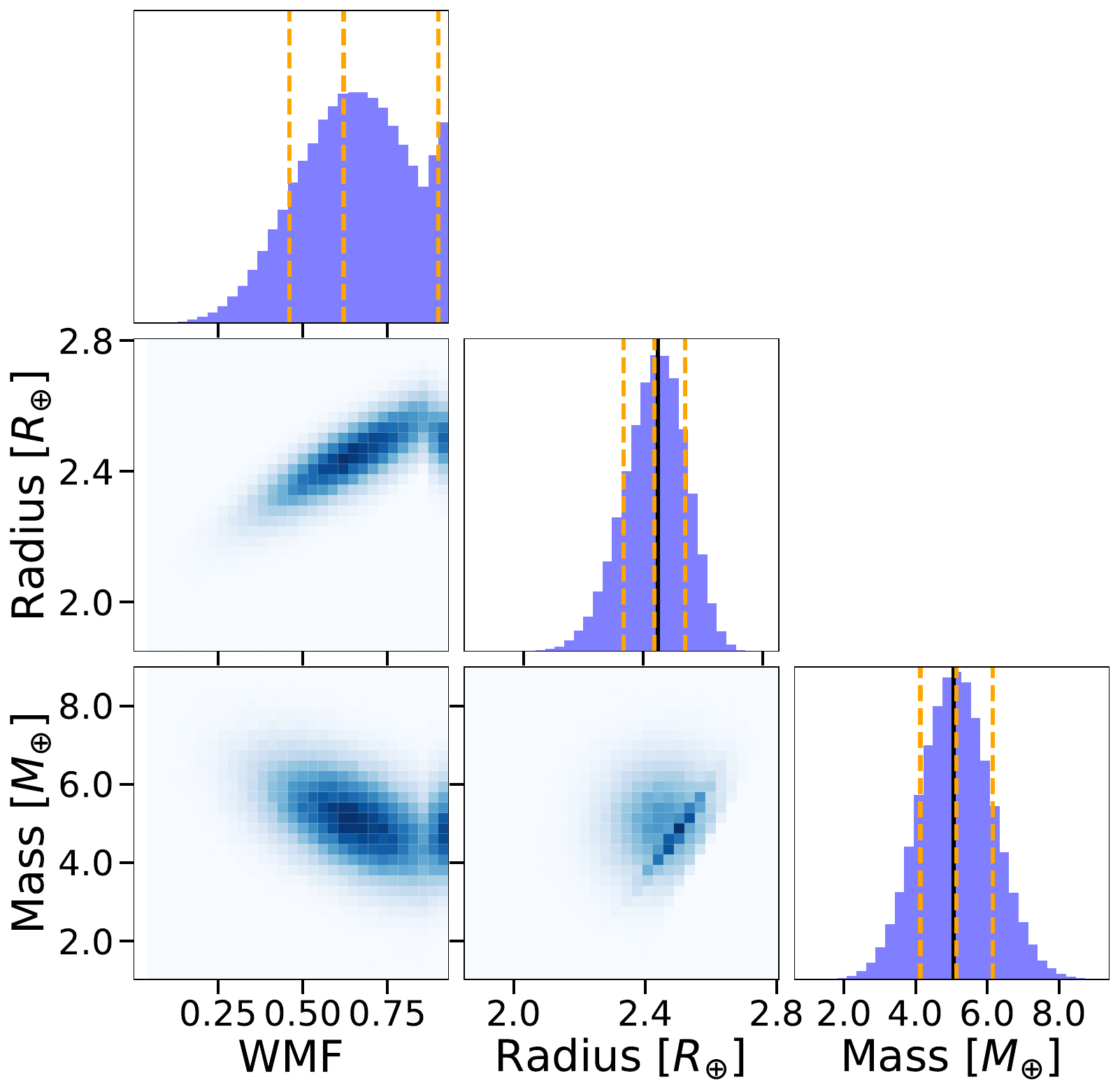}
\caption{Posterior probability distributions of the mass, radius and water mass fraction (WMF) derived from in the interior structure retrieval. The diagonal panels show the marginalized probability distributions of the individual parameters. For the observable parameters (mass and radius), the mean observed value is indicated with a black line. The mean and the 1$\sigma$ interval of the posterior distributions retrieved in the MCMC are shown as orange dashed lines for all parameters. The PDFs agree well with the mean and uncertainties of the observed mass and radius. 
}
\label{fig:interior_corner}
\end{figure}

\subsection{Prospects for atmospheric characterization}

\begin{figure}[!t]
\centering
\includegraphics[width=0.54\textwidth]{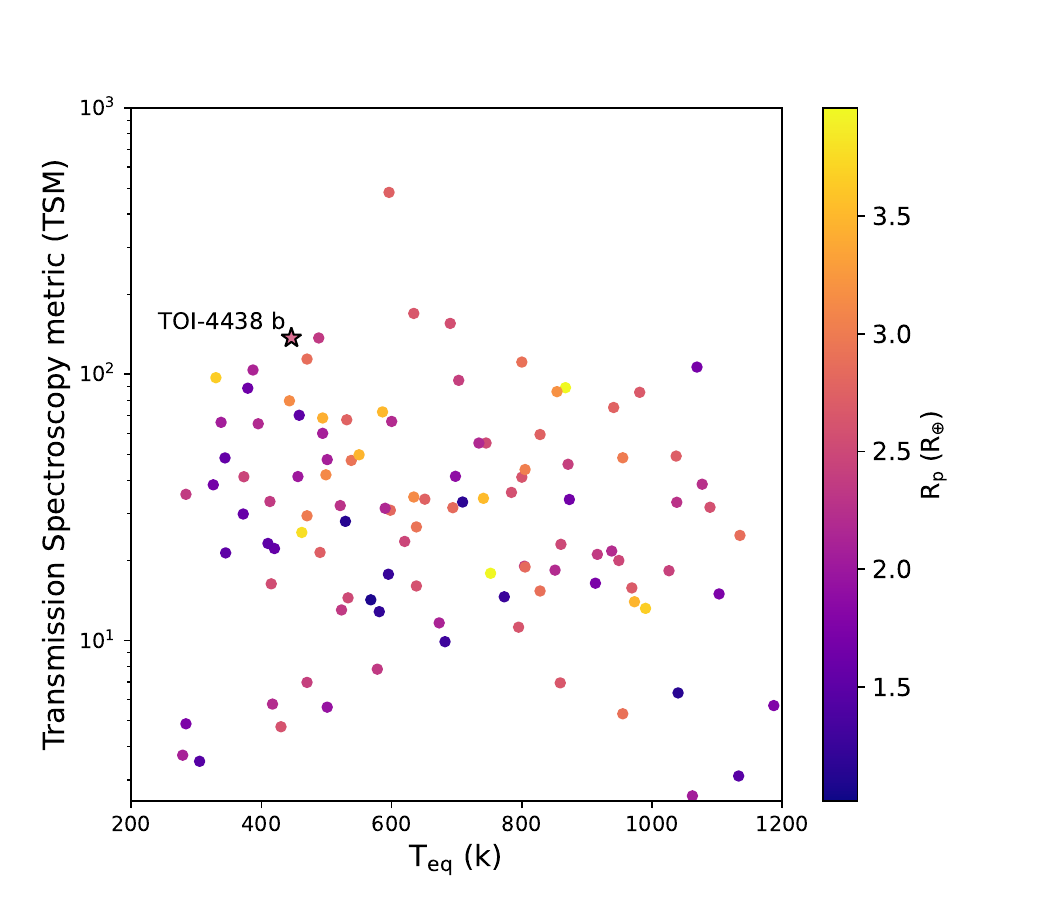}
\caption{TSM for planets with radii smaller than 4\,R$_{\oplus}$ and measured masses, whose host stars have J-band magnitudes of 8.5\,mag\,$<\,J\,<$\,11.5\,mag. TOI-4438\,b is displayed with a star symbol. The color indicates the radius of the planets.  \label{fig:TSM}
        }
\end{figure}

\begin{figure*}
\centering
\includegraphics[width=0.49\textwidth]{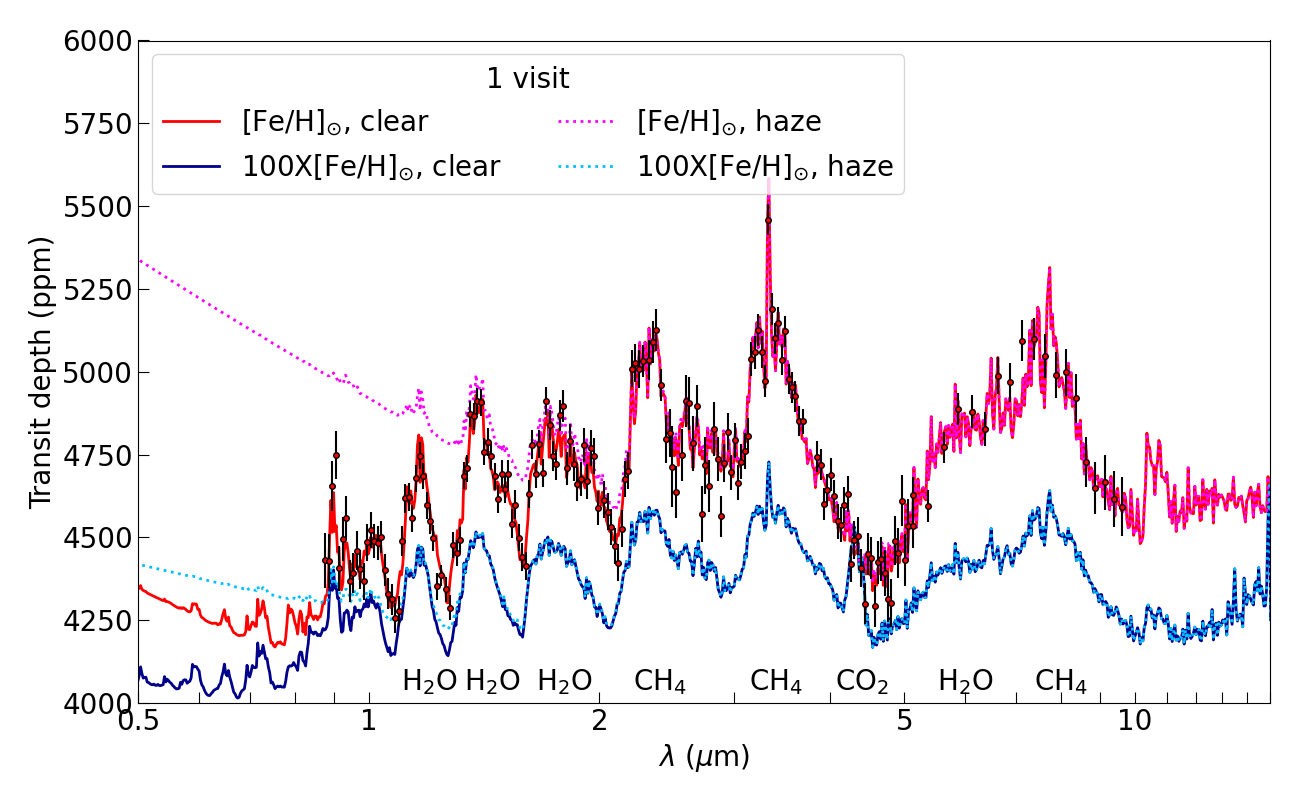}
\includegraphics[width=0.49\textwidth]{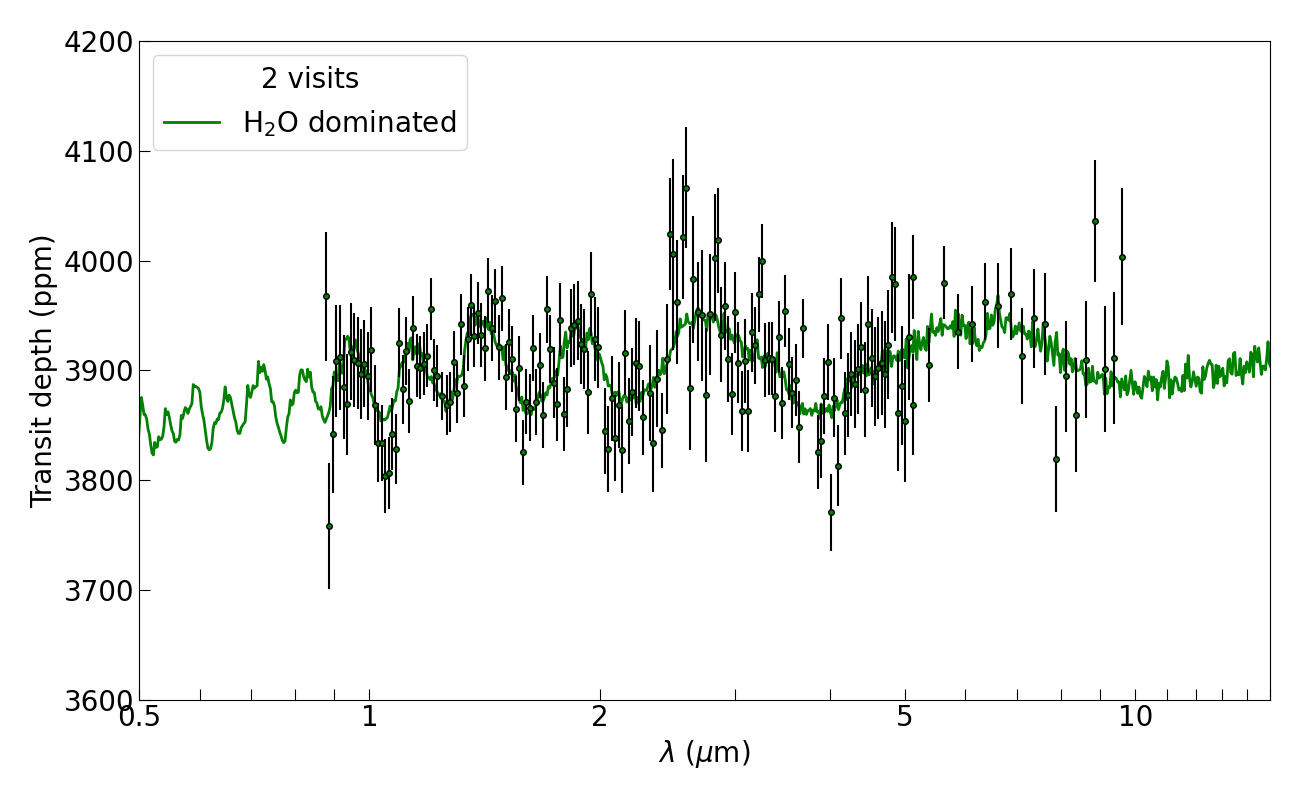}
\caption{Synthetic atmospheric transmission spectra of TOI-4438\,b. \textit{Left:} fiducial models for clear or hazy H/He atmospheres with scaled solar abundances. \textit{Right:} model for a steam H$_2$O atmosphere. 
Simulated measurements with error bars are shown for the observation of one (\textit{left}) or two (\textit{right}) transits with \textit{JWST} NIRISS-SOSS, NIRSpec-G395H, and MIRI-LRS configurations. \label{fig:jwst_atmo}
        }
\end{figure*}

We used the metric proposed by \cite{Kempton2018PASP..130k4401K} to identify the transiting planets most amenable for atmospheric characterization via transmission spectroscopy with the \textit{JWST}. For TOI-4438\,b the derived transmission spectroscopy metric (TSM) is 136\,$\pm$\,13, which is well above the threshold value of 92 defined for the respective planetary categories (small mini-Neptunes; 1.5\,$R_{\oplus}$\,$<$\,R$_p$\,$<$\,2.75\,$R_{\oplus}$) for follow-up studies \citep{Kempton2018PASP..130k4401K}. In this regard, TOI-4438\,b is placed in the second quartile (rank 25--50\%) of \textit{JWST} targets with the strongest predicted atmospheric detections, and it is among the first five planets with the highest TSM, whose host stars have $J$-band magnitudes of 8.5\,mag\,$<\,J\,<$\,11.5\,mag. Therefore, TOI-4438\,b is one of the best small mini-Neptune targets for atmospheric characterization. 
Fig.\,\ref{fig:TSM} shows the TSM value of TOI-4438\,b along with the TSMs of known transiting planets smaller than 4\,R$_{\oplus}$ with measured masses, as retrieved from the Transiting Extrasolar Planet Catalogue \citep{2011Southworth}.
We also calculated the emission spectroscopy metric (ESM) following \cite{Kempton2018PASP..130k4401K} finding an ESM of $\sim$4, which instead is below the suggested threshold value (i.e., ESM\,=\,7.5).
The planet TOI-4438\,b is not a high priority target for emission spectroscopy of terrestrial planets with JWST.

We explored the potential of TOI-4438\,b for transmission spectroscopy with the \textit{JWST} through spectral simulations for a set of model atmospheres consistent with the planetary mass, radius, and equilibrium temperature. We adopted \texttt{TauREx 3} \citep{al-refaie2021} to generate the synthetic transmission spectra. TOI-4438\,b could have a rocky core surrounded by an H/He envelope or be a waterworld (Sect\,\ref{interior}). We modeled H/He atmospheres with 1$\times$ and 100$\times$ scaled solar abundances using the atmospheric chemical equilibrium module \citep{agundez2012}, including collisionally induced absorption by H$_2$–H$_2$ and H$_2$–He \citep{abel2011,abel2012,fletcher2018}, and Rayleigh scattering. For each chemical setup, we considered the cases of clear and hazy atmospheres. The haze was modeled by Mie scattering with the formalism of \cite{lee2013}, assuming a particle size of $\alpha$\,=\,0.05\,$\mu$m, a mixing ratio of $\chi_c$\,=\,10$^{-12}$, and an extinction coefficient of $Q_0$\,=\,40. While this assumption has some limitations in reproducing the actual behavior of scattered light due to using spherical particles instead of non-spherical ones (see, e.g., \citealt{1996JQSRT..55..535M}), this strategy was motivated by its low computational cost.  
The same haze parameters were chosen in previous atmospheric simulation studies (e.g. \citealp{orell-miquel2023}). We note that a super-solar metallicity is expected from planetary formation theories (e.g., \citealp{fortney2013,thorngren2016}). The equilibrium temperature within $\sim$400--600\,K also points to a high degree of haziness due to inefficient haze removal \citep{gao2020,ohno2021,yu2021}.
Additionally, we modeled the case of a steam H$_2$O atmosphere.

We used \texttt{ExoTETHyS} \citep{Morello2021} to simulate the corresponding \textit{JWST} spectra, as observed with the NIRISS-SOSS (0.6--2.8\,$\mu$m), NIRSpec-G395H (2.88--5.20\,$\mu$m), and MIRI-LRS (5--12\,$\mu$m) instrumental modes. The spectroscopic error bars obtained with \texttt{ExoTETHyS} are consistent with those calculated using the Exoplanet Characterization Toolkit (ExoCTK, \citealp{bourque2021}) and \texttt{PandExo} \citep{batalha2017}, as proven in previous studies \citep{murgas2021,espinoza2022,luque2022a,luque2022b,chaturvedi2022,lillo-box2023,orell-miquel2023,2023A&A...678A..80P}. We conservatively increased the error estimates by 20\%. We chose the sizes of the wavelength bins following the recommendations from recent \textit{JWST} data synthesis papers that adopted a spectral resolution of $R\sim$100 for NIRISS and NIRSpec (Carter et al., under review), and a constant bin size of 0.25\,$\mu$m for MIRI-LRS observations \citep{Powell2024}.

Fig. \ref{fig:jwst_atmo} shows the synthetic transmission spectra for the atmospheric configurations above. The H/He model atmospheres exhibit strong H$_2$O and CH$_4$ absorption features of $\gtrsim$100--1000 parts per million (ppm), depending on metallicity and haze, while the steam H$_2$O atmosphere has absorption features $\lesssim$100 ppm. The predicted error bars for a single transit observation are 31--123\,ppm (mean error 53 ppm) for NIRISS-SOSS, 38--82 ppm (mean error 56 ppm) for NIRSpec-G395H, and 47--89 ppm (mean error 64 ppm) for MIRI-LRS. Our simulations suggest that a single transit observation with NIRISS-SOSS or NIRSpec-G395H is well suited to detect an H/He atmosphere, while at least two transit observations may be needed to reveal a secondary H$_2$O-- dominated atmosphere.

\subsection{Expected radio emission from star-planet interaction}

\begin{figure*}
\centering
     \begin{subfigure}[b]{0.33\textwidth}
         \centering
         \includegraphics[width=\textwidth]{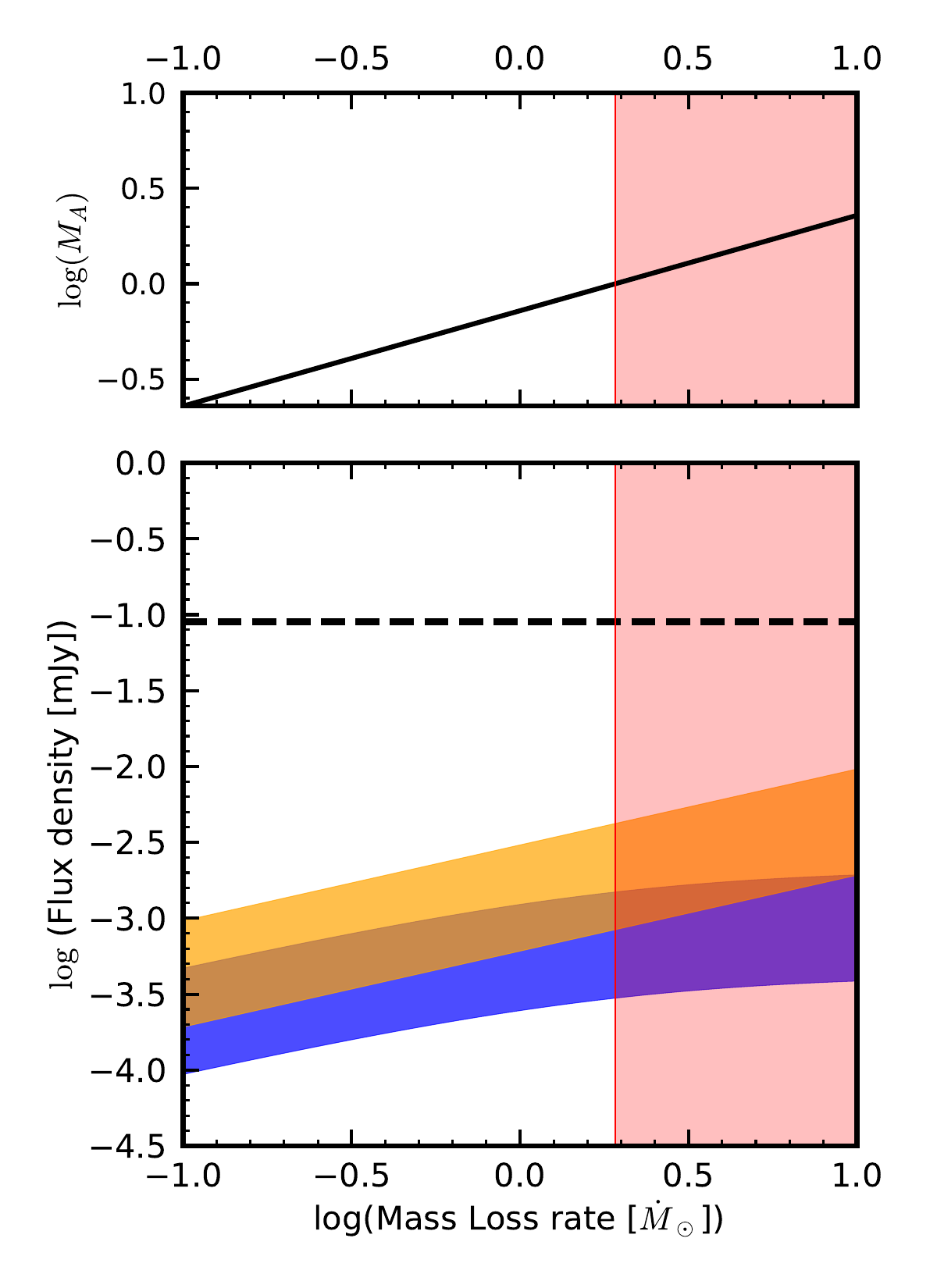}
     \end{subfigure}
     \hfill
    \begin{subfigure}[b]{0.33\textwidth}
         \centering
         \includegraphics[width=\textwidth]{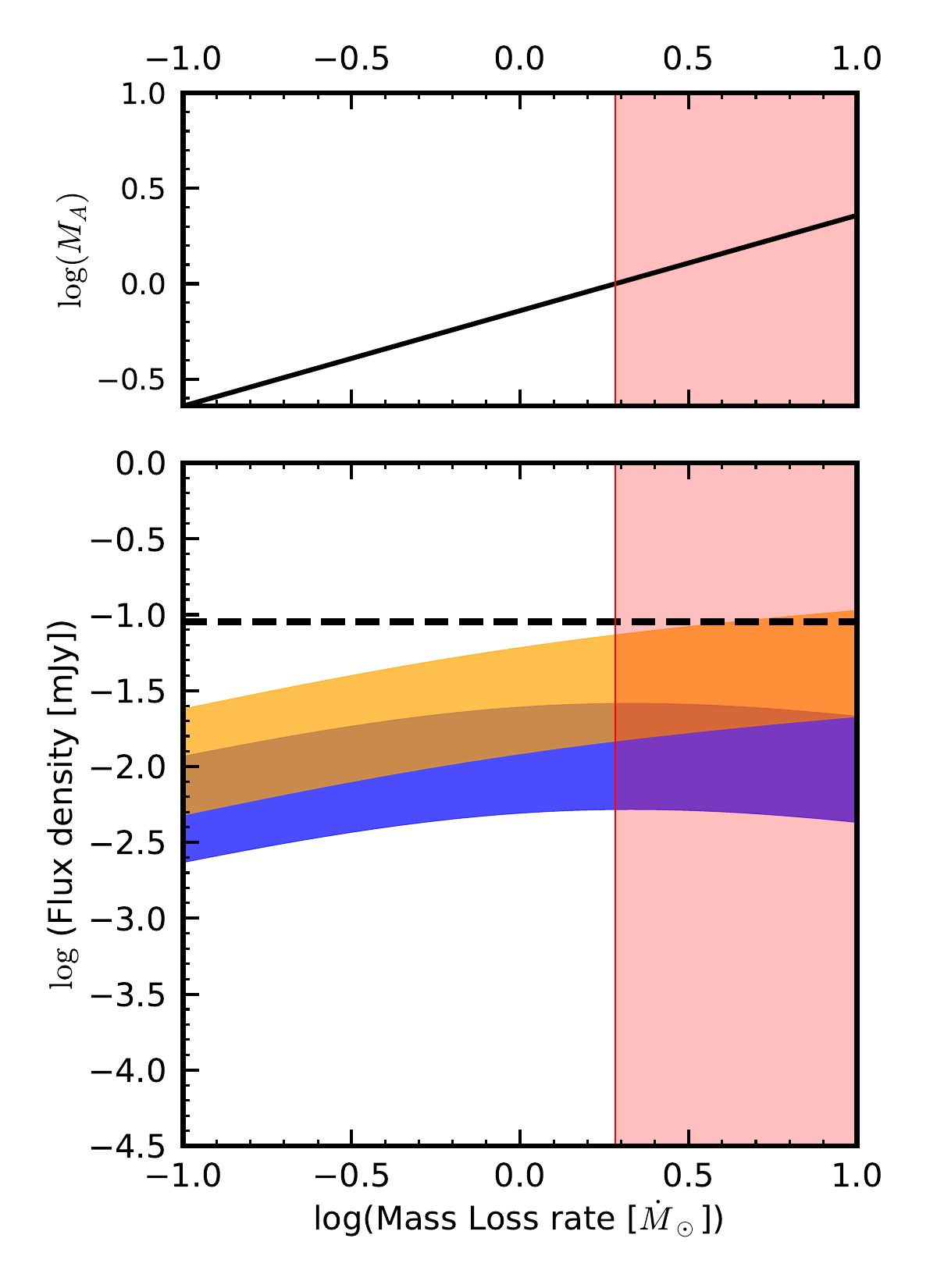}
     \end{subfigure}
     \hfill
         \begin{subfigure}[b]{0.33\textwidth}
         \centering
         \includegraphics[width=\textwidth]{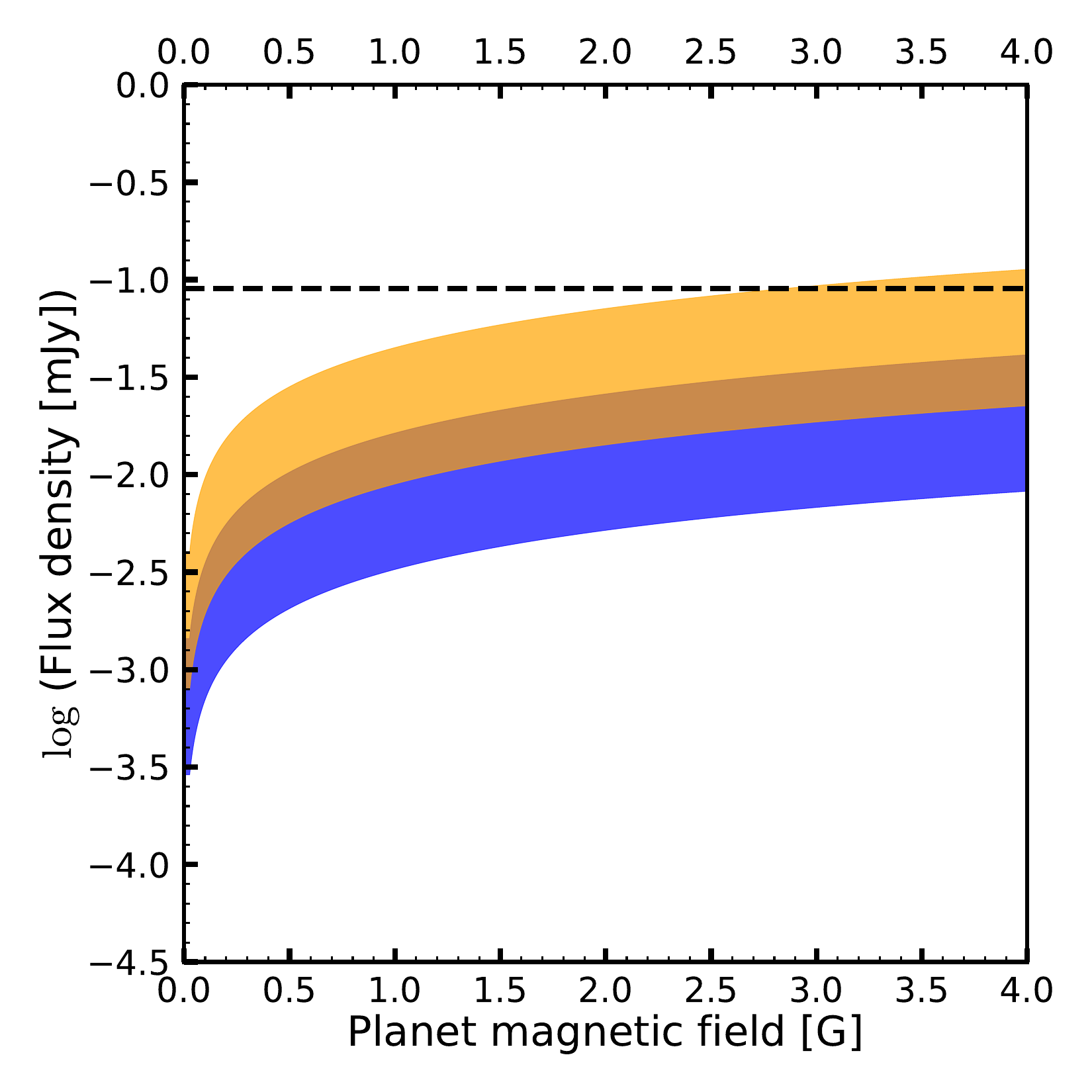}
     \end{subfigure}
\caption{Flux density arising from star planet interaction as a function of stellar mass-loss rate (\textit{left} and \textit{center}) and the magnetic field of the planet (\textit{right}). 
The emission expected from Saur--Turnpenney's model is shown in orange, while the emission expected from Zarka--Lanza's model is shown in blue. The overlap between both models is shown in brown. We show in pale red color the region when the planet is in the  super-Alfv\`enic regime. The dashed black line represents an assumed detection threshold of $100 \,\rm \mu Jy$. For the variation with respect to the magnetic field of the planet we used $\dot{M}=1.92\dot{M}_\odot$, which is the maximum value for the mass loss rate before entering the sub-Alfv\`enic regime. 
\textit{Left}: Expected flux density for an unmagnetized planet in a closed dipolar geometry, as a function of the stellar wind mass loss. The pale red colored region indicates the super-Alfvénic case, when our approach no longer applies.
\textit{Center}: The same as in the left panel, but for a magnetized planet with $B_{\rm p}$ = 2 G.
\textit{Right}: Expected flux density for a closed dipolar geometry}
\label{SPI}
\end{figure*}

Here we assess the feasibility of detecting radio emission arising from magnetic star-planet interaction (SPI) between TOI-4438b and its host star. The physical phenomenon responsible for this kind of emission is the electron cyclotron maser (ECM) instability \citep{Melrose1982}, which can generate auroral radio emission on the star and also the planet itself.

The characteristic frequency associated with ECM is given by the electron gyrofrequency, $\nu_G = 2.8 \, B$ MHz, where $B$ is the local magnetic field in Gauss. This non-thermal emission is coherent and circularly polarized (reaching up to 100\% in some instances), and may reach a large bandwidth ($\Delta\,\nu \sim \nu_G/2$).

We note, however, that any kind of SPI emission coming from the planet itself would be undetectable with current instrumentation. The reason is that the magnetic field of a planet is merely a few Gauss, so the resulting emission would be blocked by the Earth's ionosphere. We therefore studied the auroral emission on the star TOI-4438 induced by the presence of its planet b, since the global magnetic field of the star is of the order of a few hundreds of Gauss or larger. As a result, the corresponding electron gyrofrequency would be of several hundreds or thousands of MHz, a frequency range that can be studied with current observatories. This phenomenon takes place in the sub-Alfv\'enic regime, that is, when the planet is close enough to the star that the speed of the plasma wind is lower than the Alfv\'en speed, enabling energy and momentum to travel back from the planet onto the star in the form of Alfv\`en wings, triggering the radio emission close to the stellar surface.

To estimate the magnetic field of TOI-4438, we used the relation from \citet{Reiners2022}, who estimated the Rossby number as a function of the stellar mass with the relations from \citet{Wright2018}, which yields a value of $B_\star \approx 191.9$\,G.
We then used the code described in \citet{PerezTorres2021} to estimate the flux density arising from the sub-Alfvénic interaction between an exoplanet and its host star, for an isothermal wind. 
For our calculations, we assumed an isothermal wind with coronal temperature equal to that of the solar corona, $T = 2\,\times\,10^6$ K, and which is also adequate for M dwarf stars. We took the solid angle covered by the ECM emission to be 1.6 sterradians, in conformity with observations of the Jupiter-Io decametric radio emission \citep{Ray2008}. We worked with two different emission models, Saur--Turnpenney \citep{Saur2013,Turnpenney2018} and Zarka--Lanza \citep{Zarka2007,Lanza2009}, and in both cases we considered an efficiency in the conversion from Poynting flux to radio emission in the range from 0.2 to 1\%.

In the left panels of Fig.\,\ref{SPI}, we show the radio flux density from sub-Alfvénic star-planet interaction as a function of the mass-loss rate of the star for an unmagnetized planet and a planet with a magnetic field of 2\,G. In this regime, our approach does not apply and, therefore, no radio emission from SPI can reach us.  In the right panel of Fig.\,\ref{SPI} we show the emission as a function of the magnetic field of the planet. Despite the planet being rather large in size, the expected radio emission is too low to be detectable in essentially all scenarios due to the large distance to the system (30 pc). Only in the case when the planet is highly magnetized and the stellar wind is relatively powerful there could be marginal chances of detecting a signal from SPI.

\section{Conclusions}
\label{sec:Conclusions}

We have reported the confirmation and characterization of TOI-4438\,b, a mini-Neptune around an M3.5\,V star (G 182-34) on a 7.44 d orbit. We performed a joint modeling of the \textit{TESS} and MuSCAT2 light curves with CARMENES high-resolution spectroscopy measurements. We found a radius of 2.52\,$\pm$\,0.13 R$_\oplus$ and a mass of 5.4\,$\pm$\,1.1\, M$_\oplus$, resulting in a bulk density of 1.85\,$^{+0.51}_{-0.44}$\,g\,cm$^{-3}$.
The equilibrium temperature of the planet is 435\,$\pm$\,15 K.

Our interior structure retrieval with a pure water envelope yields a minimum water mass fraction of 46\% (1$\sigma$). The volatile-rich mini-Neptune TOI-4438\,b has likely H/He mixed with molecules, such as water, CO$_2$, and CH$_4$.
TOI-4438\,b presents a high transmission spectroscopy metric of 136\,$\pm$\,13, which places the planet among the most suitable targets for atmospheric observations with \textit{JWST}.
We performed spectral simulations to explore the potential for transmission spectroscopy with \textit{JWST}. A single transit observation of TOI-4438 with NIRISS-SOSS or NIRSpec-G395H should be adequate to detect an H/He atmosphere, while at least two transits may be needed to reveal a secondary H$_2$O dominated atmosphere.

\begin{acknowledgements}
      We acknowledge the use of public \textit{TESS} data from pipelines at the \textit{TESS} Science Office and at the TESS Science Processing Operations Center. TESS data presented in this paper were obtained from the Milkulski Archive for Space Telescopes (MAST) at the Space Telescope Science Institute.
      Resources supporting this work were provided by the NASA High-End Computing (HEC) Program through the NASA Advanced Supercomputing (NAS) Division at Ames Research Center for the production of the SPOC data products. 
      CARMENES is an instrument at the Centro Astron\'omico Hispano en Andaluc\'ia (CAHA) at Calar Alto (Almer\'{\i}a, Spain), operated jointly by the Junta de Andaluc\'ia and the Instituto de Astrof\'isica de Andaluc\'ia (CSIC). CARMENES was funded by the Max-Planck-Gesellschaft (MPG), the Consejo Superior de Investigaciones Cient\'{\i}ficas (CSIC), the Ministerio de Econom\'ia y Competitividad (MINECO) and the European Regional Development Fund (ERDF) through projects FICTS-2011-02, ICTS-2017-07-CAHA-4, and CAHA16-CE-3978, and the members of the CARMENES Consortium (Max-Planck-Institut f\"ur Astronomie, Instituto de Astrof\'{\i}sica de Andaluc\'{\i}a, Landessternwarte K\"onigstuhl, Institut de Ci\`encies de l'Espai, Institut f\"ur Astrophysik G\"ottingen, Universidad Complutense de Madrid, Th\"uringer Landessternwarte Tautenburg, Instituto de Astrof\'{\i}sica de Canarias, Hamburger Sternwarte, Centro de Astrobiolog\'{\i}a and Centro Astron\'omico Hispano-Alem\'an), with additional contributions by the MINECO, the Deutsche Forschungsgemeinschaft (DFG) through the Major Research Instrumentation Programme and Research Unit FOR2544 ''Blue Planets around Red Stars'', the Klaus Tschira Stiftung, the states of Baden-W\"urttemberg and Niedersachsen, and by the Junta de Andaluc\'{\i}a.
      The results reported herein benefitted from collaborations and/or information exchange within NASA’s Nexus for Exoplanet System Science (NExSS) research coordination network sponsored by NASA’s Science Mission Directorate under Agreement No. 80NSSC21K0593 for the program ``Alien Earths".
      The Joan Oró Telescope (TJO) of the Montsec Observatory (OdM) is owned by the Catalan Government and operated by the Institute for Space Studies of Catalonia (IEEC).
      G.M. has received fundings from the Ariel Postdoctoral Fellowship program of the Swedish National Space Agency (SNSA).
      S.V.J. acknowledges the support of the DFG priority program SPP 1992 “Exploring the Diversity of Extrasolar Planets (JE 701/5-1).
      M.P.T. and L.P.M. acknowledge financial support through grants CEX2021-001131-S and PID2020-117404GB-C21, funded by MCIU/AEI/ 10.13039/501100011033. L.P.M. also acknowledges funding through the grant PRE2020-095421, funded by MCIU/AEI/10.13039/501100011033 and by FSE Investing in your future.
      This work is partly supported by JSPS KAKENHI Grant Number JP18H05439, JST CREST Grant Number JPMJCR1761. This article is based on observations made with the MuSCAT2 instrument, developed by ABC, at Telescopio Carlos Sánchez operated on the island of Tenerife by the IAC in the Spanish Observatorio del Teide.
      We acknowledge financial support from the Agencia Estatal de Investigaci\'on (AEI/10.13039/501100011033) of the Ministerio de Ciencia e Innovaci\'on and the ERDF ``A way of making Europe'' through projects
  PID2022-137241NB-C4[1:4],    
  PID2021-125627OB-C31,        
  PID2019–109522GB–C5[1:4],    
and the Centre of Excellence ``Severo Ochoa'' and ``Mar\'ia de Maeztu'' awards to the Instituto de Astrof\'isica de Canarias (CEX2019-000920-S), Instituto de Astrof\'isica de Andaluc\'ia (CEX2021-001131-S) and Institut de Ci\`encies de l'Espai (CEX2020-001058-M).
  This work was also funded by the Generalitat de Catalunya/CERCA programme,
the DFG through grant HA3279/14-1 and the priority program SPP 1992 ``Exploring the Diversity of Extrasolar Planets'' (JE 701/5-1),
the JSPS KAK-ENHI through grant JP18H05439,
the JST CREST through grant JPMJCR1761,
and the Israel Science Foundation through grant 1404/22.

\end{acknowledgements}

\bibliographystyle{aa.bst} 
\bibliography{main.bib}

\begin{thebibliography}{167}
\expandafter\ifx\csname natexlab\endcsname\relax\def\natexlab#1{#1}\fi

\bibitem[{{Abel} {et~al.}(2011){Abel}, {Frommhold}, {Li}, \& {Hunt}}]{abel2011}
{Abel}, M., {Frommhold}, L., {Li}, X., \& {Hunt}, K. L.~C. 2011, Journal of
  Physical Chemistry A, 115, 6805

\bibitem[{{Abel} {et~al.}(2012){Abel}, {Frommhold}, {Li}, \& {Hunt}}]{abel2012}
{Abel}, M., {Frommhold}, L., {Li}, X., \& {Hunt}, K. L.~C. 2012, \jcp, 136,
  044319

\bibitem[{{Acu{\~n}a} {et~al.}(2023){Acu{\~n}a}, {Deleuil}, \&
  {Mousis}}]{Acuna23}
{Acu{\~n}a}, L., {Deleuil}, M., \& {Mousis}, O. 2023, \aap, 677, A14

\bibitem[{{Acu{\~n}a} {et~al.}(2021){Acu{\~n}a}, {Deleuil}, {Mousis}, {Marcq},
  {Levesque}, \& {Aguichine}}]{Acuna21}
{Acu{\~n}a}, L., {Deleuil}, M., {Mousis}, O., {et~al.} 2021, \aap, 647, A53

\bibitem[{{Aguichine} {et~al.}(2021){Aguichine}, {Mousis}, {Deleuil}, \&
  {Marcq}}]{Aguichine21}
{Aguichine}, A., {Mousis}, O., {Deleuil}, M., \& {Marcq}, E. 2021, \apj, 914,
  84

\bibitem[{{Aguichine} {et~al.}(2020){Aguichine}, {Mousis}, {Devouard}, \&
  {Ronnet}}]{Aguichine20}
{Aguichine}, A., {Mousis}, O., {Devouard}, B., \& {Ronnet}, T. 2020, \apj, 901,
  97

\bibitem[{{Ag{\'u}ndez} {et~al.}(2012){Ag{\'u}ndez}, {Venot}, {Iro}, {Selsis},
  {Hersant}, {H{\'e}brard}, \& {Dobrijevic}}]{agundez2012}
{Ag{\'u}ndez}, M., {Venot}, O., {Iro}, N., {et~al.} 2012, \aap, 548, A73

\bibitem[{{Al-Refaie} {et~al.}(2021){Al-Refaie}, {Changeat}, {Waldmann}, \&
  {Tinetti}}]{al-refaie2021}
{Al-Refaie}, A.~F., {Changeat}, Q., {Waldmann}, I.~P., \& {Tinetti}, G. 2021,
  \apj, 917, 37

\bibitem[{{Aller} {et~al.}(2020){Aller}, {Lillo-Box}, {Jones}, {Miranda}, \&
  {Barcel\'o Forteza}}]{aller2020}
{Aller}, A., {Lillo-Box}, J., {Jones}, D., {Miranda}, L.~F., \& {Barcel\'o
  Forteza}, S. 2020, \aap, 635, A128

\bibitem[{Anderson {et~al.}(2010)Anderson, Cameron, Hellier, Lendl, Maxted,
  Pollacco, Queloz, Smalley, Smith, Todd, Triaud, West, Barros, Enoch, Gillon,
  Lister, Pepe, S{\'{e} }gransan, Street, \& Udry}]{Anderson_2010}
Anderson, D.~R., Cameron, A.~C., Hellier, C., {et~al.} 2010, ApJL, 726, L19

\bibitem[{{Baglin} {et~al.}(2006){Baglin}, {Auvergne}, {Boisnard}, {Lam-Trong},
  {Barge}, {Catala}, {Deleuil}, {Michel}, \& {Weiss}}]{Baglin2006}
{Baglin}, A., {Auvergne}, M., {Boisnard}, L., {et~al.} 2006, in 36th COSPAR
  Scientific Assembly, Vol.~36, 3749

\bibitem[{{Ballerini} {et~al.}(2012){Ballerini}, {Micela}, {Lanza}, \&
  {Pagano}}]{Ballerini2012}
{Ballerini}, P., {Micela}, G., {Lanza}, A.~F., \& {Pagano}, I. 2012, \aap, 539,
  A140

\bibitem[{{Barkaoui} {et~al.}(2023){Barkaoui}, {Timmermans}, {Soubkiou},
  {Rackham}, {Burgasser}, {Chouqar}, {Pozuelos}, {Collins}, {Howell}, {Simcoe},
  {Melis}, {Stassun}, {Tregloan-Reed}, {Cointepas}, {Gillon}, {Bonfils},
  {Furlan}, {Gnilka}, {Almenara}, {Alonso}, {Benkhaldoun}, {Bonavita},
  {Bouchy}, {Burdanov}, {Chinchilla}, {Davoudi}, {Delrez}, {Demangeon},
  {Dominik}, {Demory}, {de Wit}, {Dransfield}, {Ducrot}, {Fukui}, {Hinse},
  {Hooton}, {Jehin}, {Jenkins}, {J{\o}rgensen}, {Latham}, {Garcia},
  {Carrazco-Gaxiola}, {Ghachoui}, {G{\'o}mez Maqueo Chew}, {G{\"u}nther},
  {McCormac}, {Murgas}, {Murray}, {Narita}, {Niraula}, {Pedersen}, {Queloz},
  {Rebolo-L{\'o}pez}, {Ricker}, {Sabin}, {Sajadian}, {Schanche}, {Schwarz},
  {Seager}, {Sebastian}, {Sefako}, {Sohy}, {Southworth}, {Srdoc}, {Thompson},
  {Triaud}, {Vanderspek}, {Wells}, {Winn}, \&
  {Z{\'u}{\~n}iga-Fern{\'a}ndez}}]{2023A&A...677A..38B}
{Barkaoui}, K., {Timmermans}, M., {Soubkiou}, A., {et~al.} 2023, \aap, 677, A38

\bibitem[{{Barrag{\'a}n} {et~al.}(2022){Barrag{\'a}n}, {Aigrain}, {Rajpaul}, \&
  {Zicher}}]{2022Barragan}
{Barrag{\'a}n}, O., {Aigrain}, S., {Rajpaul}, V.~M., \& {Zicher}, N. 2022,
  \mnras, 509, 866

\bibitem[{{Barrag{\'a}n} {et~al.}(2017){Barrag{\'a}n}, {Gandolfi}, \&
  {Antoniciello}}]{2018Barragan}
{Barrag{\'a}n}, O., {Gandolfi}, D., \& {Antoniciello}, G. 2017, {pyaneti:
  Multi-planet radial velocity and transit fitting}, Astrophysics Source Code
  Library, record ascl:1707.003

\bibitem[{{Batalha} {et~al.}(2017){Batalha}, {Mandell}, {Pontoppidan},
  {Stevenson}, {Lewis}, {Kalirai}, {Earl}, {Greene}, {Albert}, \&
  {Nielsen}}]{batalha2017}
{Batalha}, N.~E., {Mandell}, A., {Pontoppidan}, K., {et~al.} 2017, \pasp, 129,
  064501

\bibitem[{{Batalha} {et~al.}(2013){Batalha}, {Rowe}, {Bryson}, {Barclay},
  {Burke}, {Caldwell}, {Christiansen}, {Mullally}, {Thompson}, {Brown},
  {Dupree}, {Fabrycky}, {Ford}, {Fortney}, {Gilliland}, {Isaacson}, {Latham},
  {Marcy}, {Quinn}, {Ragozzine}, {Shporer}, {Borucki}, {Ciardi}, {Gautier},
  {Haas}, {Jenkins}, {Koch}, {Lissauer}, {Rapin}, {Basri}, {Boss}, {Buchhave},
  {Carter}, {Charbonneau}, {Christensen-Dalsgaard}, {Clarke}, {Cochran},
  {Demory}, {Desert}, {Devore}, {Doyle}, {Esquerdo}, {Everett}, {Fressin},
  {Geary}, {Girouard}, {Gould}, {Hall}, {Holman}, {Howard}, {Howell},
  {Ibrahim}, {Kinemuchi}, {Kjeldsen}, {Klaus}, {Li}, {Lucas}, {Meibom},
  {Morris}, {Pr{\v{s}}a}, {Quintana}, {Sanderfer}, {Sasselov}, {Seader},
  {Smith}, {Steffen}, {Still}, {Stumpe}, {Tarter}, {Tenenbaum}, {Torres},
  {Twicken}, {Uddin}, {Van Cleve}, {Walkowicz}, \&
  {Welsh}}]{2013ApJS..204...24B}
{Batalha}, N.~M., {Rowe}, J.~F., {Bryson}, S.~T., {et~al.} 2013, \apjs, 204, 24

\bibitem[{{Bayo} {et~al.}(2008){Bayo}, {Rodrigo}, {Barrado Y Navascu{\'e}s},
  {Solano}, {Guti{\'e}rrez}, {Morales-Calder{\'o}n}, \& {Allard}}]{VOSA}
{Bayo}, A., {Rodrigo}, C., {Barrado Y Navascu{\'e}s}, D., {et~al.} 2008, \aap,
  492, 277

\bibitem[{Bitsch {et~al.}(2021)Bitsch, Raymond, Buchhave, {Bello-Arufe},
  Rathcke, \& Schneider}]{Bitsch2021}
Bitsch, B., Raymond, S.~N., Buchhave, L.~A., {et~al.} 2021, 1

\bibitem[{{Borucki} {et~al.}(2010){Borucki}, {Koch}, {Basri}, {Batalha},
  {Brown}, {Caldwell}, {Caldwell}, {Christensen-Dalsgaard}, {Cochran},
  {DeVore}, {Dunham}, {Dupree}, {Gautier}, {Geary}, {Gilliland}, {Gould},
  {Howell}, {Jenkins}, {Kondo}, {Latham}, {Marcy}, {Meibom}, {Kjeldsen},
  {Lissauer}, {Monet}, {Morrison}, {Sasselov}, {Tarter}, {Boss}, {Brownlee},
  {Owen}, {Buzasi}, {Charbonneau}, {Doyle}, {Fortney}, {Ford}, {Holman},
  {Seager}, {Steffen}, {Welsh}, {Rowe}, {Anderson}, {Buchhave}, {Ciardi},
  {Walkowicz}, {Sherry}, {Horch}, {Isaacson}, {Everett}, {Fischer}, {Torres},
  {Johnson}, {Endl}, {MacQueen}, {Bryson}, {Dotson}, {Haas}, {Kolodziejczak},
  {Van Cleve}, {Chandrasekaran}, {Twicken}, {Quintana}, {Clarke}, {Allen},
  {Li}, {Wu}, {Tenenbaum}, {Verner}, {Bruhweiler}, {Barnes}, \&
  {Prsa}}]{Borucki2010}
{Borucki}, W.~J., {Koch}, D., {Basri}, G., {et~al.} 2010, Science, 327, 977

\bibitem[{{Bouma} {et~al.}(2023){Bouma}, {Palumbo}, \&
  {Hillenbrand}}]{2023ApJ...947L...3B}
{Bouma}, L.~G., {Palumbo}, E.~K., \& {Hillenbrand}, L.~A. 2023, \apjl, 947, L3

\bibitem[{{Bourque} {et~al.}(2021){Bourque}, {Espinoza}, {Filippazzo}, {Fix},
  {King}, {Martlin}, {Medina}, {Batalha}, {Fox}, {Fowler}, {Fraine}, {Hill},
  {Lewis}, {Stevenson}, {Valenti}, \& {Wakeford}}]{bourque2021}
{Bourque}, M., {Espinoza}, N., {Filippazzo}, J., {et~al.} 2021, {The Exoplanet
  Characterization Toolkit (ExoCTK)}, Zenodo

\bibitem[{{Brown} {et~al.}(2013){Brown}, {Baliber}, {Bianco}, {Bowman},
  {Burleson}, {Conway}, {Crellin}, {Depagne}, {De Vera}, {Dilday}, {Dragomir},
  {Dubberley}, {Eastman}, {Elphick}, {Falarski}, {Foale}, {Ford}, {Fulton},
  {Garza}, {Gomez}, {Graham}, {Greene}, {Haldeman}, {Hawkins}, {Haworth},
  {Haynes}, {Hidas}, {Hjelstrom}, {Howell}, {Hygelund}, {Lister}, {Lobdill},
  {Martinez}, {Mullins}, {Norbury}, {Parrent}, {Paulson}, {Petry}, {Pickles},
  {Posner}, {Rosing}, {Ross}, {Sand}, {Saunders}, {Shobbrook}, {Shporer},
  {Street}, {Thomas}, {Tsapras}, {Tufts}, {Valenti}, {Vander Horst}, {Walker},
  {White}, \& {Willis}}]{Brown2013}
{Brown}, T.~M., {Baliber}, N., {Bianco}, F.~B., {et~al.} 2013, \pasp, 125, 1031

\bibitem[{Burn {et~al.}(2021)Burn, Schlecker, Mordasini, Emsenhuber, Alibert,
  Henning, Klahr, \& Benz}]{Burn2021}
Burn, R., Schlecker, M., Mordasini, C., {et~al.} 2021, Astronomy \&
  Astrophysics, 656, A72

\bibitem[{{Caballero} {et~al.}(2016){Caballero}, {Gu{\`a}rdia}, {L{\'o}pez del
  Fresno}, {Zechmeister}, {de Juan}, {Alonso-Floriano}, {Amado}, {Colom{\'e}},
  {Cort{\'e}s-Contreras}, {Garc{\'\i}a-Piquer}, {Gesa}, {de Guindos}, {Hagen},
  {Helmling}, {Hern{\'a}ndez Casta{\~n}o}, {K{\"u}rster}, {L{\'o}pez-Santiago},
  {Montes}, {Morales Mu{\~n}oz}, {Pavlov}, {Quirrenbach}, {Reiners}, {Ribas},
  {Seifert}, \& {Solano}}]{2016SPIE.9910E..0EC}
{Caballero}, J.~A., {Gu{\`a}rdia}, J., {L{\'o}pez del Fresno}, M., {et~al.}
  2016, in Society of Photo-Optical Instrumentation Engineers (SPIE) Conference
  Series, Vol. 9910, Observatory Operations: Strategies, Processes, and Systems
  VI, ed. A.~B. {Peck}, R.~L. {Seaman}, \& C.~R. {Benn}, 99100E

\bibitem[{{Chaturvedi} {et~al.}(2022){Chaturvedi}, {Bluhm}, {Nagel}, {Hatzes},
  {Morello}, {Brady}, {Korth}, {Molaverdikhani}, {Kossakowski}, {Caballero},
  {Guenther}, {Pall{\'e}}, {Espinoza}, {Seifahrt}, {Lodieu}, {Cifuentes},
  {Furlan}, {Amado}, {Barclay}, {Bean}, {B{\'e}jar}, {Bergond}, {Boyle},
  {Ciardi}, {Collins}, {Collins}, {Esparza-Borges}, {Fukui}, {Gnilka}, {Goeke},
  {Guerra}, {Henning}, {Herrero}, {Howell}, {Jeffers}, {Jenkins}, {Jensen},
  {Kasper}, {Kodama}, {Latham}, {L{\'o}pez-Gonz{\'a}lez}, {Luque}, {Montes},
  {Morales}, {Mori}, {Murgas}, {Narita}, {Nowak}, {Parviainen}, {Passegger},
  {Quirrenbach}, {Reffert}, {Reiners}, {Ribas}, {Ricker}, {Rodriguez},
  {Rodr{\'\i}guez-L{\'o}pez}, {Schlecker}, {Schwarz}, {Schweitzer}, {Seager},
  {Stef{\'a}nsson}, {Stockdale}, {Tal-Or}, {Twicken}, {Vanaverbeke}, {Wang},
  {Watanabe}, {Winn}, \& {Zechmeister}}]{chaturvedi2022}
{Chaturvedi}, P., {Bluhm}, P., {Nagel}, E., {et~al.} 2022, \aap, 666, A155

\bibitem[{{Chen} {et~al.}(2021){Chen}, {Pall{\'e}}, {Parviainen}, {Wang}, {van
  Boekel}, {Murgas}, {Yan}, {B{\'e}jar}, {Casasayas-Barris}, {Crouzet},
  {Esparza-Borges}, {Fukui}, {Garai}, {Kawauchi}, {Kurita}, {Kusakabe}, {de
  Leon}, {Livingston}, {Luque}, {Madrigal-Aguado}, {Mori}, {Narita},
  {Nishiumi}, {Oshagh}, {S{\'a}nchez-Benavente}, {Tamura}, {Terada}, \&
  {Watanabe}}]{Chen2021}
{Chen}, G., {Pall{\'e}}, E., {Parviainen}, H., {et~al.} 2021, \mnras, 500, 5420

\bibitem[{{Cifuentes} {et~al.}(2020){Cifuentes}, {Caballero},
  {Cort{\'e}s-Contreras}, {Montes}, {Abell{\'a}n}, {Dorda}, {Holgado},
  {Zapatero Osorio}, {Morales}, {Amado}, {Passegger}, {Quirrenbach}, {Reiners},
  {Ribas}, {Sanz-Forcada}, {Schweitzer}, {Seifert}, \&
  {Solano}}]{cifuentes2020}
{Cifuentes}, C., {Caballero}, J.~A., {Cort{\'e}s-Contreras}, M., {et~al.} 2020,
  \aap, 642, A115

\bibitem[{{Claret} {et~al.}(2012){Claret}, {Hauschildt}, \&
  {Witte}}]{Claret2012}
{Claret}, A., {Hauschildt}, P.~H., \& {Witte}, S. 2012, \aap, 546, A14

\bibitem[{{Claret} {et~al.}(2013){Claret}, {Hauschildt}, \&
  {Witte}}]{Claret2013}
{Claret}, A., {Hauschildt}, P.~H., \& {Witte}, S. 2013, \aap, 552, A16

\bibitem[{{Cochran} \& {Hatzes}(1996)}]{1996CochranHatzes}
{Cochran}, W.~D. \& {Hatzes}, A.~P. 1996, \apss, 241, 43

\bibitem[{{Collins} {et~al.}(2017){Collins}, {Kielkopf}, {Stassun}, \&
  {Hessman}}]{2017AJ....153...77C}
{Collins}, K.~A., {Kielkopf}, J.~F., {Stassun}, K.~G., \& {Hessman}, F.~V.
  2017, \aj, 153, 77

\bibitem[{{Colom{\'e}} {et~al.}(2010){Colom{\'e}}, {Casteels}, {Ribas}, \&
  {Francisco}}]{2010SPIE.7740E..3KC}
{Colom{\'e}}, J., {Casteels}, K., {Ribas}, I., \& {Francisco}, X. 2010, in
  Society of Photo-Optical Instrumentation Engineers (SPIE) Conference Series,
  Vol. 7740, Software and Cyberinfrastructure for Astronomy, ed. N.~M.
  {Radziwill} \& A.~{Bridger}, 77403K

\bibitem[{{Colome} \& {Ribas}(2006)}]{2006IAUSS...6E..11C}
{Colome}, J. \& {Ribas}, I. 2006, IAU Special Session, 6, 11

\bibitem[{{Correia} {et~al.}(2020){Correia}, {Bourrier}, \&
  {Delisle}}]{Correiaetal20}
{Correia}, A.~C.~M., {Bourrier}, V., \& {Delisle}, J.~B. 2020, \aap, 635, A37

\bibitem[{{Cracchiolo} {et~al.}(2021){Cracchiolo}, {Micela}, {Morello}, \&
  {Peres}}]{Cracchiolo2021}
{Cracchiolo}, G., {Micela}, G., {Morello}, G., \& {Peres}, G. 2021, \mnras,
  507, 6118

\bibitem[{{Curtis} {et~al.}(2020){Curtis}, {Ag{\"u}eros}, {Matt}, {Covey},
  {Douglas}, {Angus}, {Saar}, {Cody}, {Vanderburg}, {Law}, {Kraus}, {Latham},
  {Baranec}, {Riddle}, {Ziegler}, {Lund}, {Torres}, {Meibom}, {Aguirre}, \&
  {Wright}}]{2020ApJ...904..140C}
{Curtis}, J.~L., {Ag{\"u}eros}, M.~A., {Matt}, S.~P., {et~al.} 2020, \apj, 904,
  140

\bibitem[{{Dahn} {et~al.}(1988){Dahn}, {Harrington}, {Kallarakal}, {Guetter},
  {Luginbuhl}, {Riepe}, {Walker}, {Pier}, {Vrba}, {Monet}, \&
  {Ables}}]{1988AJ.....95..237D}
{Dahn}, C.~C., {Harrington}, R.~S., {Kallarakal}, V.~V., {et~al.} 1988, \aj,
  95, 237

\bibitem[{{Dahn} {et~al.}(2002){Dahn}, {Harris}, {Vrba}, {Guetter}, {Canzian},
  {Henden}, {Levine}, {Luginbuhl}, {Monet}, {Monet}, {Pier}, {Stone}, {Walker},
  {Burgasser}, {Gizis}, {Kirkpatrick}, {Liebert}, \&
  {Reid}}]{2002AJ....124.1170D}
{Dahn}, C.~C., {Harris}, H.~C., {Vrba}, F.~J., {et~al.} 2002, \aj, 124, 1170

\bibitem[{{Delrez} {et~al.}(2022){Delrez}, {Murray}, {Pozuelos}, {Narita},
  {Ducrot}, {Timmermans}, {Watanabe}, {Burgasser}, {Hirano}, {Rackham},
  {Stassun}, {Van Grootel}, {Aganze}, {Cointepas}, {Howell}, {Kaltenegger},
  {Niraula}, {Sebastian}, {Almenara}, {Barkaoui}, {Baycroft}, {Bonfils},
  {Bouchy}, {Burdanov}, {Caldwell}, {Charbonneau}, {Ciardi}, {Collins},
  {Daylan}, {Demory}, {de Wit}, {Dransfield}, {Fajardo-Acosta}, {Fausnaugh},
  {Fukui}, {Furlan}, {Garcia}, {Gnilka}, {G{\'o}mez Maqueo Chew},
  {G{\'o}mez-Mu{\~n}oz}, {G{\"u}nther}, {Harakawa}, {Heng}, {Hooton}, {Hori},
  {Ikoma}, {Jehin}, {Jenkins}, {Kagetani}, {Kawauchi}, {Kimura}, {Kodama},
  {Kotani}, {Krishnamurthy}, {Kudo}, {Kunovac}, {Kusakabe}, {Latham},
  {Littlefield}, {McCormac}, {Melis}, {Mori}, {Murgas}, {Palle}, {Pedersen},
  {Queloz}, {Ricker}, {Sabin}, {Schanche}, {Schroffenegger}, {Seager}, {Shiao},
  {Sohy}, {Standing}, {Tamura}, {Theissen}, {Thompson}, {Triaud}, {Vanderspek},
  {Vievard}, {Wells}, {Winn}, {Zou}, {Z{\'u}{\~n}iga-Fern{\'a}ndez}, \&
  {Gillon}}]{delrez2022}
{Delrez}, L., {Murray}, C.~A., {Pozuelos}, F.~J., {et~al.} 2022, \aap, 667, A59

\bibitem[{{D{\'e}vora-Pajares} \& {Pozuelos}(2023)}]{matrix2023}
{D{\'e}vora-Pajares}, M. \& {Pozuelos}, F.~J. 2023, {MATRIX: Multi-phAse
  Transits Recovery from Injected eXoplanets toolkit}, Astrophysics Source Code
  Library, record ascl:2309.007

\bibitem[{Dorn \& Lichtenberg(2021)}]{Dorn2021}
Dorn, C. \& Lichtenberg, T. 2021, The Astrophysical Journal Letters, 922, L4

\bibitem[{{Dressing} \& {Charbonneau}(2013)}]{2013ApJ...767...95D}
{Dressing}, C.~D. \& {Charbonneau}, D. 2013, \apj, 767, 95

\bibitem[{{Engle} \& {Guinan}(2023)}]{2023ApJ...954L..50E}
{Engle}, S.~G. \& {Guinan}, E.~F. 2023, \apjl, 954, L50

\bibitem[{{Espinoza} {et~al.}(2019){Espinoza}, {Kossakowski}, \&
  {Brahm}}]{2019MNRAS.490.2262E}
{Espinoza}, N., {Kossakowski}, D., \& {Brahm}, R. 2019, \mnras, 490, 2262

\bibitem[{{Espinoza} {et~al.}(2022){Espinoza}, {Pall{\'e}}, {Kemmer}, {Luque},
  {Caballero}, {Cifuentes}, {Herrero}, {S{\'a}nchez B{\'e}jar}, {Stock},
  {Molaverdikhani}, {Morello}, {Kossakowski}, {Schlecker}, {Amado}, {Bluhm},
  {Cort{\'e}s-Contreras}, {Henning}, {Kreidberg}, {K{\"u}rster}, {Lafarga},
  {Lodieu}, {Morales}, {Oshagh}, {Passegger}, {Pavlov}, {Quirrenbach},
  {Reffert}, {Reiners}, {Ribas}, {Rodr{\'\i}guez}, {L{\'o}pez}, {Schweitzer},
  {Trifonov}, {Chaturvedi}, {Dreizler}, {Jeffers}, {Kaminski},
  {L{\'o}pez-Gonz{\'a}lez}, {Lillo-Box}, {Montes}, {Nowak}, {Pedraz},
  {Vanaverbeke}, {Zapatero Osorio}, {Zechmeister}, {Collins}, {Girardin},
  {Guerra}, {Naves}, {Crossfield}, {Matthews}, {Howell}, {Ciardi}, {Gonzales},
  {Matson}, {Beichman}, {Schlieder}, {Barclay}, {Vezie}, {Villase{\~n}or},
  {Daylan}, {Mireies}, {Dragomir}, {Twicken}, {Jenkins}, {Winn}, {Latham},
  {Ricker}, \& {Seager}}]{espinoza2022}
{Espinoza}, N., {Pall{\'e}}, E., {Kemmer}, J., {et~al.} 2022, \aj, 163, 133

\bibitem[{{Fletcher} {et~al.}(2018){Fletcher}, {Gustafsson}, \&
  {Orton}}]{fletcher2018}
{Fletcher}, L.~N., {Gustafsson}, M., \& {Orton}, G.~S. 2018, \apjs, 235, 24

\bibitem[{{Foreman-Mackey}(2016)}]{corner}
{Foreman-Mackey}, D. 2016, The Journal of Open Source Software, 1, 24

\bibitem[{{Foreman-Mackey} {et~al.}(2013){Foreman-Mackey}, {Hogg}, {Lang}, \&
  {Goodman}}]{emcee}
{Foreman-Mackey}, D., {Hogg}, D.~W., {Lang}, D., \& {Goodman}, J. 2013, \pasp,
  125, 306

\bibitem[{{Fortney} {et~al.}(2013){Fortney}, {Mordasini}, {Nettelmann},
  {Kempton}, {Greene}, \& {Zahnle}}]{fortney2013}
{Fortney}, J.~J., {Mordasini}, C., {Nettelmann}, N., {et~al.} 2013, \apj, 775,
  80

\bibitem[{{Frith} {et~al.}(2013){Frith}, {Pinfield}, {Jones}, {Barnes},
  {Pavlenko}, {Martin}, {Brown}, {Kuznetsov}, {Marocco}, {Tata}, \&
  {Cappetta}}]{2013MNRAS.435.2161F}
{Frith}, J., {Pinfield}, D.~J., {Jones}, H.~R.~A., {et~al.} 2013, \mnras, 435,
  2161

\bibitem[{{Fulton} {et~al.}(2017){Fulton}, {Petigura}, {Howard}, {Isaacson},
  {Marcy}, {Cargile}, {Hebb}, {Weiss}, {Johnson}, {Morton}, {Sinukoff},
  {Crossfield}, \& {Hirsch}}]{fulton2017rvalley}
{Fulton}, B.~J., {Petigura}, E.~A., {Howard}, A.~W., {et~al.} 2017, \aj, 154,
  109

\bibitem[{{Gaia Collaboration} {et~al.}(2021){Gaia Collaboration}, {Smart},
  {Sarro}, {Rybizki}, {Reyl{\'e}}, {Robin}, {Hambly}, {Abbas}, {Barstow}, {de
  Bruijne}, {Bucciarelli}, {Carrasco}, {Cooper}, {Hodgkin}, {Masana},
  {Michalik}, {Sahlmann}, {Sozzetti}, {Brown}, {Vallenari}, {Prusti},
  {Babusiaux}, {Biermann}, {Creevey}, {Evans}, {Eyer}, {Hutton}, {Jansen},
  {Jordi}, {Klioner}, {Lammers}, {Lindegren}, {Luri}, {Mignard}, {Panem},
  {Pourbaix}, {Randich}, {Sartoretti}, {Soubiran}, {Walton}, {Arenou},
  {Bailer-Jones}, {Bastian}, {Cropper}, {Drimmel}, {Katz}, {Lattanzi}, {van
  Leeuwen}, {Bakker}, {Casta{\~n}eda}, {De Angeli}, {Ducourant}, {Fabricius},
  {Fouesneau}, {Fr{\'e}mat}, {Guerra}, {Guerrier}, {Guiraud}, {Jean-Antoine
  Piccolo}, {Messineo}, {Mowlavi}, {Nicolas}, {Nienartowicz}, {Pailler},
  {Panuzzo}, {Riclet}, {Roux}, {Seabroke}, {Sordo}, {Tanga}, {Th{\'e}venin},
  {Gracia-Abril}, {Portell}, {Teyssier}, {Altmann}, {Andrae}, {Bellas-Velidis},
  {Benson}, {Berthier}, {Blomme}, {Brugaletta}, {Burgess}, {Busso}, {Carry},
  {Cellino}, {Cheek}, {Clementini}, {Damerdji}, {Davidson}, {Delchambre},
  {Dell'Oro}, {Fern{\'a}ndez-Hern{\'a}ndez}, {Galluccio}, {Garc{\'\i}a-Lario},
  {Garcia-Reinaldos}, {Gonz{\'a}lez-N{\'u}{\~n}ez}, {Gosset}, {Haigron},
  {Halbwachs}, {Harrison}, {Hatzidimitriou}, {Heiter}, {Hern{\'a}ndez},
  {Hestroffer}, {Holl}, {Jan{\ss}en}, {Jevardat de Fombelle}, {Jordan},
  {Krone-Martins}, {Lanzafame}, {L{\"o}ffler}, {Lorca}, {Manteiga}, {Marchal},
  {Marrese}, {Moitinho}, {Mora}, {Muinonen}, {Osborne}, {Pancino}, {Pauwels},
  {Recio-Blanco}, {Richards}, {Riello}, {Rimoldini}, {Roegiers}, {Siopis},
  {Smith}, {Ulla}, {Utrilla}, {van Leeuwen}, {van Reeven}, {Abreu Aramburu},
  {Accart}, {Aerts}, {Aguado}, {Ajaj}, {Altavilla}, {{\'A}lvarez}, {{\'A}lvarez
  Cid-Fuentes}, {Alves}, {Anderson}, {Anglada Varela}, {Antoja}, {Audard},
  {Baines}, {Baker}, {Balaguer-N{\'u}{\~n}ez}, {Balbinot}, {Balog}, {Barache},
  {Barbato}, {Barros}, {Bartolom{\'e}}, {Bassilana}, {Bauchet},
  {Baudesson-Stella}, {Becciani}, {Bellazzini}, {Bernet}, {Bertone}, {Bianchi},
  {Blanco-Cuaresma}, {Boch}, {Bombrun}, {Bossini}, {Bouquillon}, {Bragaglia},
  {Bramante}, {Breedt}, {Bressan}, {Brouillet}, {Burlacu}, {Busonero},
  {Butkevich}, {Buzzi}, {Caffau}, {Cancelliere}, {C{\'a}novas},
  {Cantat-Gaudin}, {Carballo}, {Carlucci}, {Carnerero}, {Casamiquela},
  {Castellani}, {Castro-Ginard}, {Castro Sampol}, {Chaoul}, {Charlot},
  {Chemin}, {Chiavassa}, {Cioni}, {Comoretto}, {Cornez}, {Cowell}, {Crifo},
  {Crosta}, {Crowley}, {Dafonte}, {Dapergolas}, {David}, {David}, {de Laverny},
  {De Luise}, {De March}, {De Ridder}, {de Souza}, {de Teodoro}, {de Torres},
  {del Peloso}, {del Pozo}, {Delgado}, {Delgado}, {Delisle}, {Di Matteo},
  {Diakite}, {Diener}, {Distefano}, {Dolding}, {Eappachen}, {Edvardsson},
  {Enke}, {Esquej}, {Fabre}, {Fabrizio}, {Faigler}, {Fedorets}, {Fernique},
  {Fienga}, {Figueras}, {Fouron}, {Fragkoudi}, {Fraile}, {Franke}, {Gai},
  {Garabato}, {Garcia-Gutierrez}, {Garc{\'\i}a-Torres}, {Garofalo}, {Gavras},
  {Gerlach}, {Geyer}, {Giacobbe}, {Gilmore}, {Girona}, {Giuffrida}, {Gomel},
  {Gomez}, {Gonzalez-Santamaria}, {Gonz{\'a}lez-Vidal}, {Granvik},
  {Guti{\'e}rrez-S{\'a}nchez}, {Guy}, {Hauser}, {Haywood}, {Helmi}, {Hidalgo},
  {Hilger}, {H{\l}adczuk}, {Hobbs}, {Holland}, {Huckle}, {Jasniewicz},
  {Jonker}, {Juaristi Campillo}, {Julbe}, {Karbevska}, {Kervella}, {Khanna},
  {Kochoska}, {Kontizas}, {Kordopatis}, {Korn}, {Kostrzewa-Rutkowska},
  {Kruszy{\'n}ska}, {Lambert}, {Lanza}, {Lasne}, {Le Campion}, {Le Fustec},
  {Lebreton}, {Lebzelter}, {Leccia}, {Leclerc}, {Lecoeur-Taibi}, {Liao},
  {Licata}, {Lindstr{\o}m}, {Lister}, {Livanou}, {Lobel}, {Madrero Pardo},
  {Managau}, {Mann}, {Marchant}, {Marconi}, {Marcos Santos}, {Marinoni},
  {Marocco}, {Marshall}, {Martin Polo}, {Mart{\'\i}n-Fleitas}, {Masip},
  {Massari}, {Mastrobuono-Battisti}, {Mazeh}, {McMillan}, {Messina}, {Millar},
  {Mints}, {Molina}, {Molinaro}, {Moln{\'a}r}, {Montegriffo}, {Mor},
  {Morbidelli}, {Morel}, {Morris}, {Mulone}, {Munoz}, {Muraveva}, {Murphy},
  {Musella}, {Noval}, {Ord{\'e}novic}, {Orr{\`u}}, {Osinde}, {Pagani},
  {Pagano}, {Palaversa}, {Palicio}, {Panahi}, {Pawlak}, {Pe{\~n}alosa
  Esteller}, {Penttil{\"a}}, {Piersimoni}, {Pineau}, {Plachy}, {Plum},
  {Poggio}, {Poretti}, {Poujoulet}, {Pr{\v{s}}a}, {Pulone}, {Racero},
  {Ragaini}, {Rainer}, {Raiteri}, {Rambaux}, {Ramos}, {Ramos-Lerate}, {Re
  Fiorentin}, {Regibo}, {Ripepi}, {Riva}, {Rixon}, {Robichon}, {Robin},
  {Roelens}, {Rohrbasser}, {Romero-G{\'o}mez}, {Rowell}, {Royer}, {Rybicki},
  {Sadowski}, {Sagrist{\`a} Sell{\'e}s}, {Salgado}, {Salguero}, {Samaras},
  {Sanchez Gimenez}, {Sanna}, {Santove{\~n}a}, {Sarasso}, {Schultheis},
  {Sciacca}, {Segol}, {Segovia}, {S{\'e}gransan}, {Semeux}, {Shahaf},
  {Siddiqui}, {Siebert}, {Siltala}, {Slezak}, {Solano}, {Solitro}, {Souami},
  {Souchay}, {Spagna}, {Spoto}, {Steele}, {Steidelm{\"u}ller}, {Stephenson},
  {S{\"u}veges}, {Szabados}, {Szegedi-Elek}, {Taris}, {Tauran}, {Taylor},
  {Teixeira}, {Thuillot}, {Tonello}, {Torra}, {Torra}, {Turon}, {Unger},
  {Vaillant}, {van Dillen}, {Vanel}, {Vecchiato}, {Viala}, {Vicente},
  {Voutsinas}, {Weiler}, {Wevers}, {Wyrzykowski}, {Yoldas}, {Yvard}, {Zhao},
  {Zorec}, {Zucker}, {Zurbach}, \& {Zwitter}}]{2021A&A...649A...6G}
{Gaia Collaboration}, {Smart}, R.~L., {Sarro}, L.~M., {et~al.} 2021, \aap, 649,
  A6

\bibitem[{{Gaia Collaboration} {et~al.}(2023){Gaia Collaboration}, {Vallenari},
  {Brown}, {Prusti}, {de Bruijne}, {Arenou}, {Babusiaux}, {Biermann},
  {Creevey}, {Ducourant}, {Evans}, {Eyer}, {Guerra}, {Hutton}, {Jordi},
  {Klioner}, {Lammers}, {Lindegren}, {Luri}, {Mignard}, {Panem}, {Pourbaix},
  {Randich}, {Sartoretti}, {Soubiran}, {Tanga}, {Walton}, {Bailer-Jones},
  {Bastian}, {Drimmel}, {Jansen}, {Katz}, {Lattanzi}, {van Leeuwen}, {Bakker},
  {Cacciari}, {Casta{\~n}eda}, {De Angeli}, {Fabricius}, {Fouesneau},
  {Fr{\'e}mat}, {Galluccio}, {Guerrier}, {Heiter}, {Masana}, {Messineo},
  {Mowlavi}, {Nicolas}, {Nienartowicz}, {Pailler}, {Panuzzo}, {Riclet}, {Roux},
  {Seabroke}, {Sordo}, {Th{\'e}venin}, {Gracia-Abril}, {Portell}, {Teyssier},
  {Altmann}, {Andrae}, {Audard}, {Bellas-Velidis}, {Benson}, {Berthier},
  {Blomme}, {Burgess}, {Busonero}, {Busso}, {C{\'a}novas}, {Carry}, {Cellino},
  {Cheek}, {Clementini}, {Damerdji}, {Davidson}, {de Teodoro}, {Nu{\~n}ez
  Campos}, {Delchambre}, {Dell'Oro}, {Esquej}, {Fern{\'a}ndez-Hern{\'a}ndez},
  {Fraile}, {Garabato}, {Garc{\'\i}a-Lario}, {Gosset}, {Haigron}, {Halbwachs},
  {Hambly}, {Harrison}, {Hern{\'a}ndez}, {Hestroffer}, {Hodgkin}, {Holl},
  {Jan{\ss}en}, {Jevardat de Fombelle}, {Jordan}, {Krone-Martins}, {Lanzafame},
  {L{\"o}ffler}, {Marchal}, {Marrese}, {Moitinho}, {Muinonen}, {Osborne},
  {Pancino}, {Pauwels}, {Recio-Blanco}, {Reyl{\'e}}, {Riello}, {Rimoldini},
  {Roegiers}, {Rybizki}, {Sarro}, {Siopis}, {Smith}, {Sozzetti}, {Utrilla},
  {van Leeuwen}, {Abbas}, {{\'A}brah{\'a}m}, {Abreu Aramburu}, {Aerts},
  {Aguado}, {Ajaj}, {Aldea-Montero}, {Altavilla}, {{\'A}lvarez}, {Alves},
  {Anders}, {Anderson}, {Anglada Varela}, {Antoja}, {Baines}, {Baker},
  {Balaguer-N{\'u}{\~n}ez}, {Balbinot}, {Balog}, {Barache}, {Barbato},
  {Barros}, {Barstow}, {Bartolom{\'e}}, {Bassilana}, {Bauchet}, {Becciani},
  {Bellazzini}, {Berihuete}, {Bernet}, {Bertone}, {Bianchi}, {Binnenfeld},
  {Blanco-Cuaresma}, {Blazere}, {Boch}, {Bombrun}, {Bossini}, {Bouquillon},
  {Bragaglia}, {Bramante}, {Breedt}, {Bressan}, {Brouillet}, {Brugaletta},
  {Bucciarelli}, {Burlacu}, {Butkevich}, {Buzzi}, {Caffau}, {Cancelliere},
  {Cantat-Gaudin}, {Carballo}, {Carlucci}, {Carnerero}, {Carrasco},
  {Casamiquela}, {Castellani}, {Castro-Ginard}, {Chaoul}, {Charlot}, {Chemin},
  {Chiaramida}, {Chiavassa}, {Chornay}, {Comoretto}, {Contursi}, {Cooper},
  {Cornez}, {Cowell}, {Crifo}, {Cropper}, {Crosta}, {Crowley}, {Dafonte},
  {Dapergolas}, {David}, {David}, {de Laverny}, {De Luise}, {De March}, {De
  Ridder}, {de Souza}, {de Torres}, {del Peloso}, {del Pozo}, {Delbo},
  {Delgado}, {Delisle}, {Demouchy}, {Dharmawardena}, {Di Matteo}, {Diakite},
  {Diener}, {Distefano}, {Dolding}, {Edvardsson}, {Enke}, {Fabre}, {Fabrizio},
  {Faigler}, {Fedorets}, {Fernique}, {Fienga}, {Figueras}, {Fournier},
  {Fouron}, {Fragkoudi}, {Gai}, {Garcia-Gutierrez}, {Garcia-Reinaldos},
  {Garc{\'\i}a-Torres}, {Garofalo}, {Gavel}, {Gavras}, {Gerlach}, {Geyer},
  {Giacobbe}, {Gilmore}, {Girona}, {Giuffrida}, {Gomel}, {Gomez},
  {Gonz{\'a}lez-N{\'u}{\~n}ez}, {Gonz{\'a}lez-Santamar{\'\i}a},
  {Gonz{\'a}lez-Vidal}, {Granvik}, {Guillout}, {Guiraud},
  {Guti{\'e}rrez-S{\'a}nchez}, {Guy}, {Hatzidimitriou}, {Hauser}, {Haywood},
  {Helmer}, {Helmi}, {Sarmiento}, {Hidalgo}, {Hilger}, {H{\l}adczuk}, {Hobbs},
  {Holland}, {Huckle}, {Jardine}, {Jasniewicz}, {Jean-Antoine Piccolo},
  {Jim{\'e}nez-Arranz}, {Jorissen}, {Juaristi Campillo}, {Julbe}, {Karbevska},
  {Kervella}, {Khanna}, {Kontizas}, {Kordopatis}, {Korn}, {K{\'o}sp{\'a}l},
  {Kostrzewa-Rutkowska}, {Kruszy{\'n}ska}, {Kun}, {Laizeau}, {Lambert},
  {Lanza}, {Lasne}, {Le Campion}, {Lebreton}, {Lebzelter}, {Leccia}, {Leclerc},
  {Lecoeur-Taibi}, {Liao}, {Licata}, {Lindstr{\o}m}, {Lister}, {Livanou},
  {Lobel}, {Lorca}, {Loup}, {Madrero Pardo}, {Magdaleno Romeo}, {Managau},
  {Mann}, {Manteiga}, {Marchant}, {Marconi}, {Marcos}, {Marcos Santos},
  {Mar{\'\i}n Pina}, {Marinoni}, {Marocco}, {Marshall}, {Martin Polo},
  {Mart{\'\i}n-Fleitas}, {Marton}, {Mary}, {Masip}, {Massari},
  {Mastrobuono-Battisti}, {Mazeh}, {McMillan}, {Messina}, {Michalik}, {Millar},
  {Mints}, {Molina}, {Molinaro}, {Moln{\'a}r}, {Monari}, {Mongui{\'o}},
  {Montegriffo}, {Montero}, {Mor}, {Mora}, {Morbidelli}, {Morel}, {Morris},
  {Muraveva}, {Murphy}, {Musella}, {Nagy}, {Noval}, {Oca{\~n}a}, {Ogden},
  {Ordenovic}, {Osinde}, {Pagani}, {Pagano}, {Palaversa}, {Palicio},
  {Pallas-Quintela}, {Panahi}, {Payne-Wardenaar}, {Pe{\~n}alosa Esteller},
  {Penttil{\"a}}, {Pichon}, {Piersimoni}, {Pineau}, {Plachy}, {Plum}, {Poggio},
  {Pr{\v{s}}a}, {Pulone}, {Racero}, {Ragaini}, {Rainer}, {Raiteri}, {Rambaux},
  {Ramos}, {Ramos-Lerate}, {Re Fiorentin}, {Regibo}, {Richards}, {Rios Diaz},
  {Ripepi}, {Riva}, {Rix}, {Rixon}, {Robichon}, {Robin}, {Robin}, {Roelens},
  {Rogues}, {Rohrbasser}, {Romero-G{\'o}mez}, {Rowell}, {Royer}, {Ruz Mieres},
  {Rybicki}, {Sadowski}, {S{\'a}ez N{\'u}{\~n}ez}, {Sagrist{\`a} Sell{\'e}s},
  {Sahlmann}, {Salguero}, {Samaras}, {Sanchez Gimenez}, {Sanna},
  {Santove{\~n}a}, {Sarasso}, {Schultheis}, {Sciacca}, {Segol}, {Segovia},
  {S{\'e}gransan}, {Semeux}, {Shahaf}, {Siddiqui}, {Siebert}, {Siltala},
  {Silvelo}, {Slezak}, {Slezak}, {Smart}, {Snaith}, {Solano}, {Solitro},
  {Souami}, {Souchay}, {Spagna}, {Spina}, {Spoto}, {Steele},
  {Steidelm{\"u}ller}, {Stephenson}, {S{\"u}veges}, {Surdej}, {Szabados},
  {Szegedi-Elek}, {Taris}, {Taylor}, {Teixeira}, {Tolomei}, {Tonello}, {Torra},
  {Torra}, {Torralba Elipe}, {Trabucchi}, {Tsounis}, {Turon}, {Ulla}, {Unger},
  {Vaillant}, {van Dillen}, {van Reeven}, {Vanel}, {Vecchiato}, {Viala},
  {Vicente}, {Voutsinas}, {Weiler}, {Wevers}, {Wyrzykowski}, {Yoldas}, {Yvard},
  {Zhao}, {Zorec}, {Zucker}, \& {Zwitter}}]{gaiadr3}
{Gaia Collaboration}, {Vallenari}, A., {Brown}, A.~G.~A., {et~al.} 2023, \aap,
  674, A1

\bibitem[{{Gao} \& {Zhang}(2020)}]{gao2020}
{Gao}, P. \& {Zhang}, X. 2020, \apj, 890, 93

\bibitem[{Giclas(1966)}]{GICLAS196623}
Giclas, H. 1966, Vistas in Astronomy, 8, 23

\bibitem[{{Giclas} {et~al.}(1971){Giclas}, {Burnham}, \&
  {Thomas}}]{1971lpms.book.....G}
{Giclas}, H.~L., {Burnham}, R., \& {Thomas}, N.~G. 1971, {Lowell proper motion
  survey Northern Hemisphere. The G numbered stars. 8991 stars fainter than
  magnitude 8 with motions > 0''.26/year}

\bibitem[{{Gonz{\'a}lez-{\'A}lvarez} {et~al.}(2023){Gonz{\'a}lez-{\'A}lvarez},
  {Zapatero Osorio}, {Caballero}, {B{\'e}jar}, {Cifuentes}, {Fukui}, {Herrero},
  {Kawauchi}, {Livingston}, {L{\'o}pez-Gonz{\'a}lez}, {Morello}, {Murgas},
  {Narita}, {Pall{\'e}}, {Passegger}, {Rodr{\'\i}guez},
  {Rodr{\'\i}guez-L{\'o}pez}, {Sanz-Forcada}, {Schweitzer}, {Tabernero},
  {Quirrenbach}, {Amado}, {Charbonneau}, {Ciardi}, {Cikota}, {Collins},
  {Conti}, {Fausnaugh}, {Hatzes}, {Hedges}, {Henning}, {Jenkins}, {Latham},
  {Massey}, {Moldovan}, {Montes}, {Panahi}, {Reiners}, {Ribas}, {Ricker},
  {Seager}, {Shporer}, {Srdoc}, {Tenenbaum}, {Vanderspek}, {Winn}, {Fukuda},
  {Ikoma}, {Isogai}, {Kawai}, {Mori}, {Tamura}, \&
  {Watanabe}}]{2023A&A...675A.177G}
{Gonz{\'a}lez-{\'A}lvarez}, E., {Zapatero Osorio}, M.~R., {Caballero}, J.~A.,
  {et~al.} 2023, \aap, 675, A177

\bibitem[{{Guillot} {et~al.}(2004){Guillot}, {Stevenson}, {Hubbard}, \&
  {Saumon}}]{Guillotetal04}
{Guillot}, T., {Stevenson}, D.~J., {Hubbard}, W.~B., \& {Saumon}, D. 2004, in
  Jupiter. The Planet, Satellites and Magnetosphere, ed. F.~{Bagenal}, T.~E.
  {Dowling}, \& W.~B. {McKinnon}, Vol.~1, 35--57

\bibitem[{{Hartman} \& {Bakos}(2016)}]{2016A&C....17....1H}
{Hartman}, J.~D. \& {Bakos}, G.~{\'A}. 2016, Astronomy and Computing, 17, 1

\bibitem[{{Hatzes}(2019)}]{2019dmde.book.....H}
{Hatzes}, A.~P. 2019, {The Doppler Method for the Detection of Exoplanets}

\bibitem[{Hatzes {et~al.}(2010)Hatzes, Dvorak, Wuchterl, Guterman, Hartmann,
  Fridlund, Gandolfi, Guenther, \& Pätzold}]{2010Hatzes}
Hatzes, A.~P., Dvorak, R., Wuchterl, G., {et~al.} 2010, Astronomy and
  Astrophysics, 520, A93

\bibitem[{{Henning} {et~al.}(2009){Henning}, {O'Connell}, \&
  {Sasselov}}]{2009ApJ...707.1000H}
{Henning}, W.~G., {O'Connell}, R.~J., \& {Sasselov}, D.~D. 2009, \apj, 707,
  1000

\bibitem[{{Howell} {et~al.}(2014){Howell}, {Sobeck}, {Haas}, {Still},
  {Barclay}, {Mullally}, {Troeltzsch}, {Aigrain}, {Bryson}, {Caldwell},
  {Chaplin}, {Cochran}, {Huber}, {Marcy}, {Miglio}, {Najita}, {Smith},
  {Twicken}, \& {Fortney}}]{Howell2014}
{Howell}, S.~B., {Sobeck}, C., {Haas}, M., {et~al.} 2014, \pasp, 126, 398

\bibitem[{{Husser} {et~al.}(2013){Husser}, {Wende-von Berg}, {Dreizler},
  {Homeier}, {Reiners}, {Barman}, \& {Hauschildt}}]{Husser2013}
{Husser}, T.~O., {Wende-von Berg}, S., {Dreizler}, S., {et~al.} 2013, \aap,
  553, A6

\bibitem[{{Jeffers} {et~al.}(2020){Jeffers}, {Dreizler}, {Barnes}, {Haswell},
  {Nelson}, {Rodr{\'\i}guez}, {L{\'o}pez-Gonz‧lez}, {Morales}, {Luque},
  {Zechmeister}, {Vogt}, {Jenkins}, {Palle}, {Berdi {\~n}as}, {Coleman},
  {D{\'\i}az}, {Ribas}, {Jones}, {Butler}, {Tinney}, {Bailey}, {Carter},
  {O'Toole}, {Wittenmyer}, {Crane}, {Feng}, {Shectman}, {Teske}, {Reiners},
  {Amado}, \& {Anglada-Escud{\'e}}}]{2020Sci...368.1477J}
{Jeffers}, S.~V., {Dreizler}, S., {Barnes}, J.~R., {et~al.} 2020, Science, 368,
  1477

\bibitem[{{Jeffers} {et~al.}(2018){Jeffers}, {Sch{\"o}fer}, {Lamert},
  {Reiners}, {Montes}, {Caballero}, {Cort{\'e}s-Contreras}, {Marvin},
  {Passegger}, {Zechmeister}, {Quirrenbach}, {Alonso-Floriano}, {Amado},
  {Bauer}, {Casal}, {Diez Alonso}, {Herrero}, {Morales}, {Mundt}, {Ribas}, \&
  {Sarmiento}}]{2018A&A...614A..76J}
{Jeffers}, S.~V., {Sch{\"o}fer}, P., {Lamert}, A., {et~al.} 2018, \aap, 614,
  A76

\bibitem[{{Jenkins} {et~al.}(2016){Jenkins}, {Twicken}, {McCauliff},
  {Campbell}, {Sanderfer}, {Lung}, {Mansouri-Samani}, {Girouard}, {Tenenbaum},
  {Klaus}, {Smith}, {Caldwell}, {Chacon}, {Henze}, {Heiges}, {Latham},
  {Morgan}, {Swade}, {Rinehart}, \& {Vanderspek}}]{SPOC}
{Jenkins}, J.~M., {Twicken}, J.~D., {McCauliff}, S., {et~al.} 2016, in
  \procspie, Vol. 9913, Software and Cyberinfrastructure for Astronomy IV,
  99133E

\bibitem[{Kasting(1988)}]{Kasting1988}
Kasting, J.~F. 1988, Icarus, 74, 472

\bibitem[{{Kawauchi} {et~al.}(2022){Kawauchi}, {Murgas}, {Palle}, {Narita},
  {Fukui}, {Hirano}, {Parviainen}, {Ishikawa}, {Watanabe}, {Esparaza-Borges},
  {Kuzuhara}, {Orell-Miquel}, {Krishnamurthy}, {Mori}, {Kagetani}, {Zou},
  {Isogai}, {Livingston}, {Howell}, {Crouzet}, {de Leon}, {Kimura}, {Kodama},
  {Korth}, {Kurita}, {Laza-Ramos}, {Luque}, {Madrigal-Aguado}, {Miyakawa},
  {Morello}, {Nishiumi}, {Rodr{\'\i}guez}, {S{\'a}nchez-Benavente}, {Stangret},
  {Teng}, {Terada}, {Gnilka}, {Guerrero}, {Harakawa}, {Hodapp}, {Hori},
  {Ikoma}, {Jacobson}, {Konishi}, {Kotani}, {Kudo}, {Kurokowa}, {Kusakabe},
  {Nishikawa}, {Omiya}, {Serizawa}, {Tamura}, {Ueda}, \&
  {Vievard}}]{2022A&A...666A...4K}
{Kawauchi}, K., {Murgas}, F., {Palle}, E., {et~al.} 2022, \aap, 666, A4

\bibitem[{{Kempton} {et~al.}(2018){Kempton}, {Bean}, {Louie}, {Deming}, {Koll},
  {Mansfield}, {Christiansen}, {L{\'o}pez-Morales}, {Swain}, {Zellem},
  {Ballard}, {Barclay}, {Barstow}, {Batalha}, {Beatty}, {Berta-Thompson},
  {Birkby}, {Buchhave}, {Charbonneau}, {Cowan}, {Crossfield}, {de Val-Borro},
  {Doyon}, {Dragomir}, {Gaidos}, {Heng}, {Hu}, {Kane}, {Kreidberg}, {Mallonn},
  {Morley}, {Narita}, {Nascimbeni}, {Pall{\'e}}, {Quintana}, {Rauscher},
  {Seager}, {Shkolnik}, {Sing}, {Sozzetti}, {Stassun}, {Valenti}, \& {von
  Essen}}]{Kempton2018PASP..130k4401K}
{Kempton}, E. M.~R., {Bean}, J.~L., {Louie}, D.~R., {et~al.} 2018, \pasp, 130,
  114401

\bibitem[{{Kimura} \& {Ikoma}(2020)}]{Kimura_Ikoma20}
{Kimura}, T. \& {Ikoma}, M. 2020, \mnras, 496, 3755

\bibitem[{{Kipping}(2013)}]{kipping02013}
{Kipping}, D.~M. 2013, \mnras, 435, 2152

\bibitem[{{Kochanek} {et~al.}(2017){Kochanek}, {Shappee}, {Stanek}, {Holoien},
  {Thompson}, {Prieto}, {Dong}, {Shields}, {Will}, {Britt}, {Perzanowski}, \&
  {Pojma{\'n}ski}}]{Kochanek2017}
{Kochanek}, C.~S., {Shappee}, B.~J., {Stanek}, K.~Z., {et~al.} 2017, \pasp,
  129, 104502

\bibitem[{Kopparapu {et~al.}(2014)Kopparapu, Ramirez, SchottelKotte, Kasting,
  {Domagal-Goldman}, \& Eymet}]{Kopparapu2014}
Kopparapu, R.~K., Ramirez, R.~M., SchottelKotte, J., {et~al.} 2014, The
  Astrophysical Journal Letters, 787, L29

\bibitem[{{Kossakowski} {et~al.}(2023){Kossakowski}, {K{\"u}rster}, {Trifonov},
  {Henning}, {Kemmer}, {Caballero}, {Burn}, {Sabotta}, {Crouse}, {Fauchez},
  {Nagel}, {Kaminski}, {Herrero}, {Rodr{\'\i}guez}, {Gonz{\'a}lez-{\'A}lvarez},
  {Quirrenbach}, {Amado}, {Ribas}, {Reiners}, {Aceituno}, {B{\'e}jar},
  {Baroch}, {Bastelberger}, {Chaturvedi}, {Cifuentes}, {Dreizler}, {Jeffers},
  {Kopparapu}, {Lafarga}, {L{\'o}pez-Gonz{\'a}lez}, {Mart{\'\i}n-Ruiz},
  {Montes}, {Morales}, {Pall{\'e}}, {Pavlov}, {Pedraz}, {Perdelwitz},
  {P{\'e}rez-Torres}, {Perger}, {Reffert}, {Rodr{\'\i}guez L{\'o}pez},
  {Schlecker}, {Sch{\"o}fer}, {Schweitzer}, {Shan}, {Shields}, {Stock}, {Wolf},
  {Zapatero Osorio}, \& {Zechmeister}}]{2023A&A...670A..84K}
{Kossakowski}, D., {K{\"u}rster}, M., {Trifonov}, T., {et~al.} 2023, \aap, 670,
  A84

\bibitem[{{Kov{\'a}cs} {et~al.}(2002){Kov{\'a}cs}, {Zucker}, \&
  {Mazeh}}]{2002A&A...391..369K}
{Kov{\'a}cs}, G., {Zucker}, S., \& {Mazeh}, T. 2002, \aap, 391, 369

\bibitem[{{Lamman} {et~al.}(2020){Lamman}, {Baranec}, {Berta-Thompson}, {Law},
  {Schonhut-Stasik}, {Ziegler}, {Salama}, {Jensen-Clem}, {Duev}, {Riddle},
  {Kulkarni}, {Winters}, \& {Irwin}}]{2020AJ....159..139L}
{Lamman}, C., {Baranec}, C., {Berta-Thompson}, Z.~K., {et~al.} 2020, \aj, 159,
  139

\bibitem[{{Lanza}(2009)}]{Lanza2009}
{Lanza}, A.~F. 2009, \aap, 505, 339

\bibitem[{{Leconte} {et~al.}(2010){Leconte}, {Chabrier}, {Baraffe}, \&
  {Levrard}}]{Leconteetal10}
{Leconte}, J., {Chabrier}, G., {Baraffe}, I., \& {Levrard}, B. 2010, \aap, 516,
  A64

\bibitem[{{Lee} {et~al.}(2013){Lee}, {Heng}, \& {Irwin}}]{lee2013}
{Lee}, J.-M., {Heng}, K., \& {Irwin}, P. G.~J. 2013, \apj, 778, 97

\bibitem[{{L{\'e}pine} \& {Gaidos}(2011)}]{2011AJ....142..138L}
{L{\'e}pine}, S. \& {Gaidos}, E. 2011, \aj, 142, 138

\bibitem[{{L{\'e}pine} \& {Shara}(2005)}]{2005AJ....129.1483L}
{L{\'e}pine}, S. \& {Shara}, M.~M. 2005, \aj, 129, 1483

\bibitem[{{Lillo-Box} {et~al.}(2023){Lillo-Box}, {Gandolfi}, {Armstrong},
  {Collins}, {Nielsen}, {Luque}, {Korth}, {Sousa}, {Quinn}, {Acu{\~n}a},
  {Howell}, {Morello}, {Hellier}, {Giacalone}, {Hoyer}, {Stassun}, {Palle},
  {Aguichine}, {Mousis}, {Adibekyan}, {Azevedo Silva}, {Barrado}, {Deleuil},
  {Eastman}, {Fukui}, {Hawthorn}, {Irwin}, {Jenkins}, {Latham}, {Muresan},
  {Narita}, {Persson}, {Santerne}, {Santos}, {Savel}, {Osborn}, {Teske},
  {Wheatley}, {Winn}, {Barros}, {Butler}, {Caldwell}, {Charbonneau},
  {Cloutier}, {Crane}, {Demangeon}, {D{\'\i}az}, {Dumusque}, {Esposito},
  {Falk}, {Gill}, {Hojjatpanah}, {Kreidberg}, {Mireles}, {Osborn}, {Ricker},
  {Rodriguez}, {Schwarz}, {Seager}, {Serrano Bell}, {Shectman}, {Shporer},
  {Vezie}, {Wang}, \& {Zhou}}]{lillo-box2023}
{Lillo-Box}, J., {Gandolfi}, D., {Armstrong}, D.~J., {et~al.} 2023, \aap, 669,
  A109

\bibitem[{{Lodieu} {et~al.}(2019){Lodieu}, {P{\'e}rez-Garrido}, {Smart}, \&
  {Silvotti}}]{2019A&A...628A..66L}
{Lodieu}, N., {P{\'e}rez-Garrido}, A., {Smart}, R.~L., \& {Silvotti}, R. 2019,
  \aap, 628, A66

\bibitem[{Luger \& Barnes(2015)}]{Luger2015}
Luger, R. \& Barnes, R. 2015, Astrobiology, 15, 119

\bibitem[{{Luque} {et~al.}(2022{\natexlab{a}}){Luque}, {Fulton}, {Kunimoto},
  {Amado}, {Gorrini}, {Dreizler}, {Hellier}, {Henry}, {Molaverdikhani},
  {Morello}, {Pe{\~n}a-Mo{\~n}ino}, {P{\'e}rez-Torres}, {Pozuelos}, {Shan},
  {Anglada-Escud{\'e}}, {B{\'e}jar}, {Bergond}, {Boyle}, {Caballero},
  {Charbonneau}, {Ciardi}, {Dufoer}, {Espinoza}, {Everett}, {Fischer},
  {Hatzes}, {Henning}, {Hesse}, {Howard}, {Howell}, {Isaacson}, {Jeffers},
  {Jenkins}, {Kane}, {Kemmer}, {Khalafinejad}, {Kidwell}, {Kossakowski},
  {Latham}, {Lillo-Box}, {Lissauer}, {Montes}, {Orell-Miquel}, {Pall{\'e}},
  {Pollacco}, {Quirrenbach}, {Reffert}, {Reiners}, {Ribas}, {Ricker}, {Rogers},
  {Sanz-Forcada}, {Schlecker}, {Schweitzer}, {Seager}, {Shporer}, {Stassun},
  {Stock}, {Tal-Or}, {Ting}, {Trifonov}, {Vanaverbeke}, {Vanderspek},
  {Villase{\~n}or}, {Winn}, {Winters}, \& {Zapatero Osorio}}]{luque2022a}
{Luque}, R., {Fulton}, B.~J., {Kunimoto}, M., {et~al.} 2022{\natexlab{a}},
  \aap, 664, A199

\bibitem[{{Luque} {et~al.}(2022{\natexlab{b}}){Luque}, {Nowak}, {Hirano},
  {Kossakowski}, {Pall{\'e}}, {Nixon}, {Morello}, {Amado}, {Albrecht},
  {Caballero}, {Cifuentes}, {Cochran}, {Deeg}, {Dreizler}, {Esparza-Borges},
  {Fukui}, {Gandolfi}, {Goffo}, {Guenther}, {Hatzes}, {Henning}, {Kabath},
  {Kawauchi}, {Korth}, {Kotani}, {Kudo}, {Kuzuhara}, {Lafarga}, {Lam},
  {Livingston}, {Morales}, {Muresan}, {Murgas}, {Narita}, {Osborne},
  {Parviainen}, {Passegger}, {Persson}, {Quirrenbach}, {Redfield}, {Reffert},
  {Reiners}, {Ribas}, {Serrano}, {Tamura}, {Van Eylen}, {Watanabe}, \&
  {Zapatero Osorio}}]{luque2022b}
{Luque}, R., {Nowak}, G., {Hirano}, T., {et~al.} 2022{\natexlab{b}}, \aap, 666,
  A154

\bibitem[{{Luyten}(1979)}]{1979lccs.book.....L}
{Luyten}, W.~J. 1979, {LHS catalogue. A catalogue of stars with proper motions
  exceeding 0''5 annually}

\bibitem[{{Mallorqu{\'\i}n} {et~al.}(2023){Mallorqu{\'\i}n}, {Goffo},
  {Pall{\'e}}, {Lodieu}, {B{\'e}jar}, {Isaacson}, {Zapatero Osorio},
  {Dreizler}, {Stock}, {Luque}, {Murgas}, {Pe{\~n}a}, {Sanz-Forcada},
  {Morello}, {Ciardi}, {Furlan}, {Collins}, {Herrero}, {Vanaverbeke},
  {Plavchan}, {Narita}, {Schweitzer}, {P{\'e}rez-Torres}, {Quirrenbach},
  {Kemmer}, {Hatzes}, {Howard}, {Schlecker}, {Reffert}, {Nagel}, {Morales},
  {Orell-Miquel}, {Duque-Arribas}, {Carleo}, {Cifuentes}, {Nowak}, {Ribas},
  {Reiners}, {Amado}, {Caballero}, {Henning}, {Pinter}, {Murphy}, {Beard},
  {Blunt}, {Brinkman}, {Cale}, {Chontos}, {Collins}, {Crossfield}, {Dai},
  {Dalba}, {Dufoer}, {El Mufti}, {Espinoza}, {Fetherolf}, {Fukui}, {Giacalone},
  {Gnilka}, {Gonzales}, {Grunblatt}, {Howell}, {Huber}, {Kane}, {de Le{\'o}n},
  {Lubin}, {MacDougall}, {Massey}, {Montes}, {Mori}, {Parviainen}, {Passegger},
  {Polanski}, {Robertson}, {Schwarz}, {Srdoc}, {Tabernero}, {Tanner},
  {Turtelboom}, {Van Zandt}, {Weiss}, \& {Zechmeister}}]{2023arXiv231010244M}
{Mallorqu{\'\i}n}, M., {Goffo}, E., {Pall{\'e}}, E., {et~al.} 2023, \aap, 680,
  A76

\bibitem[{Mandel \& Agol(2002)}]{2002Mandel&Agol}
Mandel, K. \& Agol, E. 2002, ApJL, 580, L171–L175

\bibitem[{{Marcq} {et~al.}(2017){Marcq}, {Salvador}, {Massol}, \&
  {Davaille}}]{Marcq17}
{Marcq}, E., {Salvador}, A., {Massol}, H., \& {Davaille}, A. 2017, Journal of
  Geophysical Research (Planets), 122, 1539

\bibitem[{{Marfil} {et~al.}(2021){Marfil}, {Tabernero}, {Montes}, {Caballero},
  {L{\'a}zaro}, {Gonz{\'a}lez Hern{\'a}ndez}, {Nagel}, {Passegger},
  {Schweitzer}, {Ribas}, {Reiners}, {Quirrenbach}, {Amado}, {Cifuentes},
  {Cort{\'e}s-Contreras}, {Dreizler}, {Duque-Arribas},
  {Galad{\'\i}-Enr{\'\i}quez}, {Henning}, {Jeffers}, {Kaminski}, {K{\"u}rster},
  {Lafarga}, {L{\'o}pez-Gallifa}, {Morales}, {Shan}, \&
  {Zechmeister}}]{marfil21}
{Marfil}, E., {Tabernero}, H.~M., {Montes}, D., {et~al.} 2021, \aap, 656, A162

\bibitem[{{Maxted} {et~al.}(2011){Maxted}, {Anderson}, {Collier Cameron},
  {Hellier}, {Queloz}, {Smalley}, {Street}, {Triaud}, {West}, {Gillon},
  {Lister}, {Pepe}, {Pollacco}, {S{\'e}gransan}, {Smith}, \&
  {Udry}}]{2011PASP..123..547M}
{Maxted}, P.~F.~L., {Anderson}, D.~R., {Collier Cameron}, A., {et~al.} 2011,
  \pasp, 123, 547

\bibitem[{{McCully} {et~al.}(2018){McCully}, {Volgenau}, {Harbeck}, {Lister},
  {Saunders}, {Turner}, {Siiverd}, \& {Bowman}}]{McCully2018}
{McCully}, C., {Volgenau}, N.~H., {Harbeck}, D.-R., {et~al.} 2018, in Society
  of Photo-Optical Instrumentation Engineers (SPIE) Conference Series, Vol.
  10707, \procspie, 107070K

\bibitem[{{McKay} {et~al.}(2019){McKay}, {DiSanti}, {Kelley}, {Knight},
  {Womack}, {Wierzchos}, {Harrington Pinto}, {Bonev}, {Villanueva}, {Dello
  Russo}, {Cochran}, {Biver}, {Bauer}, {Vervack}, {Gibb}, {Roth}, \&
  {Kawakita}}]{McKay19}
{McKay}, A.~J., {DiSanti}, M.~A., {Kelley}, M. S.~P., {et~al.} 2019, \aj, 158,
  128

\bibitem[{{Melrose} \& {Dulk}(1982)}]{Melrose1982}
{Melrose}, D.~B. \& {Dulk}, G.~A. 1982, \apj, 259, 844

\bibitem[{{Miguel} {et~al.}(2020){Miguel}, {Cridland}, {Ormel}, {Fortney}, \&
  {Ida}}]{Miguel20}
{Miguel}, Y., {Cridland}, A., {Ormel}, C.~W., {Fortney}, J.~J., \& {Ida}, S.
  2020, \mnras, 491, 1998

\bibitem[{{Mishchenko} {et~al.}(1996){Mishchenko}, {Travis}, \&
  {Mackowski}}]{1996JQSRT..55..535M}
{Mishchenko}, M.~I., {Travis}, L.~D., \& {Mackowski}, D.~W. 1996, \jqsrt, 55,
  535

\bibitem[{Mordasini(2018)}]{Mordasini2018a}
Mordasini, C. 2018, in Handbook of {{Exoplanets}} ({Cham}: {Springer
  International Publishing}), 2425--2474

\bibitem[{{Morello} {et~al.}(2023){Morello}, {Parviainen}, {Murgas},
  {Pall{\'e}}, {Oshagh}, {Fukui}, {Hirano}, {Ishikawa}, {Mori}, {Narita},
  {Collins}, {Barkaoui}, {Lewin}, {Cadieux}, {de Leon}, {Soubkiou}, {Abreu
  Garcia}, {Crouzet}, {Esparza-Borges}, {Fern{\'a}ndez Rodr{\'\i}guez},
  {Gal{\'a}n}, {Hori}, {Ikoma}, {Isogai}, {Kagetani}, {Kawauchi}, {Kimura},
  {Kodama}, {Korth}, {Kotani}, {Krishnamurthy}, {Kurita}, {Laza-Ramos},
  {Livingston}, {Luque}, {Madrigal-Aguado}, {Nishiumi}, {Orell-Miquel},
  {Puig-Subir{\`a}}, {S{\'a}nchez-Benavente}, {Stangret}, {Tamura}, {Terada},
  {Watanabe}, {Zou}, {Benkhaldoun}, {Collins}, {Doyon}, {Garcia}, {Ghachoui},
  {Gillon}, {Jehin}, {Pozuelos}, {Schwarz}, \& {Timmermans}}]{Morello2023}
{Morello}, G., {Parviainen}, H., {Murgas}, F., {et~al.} 2023, \aap, 673, A32

\bibitem[{{Morello} {et~al.}(2017){Morello}, {Tsiaras}, {Howarth}, \&
  {Homeier}}]{Morello2017}
{Morello}, G., {Tsiaras}, A., {Howarth}, I.~D., \& {Homeier}, D. 2017, \aj,
  154, 111

\bibitem[{{Morello} {et~al.}(2021){Morello}, {Zingales}, {Martin-Lagarde},
  {Gastaud}, \& {Lagage}}]{Morello2021}
{Morello}, G., {Zingales}, T., {Martin-Lagarde}, M., {Gastaud}, R., \&
  {Lagage}, P.-O. 2021, \aj, 161, 174

\bibitem[{{Murdoch} {et~al.}(1993){Murdoch}, {Hearnshaw}, \&
  {Clark}}]{murdoch_1993}
{Murdoch}, K.~A., {Hearnshaw}, J.~B., \& {Clark}, M. 1993, The Astrophysical
  Journal, 413, 349

\bibitem[{{Murgas} {et~al.}(2021){Murgas}, {Astudillo-Defru}, {Bonfils},
  {Crossfield}, {Almenara}, {Livingston}, {Stassun}, {Korth}, {Orell-Miquel},
  {Morello}, {Eastman}, {Lissauer}, {Kane}, {Morales}, {Werner}, {Gorjian},
  {Benneke}, {Dragomir}, {Matthews}, {Howell}, {Ciardi}, {Gonzales}, {Matson},
  {Beichman}, {Schlieder}, {Collins}, {Collins}, {Jensen}, {Evans}, {Pozuelos},
  {Gillon}, {Jehin}, {Barkaoui}, {Artigau}, {Bouchy}, {Charbonneau},
  {Delfosse}, {D{\'\i}az}, {Doyon}, {Figueira}, {Forveille}, {Lovis}, {Melo},
  {Gaisn{\'e}}, {Pepe}, {Santos}, {S{\'e}gransan}, {Udry}, {Goeke}, {Levine},
  {Quintana}, {Guerrero}, {Mireles}, {Caldwell}, {Tenenbaum}, {Brasseur},
  {Ricker}, {Vanderspek}, {Latham}, {Seager}, {Winn}, \&
  {Jenkins}}]{murgas2021}
{Murgas}, F., {Astudillo-Defru}, N., {Bonfils}, X., {et~al.} 2021, \aap, 653,
  A60

\bibitem[{{Murgas} {et~al.}(2023){Murgas}, {Castro-Gonz{\'a}lez}, {Pall{\'e}},
  {Pozuelos}, {Millholland}, {Foo}, {Korth}, {Marfil}, {Amado}, {Caballero},
  {Christiansen}, {Ciardi}, {Collins}, {Di Sora}, {Fukui}, {Gan}, {Gonzales},
  {Henning}, {Herrero}, {Isopi}, {Jenkins}, {Lillo-Box}, {Lodieu}, {Luque},
  {Mallia}, {Morales}, {Morello}, {Narita}, {Orell-Miquel}, {Parviainen},
  {P{\'e}rez-Torres}, {Quirrenbach}, {Reiners}, {Ribas}, {Safonov}, {Seager},
  {Schwarz}, {Schweitzer}, {Schlecker}, {Strakhov}, {Vanaverbeke}, {Watanabe},
  {Winn}, \& {Zechmeister}}]{2023A&A...677A.182M}
{Murgas}, F., {Castro-Gonz{\'a}lez}, A., {Pall{\'e}}, E., {et~al.} 2023, \aap,
  677, A182

\bibitem[{{Nagel} {et~al.}(2023){Nagel}, {Czesla}, {Kaminski}, {Zechmeister},
  {Tal-Or}, {Schmitt}, {Reiners}, {Quirrenbach}, {Garc{\'\i}a L{\'o}pez},
  {Caballero}, {Ribas}, {Amado}, {B{\'e}jar}, {Cort{\'e}s-Contreras},
  {Dreizler}, {Hatzes}, {Henning}, {Jeffers}, {K{\"u}rster}, {Lafarga},
  {L{\'o}pez-Puertas}, {Montes}, {Morales}, {Pedraz}, \&
  {Schweitzer}}]{Nagel23}
{Nagel}, E., {Czesla}, S., {Kaminski}, A., {et~al.} 2023, \aap, 680, A73

\bibitem[{Nakajima {et~al.}(1992)Nakajima, Hayashi, \& Abe}]{Nakajima1992}
Nakajima, S., Hayashi, Y.-Y., \& Abe, Y. 1992, Journal of the Atmospheric
  Sciences, 49, 2256

\bibitem[{{Narita} {et~al.}(2019){Narita}, {Fukui}, {Kusakabe}, {Watanabe},
  {Palle}, {Parviainen}, {Monta{\~n}{\'e}s-Rodr{\'\i}guez}, {Murgas},
  {Monelli}, {Aguiar}, {Perez Prieto}, {Oscoz}, {de Leon}, {Mori}, {Tamura},
  {Yamamuro}, {B{\'e}jar}, {Crouzet}, {Hidalgo}, {Klagyivik}, {Luque}, \&
  {Nishiumi}}]{Narita2019}
{Narita}, N., {Fukui}, A., {Kusakabe}, N., {et~al.} 2019, Journal of
  Astronomical Telescopes, Instruments, and Systems, 5, 015001

\bibitem[{{Ogilvie}(2014)}]{Ogilvie14}
{Ogilvie}, G.~I. 2014, \araa, 52, 171

\bibitem[{{Ohno} \& {Tanaka}(2021)}]{ohno2021}
{Ohno}, K. \& {Tanaka}, Y.~A. 2021, \apj, 920, 124

\bibitem[{{Orell-Miquel} {et~al.}(2023){Orell-Miquel}, {Nowak}, {Murgas},
  {Palle}, {Morello}, {Luque}, {Badenas-Agusti}, {Ribas}, {Lafarga},
  {Espinoza}, {Morales}, {Zechmeister}, {Alqasim}, {Cochran}, {Gandolfi},
  {Goffo}, {Kab{\'a}th}, {Korth}, {Lam}, {Livingston}, {Muresan}, {Persson}, \&
  {Van Eylen}}]{orell-miquel2023}
{Orell-Miquel}, J., {Nowak}, G., {Murgas}, F., {et~al.} 2023, \aap, 669, A40

\bibitem[{{Osborne} {et~al.}(2024){Osborne}, {Van Eylen}, {Goffo}, {Gandolfi},
  {Nowak}, {Persson}, {Livingston}, {Weeks}, {Pall{\'e}}, {Luque}, {Hellier},
  {Carleo}, {Redfield}, {Hirano}, {Garbaccio Gili}, {Alarcon}, {Barrag{\'a}n},
  {Casasayas-Barris}, {D{\'\i}az}, {Esposito}, {Knudstrup}, {Jenkins},
  {Murgas}, {Orell-Miquel}, {Rodler}, {Serrano}, {Stangret}, {Albrecht},
  {Alqasim}, {Cochran}, {Deeg}, {Fridlund}, {Hatzes}, {Korth}, \&
  {Lam}}]{2024MNRAS.527.11138}
{Osborne}, H.~L.~M., {Van Eylen}, V., {Goffo}, E., {et~al.} 2024, \mnras, 527,
  11138

\bibitem[{{Owen} \& {Wu}(2017)}]{2017ApJ...847...29O}
{Owen}, J.~E. \& {Wu}, Y. 2017, \apj, 847, 29

\bibitem[{{Palle} {et~al.}(2023){Palle}, {Orell-Miquel}, {Brady}, {Bean},
  {Hatzes}, {Morello}, {Morales}, {Murgas}, {Molaverdikhani}, {Parviainen},
  {Sanz-Forcada}, {B{\'e}jar}, {Caballero}, {Sreenivas}, {Schlecker}, {Ribas},
  {Perdelwitz}, {Tal-Or}, {P{\'e}rez-Torres}, {Luque}, {Dreizler},
  {Fuhrmeister}, {Aceituno}, {Amado}, {Anglada-Escud{\'e}}, {Caldwell},
  {Charbonneau}, {Cifuentes}, {de Leon}, {Collins}, {Dufoer}, {Espinoza},
  {Essack}, {Fukui}, {Chew}, {G{\'o}mez-Mu{\~n}oz}, {Henning}, {Herrero},
  {Jeffers}, {Jenkins}, {Kaminski}, {Kasper}, {Kunimoto}, {Latham},
  {Lillo-Box}, {L{\'o}pez-Gonz{\'a}lez}, {Montes}, {Mori}, {Narita},
  {Quirrenbach}, {Pedraz}, {Reiners}, {Rodr{\'\i}guez},
  {Rodr{\'\i}guez-L{\'o}pez}, {Sabin}, {Schanche}, {Schwarz}, {Schweitzer},
  {Seifahrt}, {Stefansson}, {Sturmer}, {Trifonov}, {Vanaverbeke}, {Wells},
  {Zapatero-Osorio}, \& {Zechmeister}}]{2023A&A...678A..80P}
{Palle}, E., {Orell-Miquel}, J., {Brady}, M., {et~al.} 2023, \aap, 678, A80

\bibitem[{Parviainen(2015)}]{Parviainen2015}
Parviainen, H. 2015, Monthly Notices of the Royal Astronomical Society, 450,
  3233

\bibitem[{{Parviainen} \& {Aigrain}(2015)}]{Parviainen2015b}
{Parviainen}, H. \& {Aigrain}, S. 2015, \mnras, 453, 3821

\bibitem[{{Parviainen} {et~al.}(2019){Parviainen}, {Tingley}, {Deeg}, {Palle},
  {Alonso}, {Montanes Rodriguez}, {Murgas}, {Narita}, {Fukui}, {Watanabe},
  {Kusakabe}, {Tamura}, {Nishiumi}, {Prieto-Arranz}, {Klagyivik}, {B{\'e}jar},
  {Crouzet}, {Mori}, {Hidalgo Soto}, {Casasayas Barris}, \&
  {Luque}}]{Parviainen2019}
{Parviainen}, H., {Tingley}, B., {Deeg}, H.~J., {et~al.} 2019, \aap, 630, A89

\bibitem[{{Pecaut} \& {Mamajek}(2013)}]{2013ApJS..208....9P}
{Pecaut}, M.~J. \& {Mamajek}, E.~E. 2013, \apjs, 208, 9

\bibitem[{{P{\'e}rez-Torres} {et~al.}(2021){P{\'e}rez-Torres}, {G{\'o}mez},
  {Ortiz}, {Leto}, {Anglada}, {G{\'o}mez}, {Rodr{\'\i}guez}, {Trigilio},
  {Amado}, {Alberdi}, {Anglada-Escud{\'e}}, {Osorio}, {Umana}, {Berdi{\~n}as},
  {L{\'o}pez-Gonz{\'a}lez}, {Morales}, {Rodr{\'\i}guez-L{\'o}pez}, \&
  {Chibueze}}]{PerezTorres2021}
{P{\'e}rez-Torres}, M., {G{\'o}mez}, J.~F., {Ortiz}, J.~L., {et~al.} 2021,
  \aap, 645, A77

\bibitem[{{Pluriel} {et~al.}(2019){Pluriel}, {Marcq}, \& {Turbet}}]{Pluriel19}
{Pluriel}, W., {Marcq}, E., \& {Turbet}, M. 2019, \icarus, 317, 583

\bibitem[{{Pollacco} {et~al.}(2006){Pollacco}, {Skillen}, {Collier Cameron},
  {Christian}, {Hellier}, {Irwin}, {Lister}, {Street}, {West}, {Anderson},
  {Clarkson}, {Deeg}, {Enoch}, {Evans}, {Fitzsimmons}, {Haswell}, {Hodgkin},
  {Horne}, {Kane}, {Keenan}, {Maxted}, {Norton}, {Osborne}, {Parley}, {Ryans},
  {Smalley}, {Wheatley}, \& {Wilson}}]{2006PASP..118.1407P}
{Pollacco}, D.~L., {Skillen}, I., {Collier Cameron}, A., {et~al.} 2006, \pasp,
  118, 1407

\bibitem[{Powell {et~al.}(2024)Powell, Feinstein, Lee, Zhang, Tsai, Taylor,
  Kirk, Bell, Barstow, Gao, Bean, Blecic, Chubb, Crossfield, Jordan, Kitzmann,
  Moran, Morello, Moses, Welbanks, Yang, Zhang, Ahrer, Bello-Arufe, Brande,
  Casewell, Crouzet, Cubillos, Demory, Dyrek, Flagg, Hu, Inglis, Jones,
  Kreidberg, L{\'o}pez-Morales, Lagage, Meier~Vald{\'e}s, Miguel, Parmentier,
  Piette, Rackham, Radica, Redfield, Stevenson, Wakeford, Aggarwal, Alam,
  Batalha, Batalha, Benneke, Berta-Thompson, Brady, Caceres, Carter,
  D{\'e}sert, Harrington, Iro, Line, Lothringer, MacDonald, Mancini,
  Molaverdikhani, Mukherjee, Nixon, Oza, Palle, Rustamkulov, Sing, Steinrueck,
  Venot, Wheatley, \& Yurchenko}]{Powell2024}
Powell, D., Feinstein, A.~D., Lee, E. K.~H., {et~al.} 2024, Nature

\bibitem[{{Pozuelos} {et~al.}(2020){Pozuelos}, {Su{\'a}rez}, {de El{\'\i}a},
  {Berdi{\~n}as}, {Bonfanti}, {Dugaro}, {Gillon}, {Jehin}, {G{\"u}nther}, {Van
  Grootel}, {Garcia}, {Thuillier}, {Delrez}, \& {Rod{\'o}n}}]{pozuelos2020}
{Pozuelos}, F.~J., {Su{\'a}rez}, J.~C., {de El{\'\i}a}, G.~C., {et~al.} 2020,
  \aap, 641, A23

\bibitem[{{Pozuelos} {et~al.}(2023){Pozuelos}, {Timmermans}, {Rackham},
  {Garcia}, {Burgasser}, {Kane}, {G{\"u}nther}, {Stassun}, {Van Grootel},
  {D{\'e}vora-Pajares}, {Luque}, {Edwards}, {Niraula}, {Schanche}, {Wells},
  {Ducrot}, {Howell}, {Sebastian}, {Barkaoui}, {Waalkes}, {Cadieux}, {Doyon},
  {Boyle}, {Dietrich}, {Burdanov}, {Delrez}, {Demory}, {de Wit}, {Dransfield},
  {Gillon}, {G{\'o}mez Maqueo Chew}, {Hooton}, {Jehin}, {Murray}, {Pedersen},
  {Queloz}, {Thompson}, {Triaud}, {Z{\'u}{\~n}iga-Fern{\'a}ndez}, {Collins},
  {Fausnaugh}, {Hedges}, {Hesse}, {Jenkins}, {Kunimoto}, {Latham}, {Shporer},
  {Ting}, {Torres}, {Amado}, {Rod{\'o}n}, {Rodr{\'\i}guez-L{\'o}pez},
  {Su{\'a}rez}, {Alonso}, {Benkhaldoun}, {Berta-Thompson}, {Chinchilla},
  {Ghachoui}, {G{\'o}mez-Mu{\~n}oz}, {Rebolo}, {Sabin}, {Schroffenegger},
  {Furlan}, {Gnilka}, {Lester}, {Scott}, {Aganze}, {Gerasimov}, {Hsu},
  {Theissen}, {Apai}, {Chen}, {Gabor}, {Henning}, \& {Mancini}}]{pozuelos2023}
{Pozuelos}, F.~J., {Timmermans}, M., {Rackham}, B.~V., {et~al.} 2023, \aap,
  672, A70

\bibitem[{{Quirrenbach} {et~al.}(2014){Quirrenbach}, {Amado}, {Caballero},
  {Mundt}, {Reiners}, {Ribas}, {Seifert}, {Abril}, {Aceituno},
  {Alonso-Floriano}, {Ammler-von Eiff}, {Antona Jim{\'e}nez},
  {Anwand-Heerwart}, {Azzaro}, {Bauer}, {Barrado}, {Becerril}, {B{\'e}jar},
  {Ben{\'{\i}}tez}, {Berdi{\~n}as}, {C{\'a}rdenas}, {Casal}, {Claret},
  {Colom{\'e}}, {Cort{\'e}s-Contreras}, {Czesla}, {Doellinger}, {Dreizler},
  {Feiz}, {Fern{\'a}ndez}, {Galad{\'{\i}}}, {G{\'a}lvez-Ortiz},
  {Garc{\'{\i}}a-Piquer}, {Garc{\'{\i}}a-Vargas}, {Garrido}, {Gesa}, {G{\'o}mez
  Galera}, {Gonz{\'a}lez {\'A}lvarez}, {Gonz{\'a}lez Hern{\'a}ndez},
  {Gr{\"o}zinger}, {Gu{\`a}rdia}, {Guenther}, {de Guindos},
  {Guti{\'e}rrez-Soto}, {Hagen}, {Hatzes}, {Hauschildt}, {Helmling}, {Henning},
  {Hermann}, {Hern{\'a}ndez Casta{\~n}o}, {Herrero}, {Hidalgo}, {Holgado},
  {Huber}, {Huber}, {Jeffers}, {Joergens}, {de Juan}, {Kehr}, {Klein},
  {K{\"u}rster}, {Lamert}, {Lalitha}, {Laun}, {Lemke}, {Lenzen}, {L{\'o}pez del
  Fresno}, {L{\'o}pez Mart{\'{\i}}}, {L{\'o}pez-Santiago}, {Mall}, {Mandel},
  {Mart{\'{\i}}n}, {Mart{\'{\i}}n-Ruiz}, {Mart{\'{\i}}nez-Rodr{\'{\i}}guez},
  {Marvin}, {Mathar}, {Mirabet}, {Montes}, {Morales Mu{\~n}oz}, {Moya},
  {Naranjo}, {Ofir}, {Oreiro}, {Pall{\'e}}, {Panduro}, {Passegger},
  {P{\'e}rez-Calpena}, {P{\'e}rez Medialdea}, {Perger}, {Pluto}, {Ram{\'o}n},
  {Rebolo}, {Redondo}, {Reffert}, {Reinhardt}, {Rhode}, {Rix}, {Rodler},
  {Rodr{\'{\i}}guez}, {Rodr{\'{\i}}guez-L{\'o}pez},
  {Rodr{\'{\i}}guez-P{\'e}rez}, {Rohloff}, {Rosich}, {S{\'a}nchez-Blanco},
  {S{\'a}nchez Carrasco}, {Sanz-Forcada}, {Sarmiento}, {Sch{\"a}fer},
  {Schiller}, {Schmidt}, {Schmitt}, {Solano}, {Stahl}, {Storz}, {St{\"u}rmer},
  {Su{\'a}rez}, {Ulbrich}, {Veredas}, {Wagner}, {Winkler}, {Zapatero Osorio},
  {Zechmeister}, {Abell{\'a}n de Paco}, {Anglada-Escud{\'e}}, {del Burgo},
  {Klutsch}, {Lizon}, {L{\'o}pez-Morales}, {Morales}, {Perryman}, {Tulloch}, \&
  {Xu}}]{CARMENES}
{Quirrenbach}, A., {Amado}, P.~J., {Caballero}, J.~A., {et~al.} 2014, in
  \procspie, Vol. 9147, Ground-based and Airborne Instrumentation for Astronomy
  V, 91471F

\bibitem[{{Quirrenbach} {et~al.}(2018){Quirrenbach}, {Amado}, {Ribas},
  {Reiners}, {Caballero}, {Seifert}, {Aceituno}, {Azzaro}, {Baroch}, {Barrado},
  \& et~al.}]{CARMENES18}
{Quirrenbach}, A., {Amado}, P.~J., {Ribas}, I., {et~al.} 2018, in Society of
  Photo-Optical Instrumentation Engineers (SPIE) Conference Series, Vol. 10702,
  Ground-based and Airborne Instrumentation for Astronomy VII, 107020W

\bibitem[{{Ray} \& {Hess}(2008)}]{Ray2008}
{Ray}, L.~C. \& {Hess}, S. 2008, Journal of Geophysical Research (Space
  Physics), 113, A11218

\bibitem[{{Reid} \& {Cruz}(2002)}]{2002AJ....123.2806R}
{Reid}, I.~N. \& {Cruz}, K.~L. 2002, \aj, 123, 2806

\bibitem[{{Reid} {et~al.}(2003){Reid}, {Cruz}, {Allen}, {Mungall}, {Kilkenny},
  {Liebert}, {Hawley}, {Fraser}, {Covey}, \& {Lowrance}}]{2003AJ....126.3007R}
{Reid}, I.~N., {Cruz}, K.~L., {Allen}, P., {et~al.} 2003, \aj, 126, 3007

\bibitem[{{Reiners} {et~al.}(2022){Reiners}, {Shulyak}, {K{\"a}pyl{\"a}},
  {Ribas}, {Nagel}, {Zechmeister}, {Caballero}, {Shan}, {Fuhrmeister},
  {Quirrenbach}, {Amado}, {Montes}, {Jeffers}, {Azzaro}, {B{\'e}jar},
  {Chaturvedi}, {Henning}, {K{\"u}rster}, \& {Pall{\'e}}}]{Reiners2022}
{Reiners}, A., {Shulyak}, D., {K{\"a}pyl{\"a}}, P.~J., {et~al.} 2022, \aap,
  662, A41

\bibitem[{{Ribas} {et~al.}(2023){Ribas}, {Reiners}, {Zechmeister}, {Caballero},
  {Morales}, {Sabotta}, {Baroch}, {Amado}, {Quirrenbach}, {Abril}, {Aceituno},
  {Anglada-Escud{\'e}}, {Azzaro}, {Barrado}, {B{\'e}jar}, {Ben{\'\i}tez de
  Haro}, {Bergond}, {Bluhm}, {Calvo Ortega}, {Cardona Guill{\'e}n},
  {Chaturvedi}, {Cifuentes}, {Colom{\'e}}, {Cont}, {Cort{\'e}s-Contreras},
  {Czesla}, {D{\'\i}ez-Alonso}, {Dreizler}, {Duque-Arribas}, {Espinoza},
  {Fern{\'a}ndez}, {Fuhrmeister}, {Galad{\'\i}-Enr{\'\i}quez},
  {Garc{\'\i}a-L{\'o}pez}, {Gonz{\'a}lez-{\'A}lvarez}, {Gonz{\'a}lez
  Hern{\'a}ndez}, {Guenther}, {de Guindos}, {Hatzes}, {Henning}, {Herrero},
  {Hintz}, {Huelmo}, {Jeffers}, {Johnson}, {de Juan}, {Kaminski}, {Kemmer},
  {Khaimova}, {Khalafinejad}, {Kossakowski}, {K{\"u}rster}, {Labarga},
  {Lafarga}, {Lalitha}, {Lamp{\'o}n}, {Lillo-Box}, {Lodieu}, {L{\'o}pez
  Gonz{\'a}lez}, {L{\'o}pez-Puertas}, {Luque}, {Mag{\'a}n}, {Mancini},
  {Marfil}, {Mart{\'\i}n}, {Mart{\'\i}n-Ruiz}, {Molaverdikhani}, {Montes},
  {Nagel}, {Nortmann}, {Nowak}, {Pall{\'e}}, {Passegger}, {Pavlov}, {Pedraz},
  {Perdelwitz}, {Perger}, {Ram{\'o}n-Ballesta}, {Reffert}, {Revilla},
  {Rodr{\'\i}guez}, {Rodr{\'\i}guez-L{\'o}pez}, {Sadegi}, {S{\'a}nchez
  Carrasco}, {S{\'a}nchez-L{\'o}pez}, {Sanz-Forcada}, {Sch{\"a}fer},
  {Schlecker}, {Schmitt}, {Sch{\"o}fer}, {Schweitzer}, {Seifert}, {Shan},
  {Skrzypinski}, {Solano}, {Stahl}, {Stangret}, {Stock}, {St{\"u}rmer},
  {Tabernero}, {Tal-Or}, {Trifonov}, {Vanaverbeke}, {Yan}, \& {Zapatero
  Osorio}}]{2023A&A...670A.139R}
{Ribas}, I., {Reiners}, A., {Zechmeister}, M., {et~al.} 2023, \aap, 670, A139

\bibitem[{{Ricker} {et~al.}(2015){Ricker}, {Winn}, {Vanderspek}, {Latham},
  {Bakos}, {Bean}, {Berta-Thompson}, {Brown}, {Buchhave}, {Butler}, {Butler},
  {Chaplin}, {Charbonneau}, {Christensen-Dalsgaard}, {Clampin}, {Deming},
  {Doty}, {De Lee}, {Dressing}, {Dunham}, {Endl}, {Fressin}, {Ge}, {Henning},
  {Holman}, {Howard}, {Ida}, {Jenkins}, {Jernigan}, {Johnson}, {Kaltenegger},
  {Kawai}, {Kjeldsen}, {Laughlin}, {Levine}, {Lin}, {Lissauer}, {MacQueen},
  {Marcy}, {McCullough}, {Morton}, {Narita}, {Paegert}, {Palle}, {Pepe},
  {Pepper}, {Quirrenbach}, {Rinehart}, {Sasselov}, {Sato}, {Seager},
  {Sozzetti}, {Stassun}, {Sullivan}, {Szentgyorgyi}, {Torres}, {Udry}, \&
  {Villasenor}}]{Ricker2015}
{Ricker}, G.~R., {Winn}, J.~N., {Vanderspek}, R., {et~al.} 2015, Journal of
  Astronomical Telescopes, Instruments, and Systems, 1, 014003

\bibitem[{{Sabotta} {et~al.}(2021){Sabotta}, {Schlecker}, {Chaturvedi},
  {Guenther}, {Mu{\~n}oz Rodr{\'\i}guez}, {Mu{\~n}oz S{\'a}nchez}, {Caballero},
  {Shan}, {Reffert}, {Ribas}, {Reiners}, {Hatzes}, {Amado}, {Klahr}, {Morales},
  {Quirrenbach}, {Henning}, {Dreizler}, {Pall{\'e}}, {Perger}, {Azzaro},
  {Jeffers}, {Kaminski}, {K{\"u}rster}, {Lafarga}, {Montes}, {Passegger}, \&
  {Zechmeister}}]{2021A&A...653A.114S}
{Sabotta}, S., {Schlecker}, M., {Chaturvedi}, P., {et~al.} 2021, \aap, 653,
  A114

\bibitem[{{Saur} {et~al.}(2013){Saur}, {Grambusch}, {Duling}, {Neubauer}, \&
  {Simon}}]{Saur2013}
{Saur}, J., {Grambusch}, T., {Duling}, S., {Neubauer}, F.~M., \& {Simon}, S.
  2013, \aap, 552, A119

\bibitem[{{Schanche} {et~al.}(2022){Schanche}, {Pozuelos}, {G{\"u}nther},
  {Wells}, {Burgasser}, {Chinchilla}, {Delrez}, {Ducrot}, {Garcia}, {G{\'o}mez
  Maqueo Chew}, {Jofr{\'e}}, {Rackham}, {Sebastian}, {Stassun}, {Stern},
  {Timmermans}, {Barkaoui}, {Belinski}, {Benkhaldoun}, {Benz}, {Bieryla},
  {Bouchy}, {Burdanov}, {Charbonneau}, {Christiansen}, {Collins}, {Demory},
  {D{\'e}vora-Pajares}, {de Wit}, {Dragomir}, {Dransfield}, {Furlan},
  {Ghachoui}, {Gillon}, {Gnilka}, {G{\'o}mez-Mu{\~n}oz}, {Guerrero}, {Harris},
  {Heng}, {Henze}, {Hesse}, {Howell}, {Jehin}, {Jenkins}, {Jensen}, {Kunimoto},
  {Latham}, {Lester}, {McLeod}, {Mireles}, {Murray}, {Niraula}, {Pedersen},
  {Queloz}, {Quintana}, {Ricker}, {Rudat}, {Sabin}, {Safonov},
  {Schroffenegger}, {Scott}, {Seager}, {Strakhov}, {Triaud}, {Vanderspek},
  {Vezie}, \& {Winn}}]{schanche2022}
{Schanche}, N., {Pozuelos}, F.~J., {G{\"u}nther}, M.~N., {et~al.} 2022, \aap,
  657, A45

\bibitem[{Schlecker {et~al.}(2023)Schlecker, Apai, Lichtenberg, Bergsten,
  Salvador, \& {Hardegree-Ullman}}]{Schlecker2023}
Schlecker, M., Apai, D., Lichtenberg, T., {et~al.} 2023, Bioverse: {{The
  Habitable Zone Inner Edge Discontinuity}} as an {{Imprint}} of {{Runaway
  Greenhouse Climates}} on {{Exoplanet Demographics}}

\bibitem[{Schlecker {et~al.}(2021{\natexlab{a}})Schlecker, Mordasini,
  Emsenhuber, Klahr, Henning, Burn, Alibert, \& Benz}]{Schlecker2021}
Schlecker, M., Mordasini, C., Emsenhuber, A., {et~al.} 2021{\natexlab{a}},
  Astronomy \& Astrophysics, 656, A71

\bibitem[{Schlecker {et~al.}(2021{\natexlab{b}})Schlecker, Pham, Burn, Alibert,
  Mordasini, Emsenhuber, Klahr, Henning, \& Mishra}]{Schlecker2021b}
Schlecker, M., Pham, D., Burn, R., {et~al.} 2021{\natexlab{b}}, Astronomy \&
  Astrophysics, 656, A73

\bibitem[{{Schneider} {et~al.}(2016){Schneider}, {Greco}, {Cushing},
  {Kirkpatrick}, {Mainzer}, {Gelino}, {Fajardo-Acosta}, \&
  {Bauer}}]{2016ApJ...817..112S}
{Schneider}, A.~C., {Greco}, J., {Cushing}, M.~C., {et~al.} 2016, \apj, 817,
  112

\bibitem[{{Sch{\"o}fer} {et~al.}(2019){Sch{\"o}fer}, {Jeffers}, {Reiners},
  {Shulyak}, {Fuhrmeister}, {Johnson}, {Zechmeister}, {Ribas}, {Quirrenbach},
  {Amado}, {Caballero}, {Anglada-Escud{\'e}}, {Bauer}, {B{\'e}jar},
  {Cort{\'e}s-Contreras}, {Dreizler}, {Guenther}, {Kaminski}, {K{\"u}rster},
  {Lafarga}, {Montes}, {Morales}, {Pedraz}, \& {Tal-Or}}]{2019A&A...623A..44S}
{Sch{\"o}fer}, P., {Jeffers}, S.~V., {Reiners}, A., {et~al.} 2019, \aap, 623,
  A44

\bibitem[{{Schweitzer} {et~al.}(2019){Schweitzer}, {Passegger}, {Cifuentes},
  {B{\'e}jar}, {Cort{\'e}s-Contreras}, {Caballero}, {del Burgo}, {Czesla},
  {K{\"u}rster}, {Montes}, {Zapatero Osorio}, {Ribas}, {Reiners},
  {Quirrenbach}, {Amado}, {Aceituno}, {Anglada-Escud{\'e}}, {Bauer},
  {Dreizler}, {Jeffers}, {Guenther}, {Henning}, {Kaminski}, {Lafarga},
  {Marfil}, {Morales}, {Schmitt}, {Seifert}, {Solano}, {Tabernero}, \&
  {Zechmeister}}]{schweitzer2019}
{Schweitzer}, A., {Passegger}, V.~M., {Cifuentes}, C., {et~al.} 2019, \aap,
  625, A68

\bibitem[{{Shappee} {et~al.}(2014){Shappee}, {Prieto}, {Grupe}, {Kochanek},
  {Stanek}, {De Rosa}, {Mathur}, {Zu}, {Peterson}, {Pogge}, {Komossa}, {Im},
  {Jencson}, {Holoien}, {Basu}, {Beacom}, {Szczygie{\l}}, {Brimacombe},
  {Adams}, {Campillay}, {Choi}, {Contreras}, {Dietrich}, {Dubberley},
  {Elphick}, {Foale}, {Giustini}, {Gonzalez}, {Hawkins}, {Howell}, {Hsiao},
  {Koss}, {Leighly}, {Morrell}, {Mudd}, {Mullins}, {Nugent}, {Parrent},
  {Phillips}, {Pojmanski}, {Rosing}, {Ross}, {Sand}, {Terndrup}, {Valenti},
  {Walker}, \& {Yoon}}]{Shappee2014}
{Shappee}, B.~J., {Prieto}, J.~L., {Grupe}, D., {et~al.} 2014, \apj, 788, 48

\bibitem[{{Skrutskie} {et~al.}(2006){Skrutskie}, {Cutri}, {Stiening},
  {Weinberg}, {Schneider}, {Carpenter}, {Beichman}, {Capps}, {Chester},
  {Elias}, {Huchra}, {Liebert}, {Lonsdale}, {Monet}, {Price}, {Seitzer},
  {Jarrett}, {Kirkpatrick}, {Gizis}, {Howard}, {Evans}, {Fowler}, {Fullmer},
  {Hurt}, {Light}, {Kopan}, {Marsh}, {McCallon}, {Tam}, {Van Dyk}, \&
  {Wheelock}}]{skrutskie2006}
{Skrutskie}, M.~F., {Cutri}, R.~M., {Stiening}, R., {et~al.} 2006, \aj, 131,
  1163

\bibitem[{{Smith} {et~al.}(2012){Smith}, {Stumpe}, {Van Cleve}, {Jenkins},
  {Barclay}, {Fanelli}, {Girouard}, {Kolodziejczak}, {McCauliff}, {Morris}, \&
  {Twicken}}]{2012Smith}
{Smith}, J.~C., {Stumpe}, M.~C., {Van Cleve}, J.~E., {et~al.} 2012, \pasp, 124,
  1000

\bibitem[{{Southworth}(2011)}]{2011Southworth}
{Southworth}, J. 2011, \mnras, 417, 2166

\bibitem[{{Stassun} {et~al.}(2019){Stassun}, {Oelkers}, {Paegert}, {Torres},
  {Pepper}, {De Lee}, {Collins}, {Latham}, {Muirhead}, {Chittidi},
  {Rojas-Ayala}, {Fleming}, {Rose}, {Tenenbaum}, {Ting}, {Kane}, {Barclay},
  {Bean}, {Brassuer}, {Charbonneau}, {Ge}, {Lissauer}, {Mann}, {McLean},
  {Mullally}, {Narita}, {Plavchan}, {Ricker}, {Sasselov}, {Seager}, {Sharma},
  {Shiao}, {Sozzetti}, {Stello}, {Vanderspek}, {Wallace}, \&
  {Winn}}]{Stassun2019}
{Stassun}, K.~G., {Oelkers}, R.~J., {Paegert}, M., {et~al.} 2019, \aj, 158, 138

\bibitem[{{Stumpe} {et~al.}(2014){Stumpe}, {Smith}, {Catanzarite}, {Van Cleve},
  {Jenkins}, {Twicken}, \& {Girouard}}]{2014PASP..126..100S}
{Stumpe}, M.~C., {Smith}, J.~C., {Catanzarite}, J.~H., {et~al.} 2014, \pasp,
  126, 100

\bibitem[{{Stumpe} {et~al.}(2012){Stumpe}, {Smith}, {Van Cleve}, {Twicken},
  {Barclay}, {Fanelli}, {Girouard}, {Jenkins}, {Kolodziejczak}, {McCauliff}, \&
  {Morris}}]{2012Stumpe}
{Stumpe}, M.~C., {Smith}, J.~C., {Van Cleve}, J.~E., {et~al.} 2012, \pasp, 124,
  985

\bibitem[{{Tabernero} {et~al.}(2022){Tabernero}, {Marfil}, {Montes}, \&
  {Gonz{\'a}lez Hern{\'a}ndez}}]{tab22}
{Tabernero}, H.~M., {Marfil}, E., {Montes}, D., \& {Gonz{\'a}lez
  Hern{\'a}ndez}, J.~I. 2022, \aap, 657, A66

\bibitem[{{Thompson} {et~al.}(2023){Thompson}, {Biagini}, {Cracchiolo},
  {Petralia}, {Changeat}, {Saba}, {Morello}, {Morvan}, {Micela}, \&
  {Tinetti}}]{Thompson2023}
{Thompson}, A., {Biagini}, A., {Cracchiolo}, G., {et~al.} 2023, arXiv e-prints,
  arXiv:2302.04574

\bibitem[{{Thorngren} {et~al.}(2016){Thorngren}, {Fortney}, {Murray-Clay}, \&
  {Lopez}}]{thorngren2016}
{Thorngren}, D.~P., {Fortney}, J.~J., {Murray-Clay}, R.~A., \& {Lopez}, E.~D.
  2016, \apj, 831, 64

\bibitem[{{Tittemore} \& {Wisdom}(1990)}]{TittemoreWisdom90}
{Tittemore}, W.~C. \& {Wisdom}, J. 1990, \icarus, 85, 394

\bibitem[{Turbet {et~al.}(2020)Turbet, Bolmont, Ehrenreich, Gratier, Leconte,
  Selsis, Hara, \& Lovis}]{Turbet2020}
Turbet, M., Bolmont, E., Ehrenreich, D., {et~al.} 2020, Astronomy \&
  Astrophysics, Volume 638, id.A41, 638, A41

\bibitem[{Turbet {et~al.}(2019)Turbet, Ehrenreich, Lovis, Bolmont, \&
  Fauchez}]{Turbet2019}
Turbet, M., Ehrenreich, D., Lovis, C., Bolmont, E., \& Fauchez, T. 2019,
  Astronomy \& Astrophysics, 628, A12

\bibitem[{{Turnpenney} {et~al.}(2018){Turnpenney}, {Nichols}, {Wynn}, \&
  {Burleigh}}]{Turnpenney2018}
{Turnpenney}, S., {Nichols}, J.~D., {Wynn}, G.~A., \& {Burleigh}, M.~R. 2018,
  \apj, 854, 72

\bibitem[{{van Altena} {et~al.}(1995){van Altena}, {Lee}, \&
  {Hoffleit}}]{1995gcts.book.....V}
{van Altena}, W.~F., {Lee}, J.~T., \& {Hoffleit}, E.~D. 1995, {The general
  catalogue of trigonometric [stellar] parallaxes}

\bibitem[{Venturini {et~al.}(2020)Venturini, Guilera, Haldemann, Ronco, \&
  Mordasini}]{Venturini2020}
Venturini, J., Guilera, O.~M., Haldemann, J., Ronco, M.~P., \& Mordasini, C.
  2020, Astronomy \& Astrophysics, 643, L1

\bibitem[{{Wells} {et~al.}(2021){Wells}, {Rackham}, {Schanche}, {Petrucci},
  {G{\'o}mez Maqueo Chew}, {Demory}, {Burgasser}, {Burn}, {Pozuelos},
  {G{\"u}nther}, {Sabin}, {Schroffenegger}, {G{\'o}mez-Mu{\~n}oz}, {Stassun},
  {Van Grootel}, {Howell}, {Sebastian}, {Triaud}, {Apai}, {Plauchu-Frayn},
  {Guerrero}, {Guill{\'e}n}, {Landa}, {Melgoza}, {Montalvo}, {Serrano},
  {Riesgo}, {Barkaoui}, {Bixel}, {Burdanov}, {Chen}, {Chinchilla}, {Collins},
  {Daylan}, {de Wit}, {Delrez}, {D{\'e}vora-Pajares}, {Dietrich}, {Dransfield},
  {Ducrot}, {Fausnaugh}, {Furlan}, {Gabor}, {Gan}, {Garcia}, {Ghachoui},
  {Giacalone}, {Gibbs}, {Gillon}, {Gnilka}, {Gore}, {Guerrero}, {Henning},
  {Hesse}, {Jehin}, {Jenkins}, {Latham}, {Lester}, {McCormac}, {Murray},
  {Niraula}, {Pedersen}, {Queloz}, {Ricker}, {Rodriguez}, {Schroeder},
  {Schwarz}, {Scott}, {Seager}, {Theissen}, {Thompson}, {Timmermans},
  {Twicken}, \& {Winn}}]{wells2021}
{Wells}, R.~D., {Rackham}, B.~V., {Schanche}, N., {et~al.} 2021, \aap, 653, A97

\bibitem[{Wright {et~al.}(2018)Wright, Newton, Williams, Drake, \&
  Yadav}]{Wright2018}
Wright, N.~J., Newton, E.~R., Williams, P. K.~G., Drake, J.~J., \& Yadav, R.~K.
  2018, Monthly Notices of the Royal Astronomical Society, 479, 2351

\bibitem[{{Yu} {et~al.}(2021){Yu}, {He}, {Zhang}, {H{\"o}rst}, {Dymont},
  {McGuiggan}, {Moses}, {Lewis}, {Fortney}, {Gao}, {Kempton}, {Moran},
  {Morley}, {Powell}, {Valenti}, \& {Vuitton}}]{yu2021}
{Yu}, X., {He}, C., {Zhang}, X., {et~al.} 2021, Nature Astronomy, 5, 822

\bibitem[{{Zarka}(2007)}]{Zarka2007}
{Zarka}, P. 2007, \planss, 55, 598

\bibitem[{{Zechmeister} {et~al.}(2014){Zechmeister}, {Anglada-Escud{\'e}}, \&
  {Reiners}}]{2014A&A...561A..59Z}
{Zechmeister}, M., {Anglada-Escud{\'e}}, G., \& {Reiners}, A. 2014, \aap, 561,
  A59

\bibitem[{Zechmeister \& Kürster(2009)}]{2009Zechmeister}
Zechmeister, M. \& Kürster, M. 2009, Astronomy \& Astrophysics, 496, 577–584

\bibitem[{{Zechmeister} {et~al.}(2018){Zechmeister}, {Reiners}, {Amado},
  {Azzaro}, {Bauer}, {B{\'e}jar}, {Caballero}, {Guenther}, {Hagen}, {Jeffers},
  {Kaminski}, {K{\"u}rster}, {Launhardt}, {Montes}, {Morales}, {Quirrenbach},
  {Reffert}, {Ribas}, {Seifert}, {Tal-Or}, \& {Wolthoff}}]{2018A&A...609A..12Z}
{Zechmeister}, M., {Reiners}, A., {Amado}, P.~J., {et~al.} 2018, \aap, 609, A12

\bibitem[{{Zeng} {et~al.}(2019){Zeng}, {Jacobsen}, {Sasselov}, {Petaev},
  {Vanderburg}, {Lopez-Morales}, {Perez-Mercader}, {Mattsson}, {Li}, {Heising},
  {Bonomo}, {Damasso}, {Berger}, {Cao}, {Levi}, \&
  {Wordsworth}}]{Zeng2019PNAS..116.9723Z}
{Zeng}, L., {Jacobsen}, S.~B., {Sasselov}, D.~D., {et~al.} 2019, Proceedings of
  the National Academy of Science, 116, 9723

\bibitem[{{Zeng} {et~al.}(2016){Zeng}, {Sasselov}, \& {Jacobsen}}]{2016Zeng}
{Zeng}, L., {Sasselov}, D.~D., \& {Jacobsen}, S.~B. 2016, \apj, 819, 127

\end{thebibliography}

\begin{appendix}

\section{Radial velocity data}

\begin{table}[ht!] 
\caption{RV data from CARMENES VIS.}
\label{tab:CARMV_RV_VIS} 
\begin{adjustbox}{width=0.6\linewidth,center}
\begin{tabular}{crr}        
\hline\hline                 
\noalign{\smallskip}
Time & \multicolumn{1}{c}{RV} & \multicolumn{1}{c}{$\sigma$} \\ 

[BJD$_{\mathrm{TDB}}$] & [m\,s$^{-1}$] & [m\,s$^{-1}$] \\ 
\hline                 
\noalign{\smallskip}
 2459720.6490 &            1.90 &           3.99\\
 2459726.5607 &           --6.93 &           2.02\\
 2459730.5751 &           --1.11 &           2.20\\
 2459736.5664 &            2.74 &           1.81\\
 2459738.6403 &           --3.77 &           1.97\\
 2459740.6490 &           --4.64 &           2.32\\
 2459748.5283 &           --0.98 &           3.35\\
 2459768.4734 &           --1.05 &           2.57\\
 2459770.4685 &           --0.97 &           1.94\\
 2459772.4574 &           --4.01 &           2.31\\
 2459776.4443 &            0.86 &           2.09\\
 2459778.4424 &           --3.92 &           2.31\\
 2459780.6099 &            1.97 &           2.25\\
 2459784.6203 &            0.36 &           3.21\\
 2459786.5898 &           --2.26 &           3.63\\
 2459788.6297 &            4.97 &           1.98\\
 2459796.5221 &            4.49 &           2.64\\
 2459812.4809 &           --2.08 &           2.23\\
 2459812.5016 &            2.40 &           1.98\\
 2459813.4895 &            3.57 &           3.72\\
 2459813.5105 &           --2.29 &           3.12\\
 2459814.4412 &           --6.17 &           2.09\\
 2459826.4745 &            3.32 &           2.61\\
 2459831.3528 &           --3.44 &           2.42\\
 2460007.6296 &           --0.85 &           2.37\\
 2460015.6770 &           --4.80 &           1.72\\
 2460033.6556 &            3.51 &           3.38\\
 2460037.5939 &            4.58 &           3.48\\
 2460041.6186 &            2.86 &           1.92\\
 2460057.6177 &            6.53 &           2.07\\
 2460060.6283 &            1.38 &           1.98\\
 2460066.6401 &            2.64 &           2.30\\

\noalign{\smallskip}
\hline
\noalign{\smallskip}
\end{tabular}
\end{adjustbox}
\end{table}

\section{Photometric fit with unocculted star spots}
\label{app:starspot_transpec}

We fitted the color dependent transit depths that appear when considering the measurements from TESS and all MuSCAT2 filters, including $g'$ and $r'$, following the method described in Sect.\,\ref{tronly}. For simplicity, we fixed the stellar parameters of the quiet photosphere to those of TOI-4438, as reported in Table \ref{tab:stellar_parameters}. The free parameters in our fit were the geometric radius ratio $R_b/R_{\star}$, the star spot photospheric parameters $T_{\mathrm{spot}}$ and $\log{g_{\mathrm{spot}}}$, and filling factor $f_{\mathrm{spot}}$. We run \texttt{emcee} \citep{emcee} with 16 walkers and 30\,000 iterations to sample the posterior distribution, then applying a conservative burn-in of 10\,000 iterations, which is more than 20 times the maximum autocorrelation length. For each parameters set from the chains, we also computed the apparent transit depth in each passband using the formulae from \cite{Ballerini2012}. 

Fig. \ref{fig:contam_transpec} compares the measured transit depths in each passband with those derived from the fit with unocculted star spots and the posterior distribution for the model parameters. While the unocculted star spots may explain the larger transit depths in $g'$ and $r'$, they would also cause a difference between the $i'$ and $z_s$ that is not seen in the data. As a compromise, the model tends to underestimate the $g'$ and $r'$ transit depths to keep a small difference between $i'$ and $z_s$. The modelled transit depths are consistent within 1$\sigma$ with the measured ones, owing to the large error bars in $g'$ and $r'$. The spot temperature contrast is -490$_{-640}^{+380}$\,K and the filling factor is 0.44$_{-0.17}^{+0.24}$. In this scenario, the planet-to-star radius ratio is $R_b/R_{\star} = 0.055_{-0.011}^{+0.004}$, that is significantly smaller and with order-of-magnitude larger error bars than those obtained discarding the $g'$ and $r'$ bands and neglecting star spots.

If we accepted this solution, the planetary radius would be $R_b = 2.24_{-0.55}^{+0.28} \, R_{\oplus}$. This slightly smaller radius would not significantly change the predicted composition, which is volatile-rich. With a smaller TSM of 96, TOI-4438\,b would remain a top target for atmospheric characterization with \textit{JWST}.

\begin{figure*}[!t]
\centering
\includegraphics[width=0.49\textwidth]{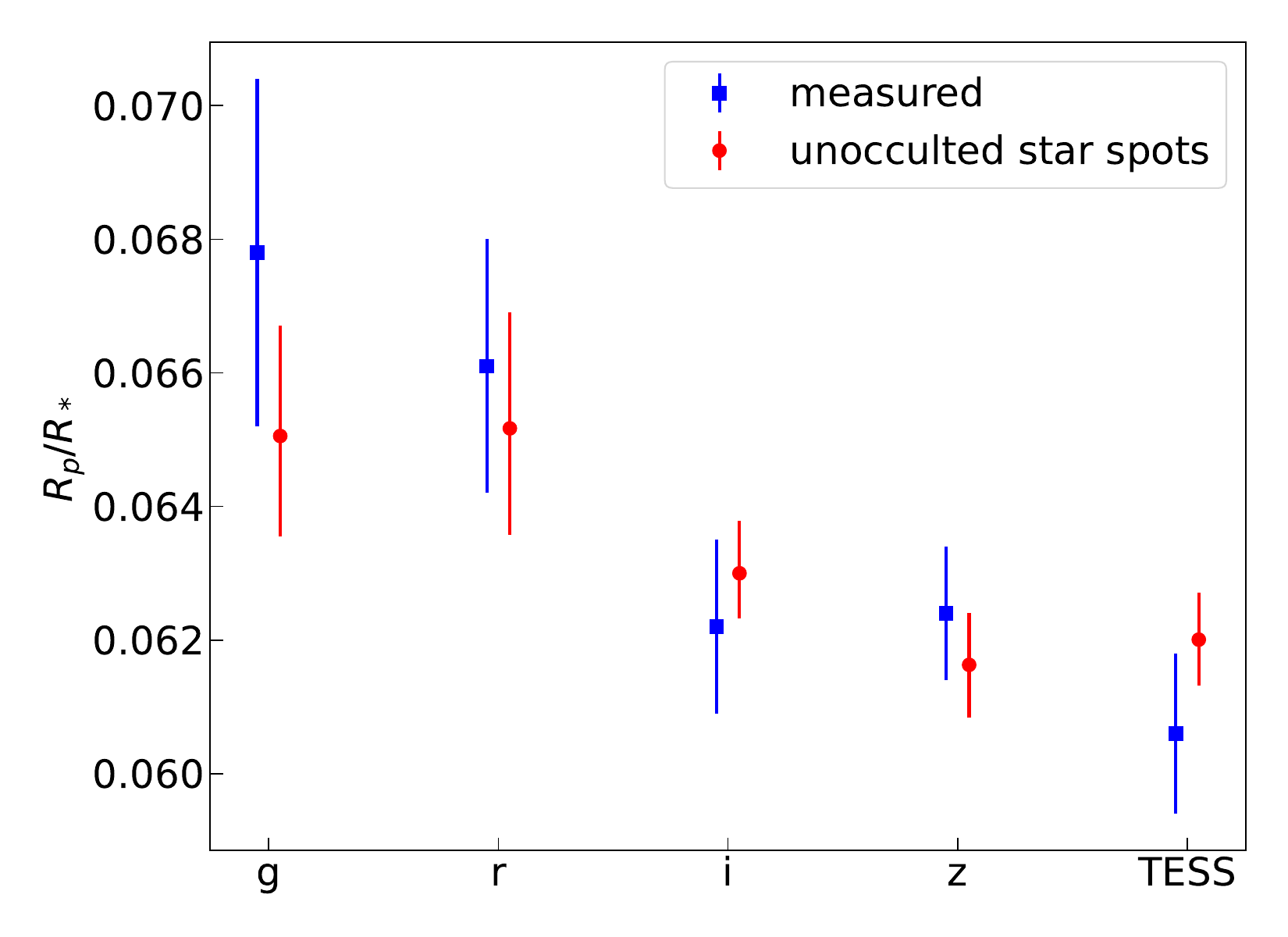}
\includegraphics[width=0.49\textwidth]{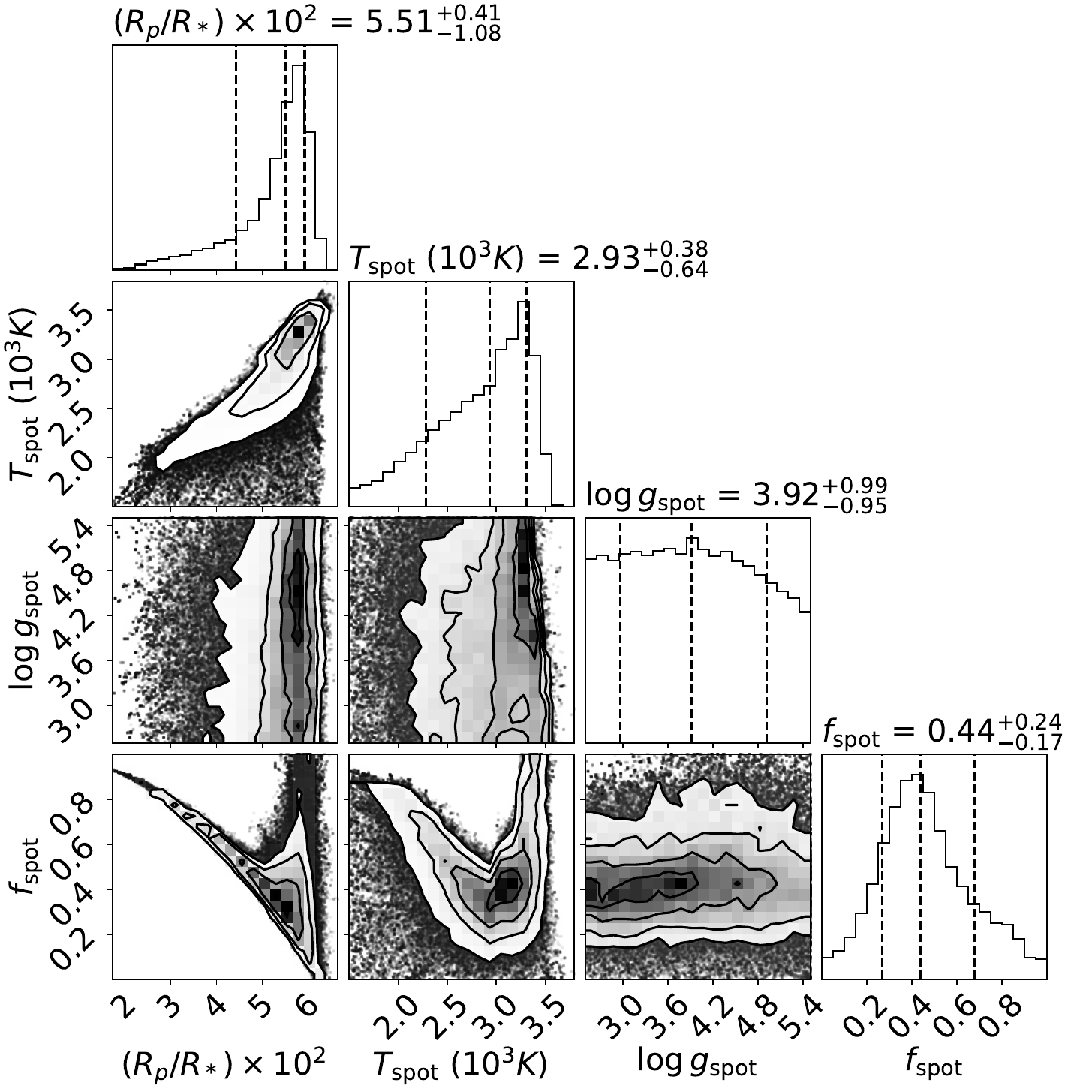}
\caption{Results of the photometric fit with unocculted star spots. \textit{Left}: Individual transit depths derived from photometric observations (blue squares) and best-fit model assuming that the color-dependence is due to unocculted star spots. \textit{Right}: corner plot with the posterior distribution of the model parameters, generated with the \texttt{corner} package \citep{corner}. \label{fig:contam_transpec}
        }
\end{figure*}

\end{appendix}

\end{document}